\UseRawInputEncoding
\documentclass[11pt,aps,amssymb,prd,a4paper,nofootinbib]{revtex4-1}
\pdfoutput=1
\usepackage{float}
\usepackage{cancel}
\usepackage{amsfonts}
\usepackage{amsmath}
\usepackage{amssymb}
\usepackage{color}
\usepackage{graphicx}
\usepackage{epsfig}
\usepackage{bm}
\def\d{{\rm d}}
{}
{}
  \newcommand{\sym}{$\mathcal {Z}_2 \times \mathcal {Z}_2^ \prime $ }
  \def\beq{\begin{eqnarray}}
\def\eeq{\end{eqnarray}}

\def \EMET{E{\!\!\!/}_T}
\def \G {\mathcal{G}}

\newcommand{\eVq}  {\text{eV}^2}
\begin{document}

\title{Magnetic Moments of Leptons, Charged Lepton Flavor Violations and Dark Matter Phenomenology of
  a Minimal Radiative Dirac Neutrino Mass Model}

\author{Bibhabasu De}
\email{bibhabasu.d@iopb.res.in}
\affiliation{ Institute of Physics, Sachivalaya Marg, Bhubaneswar, 751 005, India}
\affiliation{Homi Bhabha National Institute, Training School Complex, Anushakti Nagar, Mumbai 400 094, India}
\author{Debottam Das}
\email{debottam@iopb.res.in}
\affiliation{ Institute of Physics, Sachivalaya Marg, Bhubaneswar, 751 005, India}
\affiliation{Homi Bhabha National Institute, Training School Complex, Anushakti Nagar, Mumbai 400 094, India}
\author{Manimala Mitra}
\email{manimala@iopb.res.in}
\affiliation{ Institute of Physics, Sachivalaya Marg, Bhubaneswar, 751 005, India}
\affiliation{Homi Bhabha National Institute, Training School Complex, Anushakti Nagar, Mumbai 400 094, India}
\author{Nirakar Sahoo}
\email{nirakar.pintu.sahoo@gmail.com}
\affiliation{Center of Excellence, High Energy and Condensed Matter Physics, Department of Physics, Utkal University, Bhubaneswar- 751004, India}

\preprint{}

\begin{abstract} 
  In a simple extension of the standard model (SM), a pair of vector like lepton
  doublets~($L_1$ and $L_2$) and a $SU(2)_L$ scalar doublet ($\eta$) have been introduced to help in accommodating
  the discrepancy in
  determination of the anomalous magnetic moments of the light leptons, namely,
  $e$ and $\mu$. Moreover, to make
  our scenario friendly to a Dirac like neutrino and also for a consistent dark matter phenomenology,
  we specifically add a singlet scalar ($S$) and a singlet fermion ($\psi$) in the set-up. However, the singlet states
also induce a meaningful contribution in other charged lepton 
processes.
  A discrete symmetry \sym has been imposed under which all the SM particles are even while
  the new particles may be assumed to
  have odd charges. In a bottom-up approach, with a minimal particle content,
  we systematically explore the available parameter space
  in terms of couplings and masses of the new particles. Here a number of observables
  associated with the SM leptons have been considered, e.g.,
  masses and mixings of neutrinos, $(g-2)$ anomalies of $e$, $\mu$,
  charged lepton flavor
  violating (cLFV) observables and the dark matter (DM) phenomenology of a singlet-doublet dark matter.
  Neutrinos, promoted
  as the Dirac type states, acquire mass at
  one loop level after the discrete $\mathcal{Z}_2^\prime$ symmetry gets softly broken, while the
  unbroken $\mathcal{Z}_2$ keeps the dark matter stable. The mixing between the singlet
  $\psi$ and the doublet vector lepton can be constrained to satisfy the
  electroweak precision observables
  and the 
  spin independent~(SI) direct detection (DD) cross section of the dark matter. In this analysis,
  potentially important LHC bounds have also been discussed.
\end{abstract}

\maketitle

\section{Introduction}
The standard model (SM) of particle physics has been quite successful in explaining the interactions of elementary particles~\cite{pdg}. 
The recent
discovery of a Higgs boson with a mass of 125 GeV at the Large Hadron Collider
\cite{Aad:2012tfa,Chatrchyan:2012xdj} has been showing good agreements with the SM
expectations \cite{Khachatryan:2016vau,Aad:2019mbh}. However, there exists a few
experimental and theoretical issues, which cannot be explained
in the SM paradigm, 
thus, hint towards a more complete theory --- beyond SM physics (BSM) at the TeV scale.
Among these signatures, the precise measurement of the dark matter (DM) abundance and the
non-zero values of the neutrino masses and mixings are
of particular interests to us. Here, one may broadly recall the issues at hand.
$(i)$ Assuming the origin of the dark matter is related to a new kind of particle, the
simplest and most compelling candidate has been considered as a 
weakly interacting massive particle (WIMP). The 
experiments like PLANCK \cite{Aghanim:2015xee} and WMAP \cite{Hinshaw:2012aka} have already
provided precise measurements
of DM relic density. WIMPs with masses
$\sim$ 1 TeV can lead to the correct relic density through its annihilations to SM
particles. Such a mass scale can be probed at the high-energy collider experiments like
the LHC and also at the dark matter direct detection experiments.
$(ii)$ Non-zero neutrino masses and substantial
mixing among the three light neutrino states
require specific extensions of the SM. In the simplest case, one may introduce
right handed neutrinos $\nu_R$ and assumes a
Dirac mass term $m_D$ for the neutrinos. 
But, then the neutrino Yukawa couplings are assumed to be $ \simeq 10^{-11}$
to generate a neutrino mass $\sim 0.1$~eV. However, being a singlet under the
SM gauge group, $\nu_R$ can also accommodate a 
large Majorana mass parameter $M$ which violates the lepton number by 2 units. Such a
mass term
leads to an attractive possibility --- called ``seesaw mechanism" where the light neutrinos
$\nu_L$ obtain an
effective small Majorana mass 
term \cite{Minkowski:1977,Yanagida:1979as,GellMann:1980vs}. 
The tinyness of neutrino masses can be explained naturally without requiring a tiny
Yukawa coupling. Though seesaw mechanism is more favoured, 
experimentally, the searches to probe the Majorana nature of neutrinos through neutrinoless
double beta decay experiments
have not yet lead to any conclusive evidence. So the simple idea of considering neutrino as
a Dirac particle has been still quite popular. 

There have already been many
proposals which may incorporate new particles and appropriate mixings, thus,
explains the masses for neutrinos and the dark matter abundance in the extensions of
the SM. However, it is more natural to consider that there
exists a tie-up between these two important pieces which may lead to a somewhat
economical and an attractive extension of the SM to deal with. Driven
by the same pursuit, here we will also
furnish a connection between these two important issues assuming neutrino as a Dirac
particle. Interestingly and more importantly, we will observe that the precision
observables like
anomalous magnetic moments of $\mu$ ($a_\mu = \frac{(g-2)_\mu}{2}$) and $e$
($a_e = \frac{(g-2)_e}{2}$) can be accommodated along with the experimental constraints
related to the charged lepton flavor violations. 

The idea of neutrino as a Dirac particle has revived in the recent past when the main
theoretical objection of having a very tiny tree level 
Yukawa coupling has been addressed through the radiative generation of neutrino
masses \cite{Kanemura:2011jj,Kanemura:2011mw,Farzan:2012sa,Ma:2016mwh,Kanemura:2016ixx,
  Yao:2017vtm,Ma:2017kgb,Singirala:2017see,Yao:2017vtm,Calle:2018ovc,Yao:2018ekp,
  Borah:2018gjk,Ma:2019yfo,Dasgupta:2019rmf,Das:2017ski} (for a review see \cite{Cai:2017jrq}). 
The main idea is simple and can be realized through an additional \sym symmetry in the
SM set up :
(1) one may
assume a discrete symmetry (here $\mathcal{Z}_2^\prime$) to forbid a
tree-level Dirac neutrino masses.
This symmetry would be finally broken
softly to generate a tiny neutrino mass through a radiative mechanism. (2) 
New fields may be introduced; in the simplest case, an inert scalar doublet
$(\eta^+\quad\eta^0)^T$ and neutral singlet fermions can be considered (see below)
to radiatively
induce neutrino masses in the loop. The new fields may transform odd under
the another $\mathcal{Z}_2$ symmetry
to prohibit their couplings with the other SM fermions, thus, offers an interesting
possibility where
the lightest state~(a new 
$\mathcal{Z}_2$ odd fermion or a neutral scalar) may become the cold dark matter
(CDM) of the universe.
This class of models where neutrinos acquire
masses through dark matter in the loop,
thereby connects the two important BSM aspects of the particle physics has been
dubbed as ``scotogenic'' model~\cite{Ma:2006km}. In the original idea, the neutrino
masses
have been assumed to be of Majorana type. However, one may employ the same idea to
generate the masses for the neutrinos radiatively considering them as the Dirac particle,
if a symmetry like global
or gauge $U(1)$ symmetry is assumed to prohibit the Majorana mass term in the Lagrangian~\cite{Ma:2016mwh}.

Assuming the lepton number as a good symmetry of
the Lagrangian at the backdrop of our work we start our discussion with a simple
realization\footnote{for some recent works, see
  \cite{Calle:2018ovc,Calle:2019mxn,Jana:2019mgj,Nanda:2019nqy,Leite:2020wjl,
    Escribano:2020iqq,Guo:2020qin}.}.
We consider new leptons/scalars at the electroweak (EW) scale
in addition to the usual right handed neutrinos $\nu_R$: 
singlet Dirac fermion(s) ($N$), two
scalars --- an inert scalar doublet $\eta$ and a real singlet scalar $S$ in the
particle content of the SM. 
A perturbative value of the coupling 
$Y_{\ell} \bar{N}_R \ell \eta$ ($\ell\in e,\mu,\tau$)
may help to realize tiny nature of the neutrino Yukawa couplings radiatively,  
if the other interaction terms $Y_R \bar{N}S \nu _R$ and $\mu^\prime \eta^\dagger H S$ are included in the interacting Lagrangian.
Here the last term $\mu^\prime$ can be regarded as the soft symmetry breaking parameter.
As in the case of a ``scotogenic" model,
with proper charge assignments under \sym symmetry, Dirac masses for
the SM neutrinos, proportional to the soft breaking scale $\mu^\prime$,
would be generated radiatively through a $N-\eta-S$ loop.
 Similarly,
 observable abundance of the dark matter $N$ would follow naturally. 
However, this simple model fall short to account for the BSM
contributions in the measurement of the anomalous magnetic moment of muon
$a_\mu$ \cite{Ma:2001mr}, though can help to acclimatize the measurement
of $a_e$. Primarily, the non-SM contribution,
controlled by the $N-\eta^\pm$ loop,
comes out to be negative while the discrepancy
in the muon anomalous magnetic moment $\Delta a_\mu$ requires a
positive boost, thus, disfavours this simple set-up (for a
generic discussion on the new physics contributions
to $a_\mu$, see~\cite{Queiroz:2014zfa,Lindner:2016bgg,Kowalska:2017iqv}). 

We next consider the vector like (VL) leptons in place of singlet Dirac like
state $N$ in the
SM set-up, without changing the basic structure of the model. For a color singlet
VL, left and
right handed components transform similarly under the SM gauge symmetries,
and one may observe that $\Delta a_\mu$ can be accommodated through the mixings
with the SM leptons
\cite{Dermisek:2013gta,Poh:2017tfo,Barman:2018jhz,Chen:2019nud,deJesus:2020upp}. 
However, addressing $a_e$ along with $a_\mu$ invites a further modification. 
We, thus introduce a pair of $SU(2)$ vector like leptons $L_1\equiv (L_1^0\quad L_1^-)^T$,
$L_2\equiv (L_2^0\quad L_2^-)^T$ with same hypercharge
(but charged differently under \sym symmetry) which can be found to be suitable
when coupled to
new states; e.g., an inert Higgs doublet $\eta$, a real singlet scalar $S$ and a SM
singlet fermion
$\psi$
in the present context. As in the previous case, $S$ acts to realize the soft breaking
of $\mathcal{Z}_2^\prime$ symmetry; thus to generate Dirac masses for the
neutrinos while $\psi$ has
its role 
to realize the proper dark matter abundance. 
In fact, $L_1$ and $\psi$ can enjoy the same transformation properties
under the \sym symmetry; 
thus the neutral $L^0_1$ and $\psi$ can mix to provide with a suitable
candidate for dark matter~($\chi_0$) and to accommodate 
$(g-2)_e$ anomaly through neutral fermions and charged scalars
running in the loop. The charged components of the new leptons help
to explain the
other anomaly in $(g-2)_\mu$. Naturally, neutrino mass as well as
cLFV processes
receive contributions from
the diagrams that involve both of the VL leptons in the loops.
In \cite{Chen:2019nud}, authors find that a vector like lepton doublet in 
presence of a right handed neutrino and inert Higgs doublet may indeed be helpful in 
explaining $(g-2)_\mu$ deviation while the tiny Majorana masses for
the neutrinos can also be generated in a ``scotogenic" model. Here
we will try to find if the both anomalous $(g-2)_\mu$ and $(g-2)_e$ can be accommodated
with the said particle contents while neutrinos acquire Dirac masses
through dark matter $\chi_{0}$ in the loop.  

In dark matter phenomenology, singlet-doublet DM $\chi_{0}$ comprised of $L^0_1$ and singlet $\psi$,
could just be able to produce the correct relic abundance  
\cite{Cohen:2011ec,Cheung:2013dua,Vicente:2014wga,Restrepo:2015ura,
  Calibbi:2015nha,Bhattacharya:2015qpa,Yaguna:2015mva,Arcadi:2018pfo,Konar:2020wvl}.
Admitting only VL doublet lepton $L^0_1$, one finds a large DM-nucleon elastic
cross-section
through $Z$
mediated processes, thus has essentially been ruled out by the experiments
such as XENON1T \cite{Aprile:2018dbl} or LUX \cite{Akerib:2017kat}.
As a natural deviation, one finds that a singlet-doublet fermion dark matter,
through its SM singlet component may escape the
stringent direct detection bounds. %
For practical purposes, the
dark matter particle has to be essentially dominated by the singlet component,
while only a very small doublet part can be allowed. For the same reason, we purposefully introduce
$\psi$ in the particle content.

We organise our paper as follows.  In sec~\ref{sec:model}, we explain the
details of our model including the new particles and their charges under the
complete gauge group which would be considered. After electroweak symmetry
breaking (EWSB), our model predicts additional neutral and charged leptons.
Consequently, relevant interactions of the new particles with the
SM particles can be realized.
Theoretical and experimental bounds on their couplings
and masses have been summarized in \ref{sec:bounds}. These include
~$(i)$ anomalous magnetic moments and different charged lepton
flavor violating decays of the SM leptons,~$(ii)$ 
vacuum stability of the tree level scalar potential,~$(iii)$
Electroweak precision observables (EWPO) and $(iv)$ collider physics
constraints. In the results sections, we present
radiative generation of the neutrino masses and mixing angles
in sec~\ref{sec:nu_mass}. As discussed, one of the motivations is
to show that our model can accommodate anomalous magnetic moments of the lighter
charged leptons. We depict the parameter space of our model in sec~\ref{sec:LFV},
where
discrepancies in $a_{\mu/e}$ can simultaneously be satisfied. Subsequently,
we probe our model parameters with different charged lepton flavor violating (cLFV)
observables, namely
$\ell_\alpha\to \ell_\beta \gamma$, $\ell_\alpha \to 3 \ell_\beta$ and flavor violating decays of $Z$ boson. 
DM phenomenology including the
relic
density and the direct detection of a singlet-doublet fermionic DM have been covered
in sec~\ref{sec:DM}. Finally, we
conclude this work in sec~\ref{sec:conc}.

\section{The Model: Relevant Lagrangian and Scalar potential at the tree level} \label{sec:model}
As stated, the proposed model is a simple extension of the Standard Model where we augment 
two scalars, namely a real singlet ($S$) and a $SU(2)_L$ doublet
$\eta \equiv (\eta^+\quad \eta^0)^T$, two vector like lepton doublets
$L_1\equiv (L_1^0\quad L_1^-)^T$, $L_2\equiv (L_2^0\quad L_2^-)^T$,
a singlet fermion $\psi$ and the usual SM singlet right handed neutrinos $\nu_R$. 
All the new states are charged
under an additional \sym symmetry (see Table~\ref{tab:fields}).

The allowed interactions of the new fields and the SM fields can be
read from the following Lagrangian~:
\begin{equation}
\mathcal{L}=\mathcal{L}_{SM} + \mathcal{L}_{new}~,
\end{equation}
where $\mathcal{L}_{new}$, the new physics Lagrangian is given by,
\begin{eqnarray}
\mathcal{L}_{new} = & i\bar{L}_1 \cancel{D} L_1 - M_{L_1} \bar{L}_1 L_1 + i\bar{L}_2 \cancel{D} L_2 - M_{L_2} \bar{L}_2L_2+ i\bar{\psi} \cancel{\partial} \psi - M_{\psi} \bar{\psi}\psi - \nonumber\\
& \Big[Y_{1(1i)} \bar{L}_{1L}\eta\ell_{Ri}  +  {Y_{2(1i)} \bar{L}_{2L}\tilde{\eta}\nu_{Ri} }+ Y_{3(i1)} \bar{\ell}_i {L}_{2} S + Y_{4 (i1)} \bar{\ell}_i \tilde{\eta} \psi + Y_5 \bar{L}_1\tilde{H}\psi + Y_{6(1i)} S\bar{\psi}_L\nu_{Ri} + h.c. \Big]+  \nonumber \\
& + (\partial ^ \mu S)^\dagger (\partial _ \mu S) + (D ^ \mu \eta)^\dagger (D _ \mu \eta) - V(\eta,H,S).
\label{eq:L_new}
\end{eqnarray}
Here, $D_\mu$ is the $SU(2)_L\times U(1)_Y$ covariant derivative and $V(\eta,H,S)$ is
the scalar potential. We define field $\tilde{\Phi}$ as $i\tau_2 \Phi^*$. We are
following the convention $Q_{EM}=T_3+Y$. For clarity, we refrain from explicit showing
of $SU(2)$ contractions. Except for the right handed neutrinos,
single generation of all the other new states would suffice for our purpose
(see Table \ref{tab:fields} for details). Here we note that, $L_1$ and $L_2$
are assumed to have different charges under \sym symmetry.
Interacting Lagrangian is realized through the new Yukawa couplings
$Y_1 \cdot\cdot\cdot Y_6$ where in the parenthesis, number of the fermion
generations that are involved, 
are presented. All the Yukawa couplings are assumed to be real. 
The new fermion states $L_1$, $L_2$, $\psi$ and also the RH neutrino $\nu_R$ have
one unit of lepton number to preserve the lepton number conservation.
Moreover, in this work, VLs can only couple to the SM leptons through the new scalar
states which do not acquire any vacuum expectation values (VEV); thus the masses and mixings of the SM charged
leptons would remain unaffected. In Eq.~\eqref{eq:L_new}, the interaction between
$L_1$ and SM singlet $\psi$ is felicitated through the SM Higgs $H$ which drives
the DM phenomenology.

Finally, we may express the scalar potential $V(\eta,H,S)$ in Eq.~\eqref{eq:L_new}
\noindent
which adheres the proposed symmetry as follows:
\begin{eqnarray}
V(\eta,H,S) & = & \mu_H^2 H^\dagger H + \mu_\eta ^2 \eta^\dagger \eta + \mu_S^2 S^\dagger S + \lambda_H (H^\dagger H)^2+          
      \lambda_\eta (\eta^\dagger \eta)^2 + \lambda_S (S^\dagger S) ^2 +\lambda_{\eta H} ( \eta^\dagger \eta
       ) (H^\dagger H) \nonumber \\ 
&   + &  \lambda^\prime _{\eta H} ( \eta^\dagger H )   
   (H^\dagger \eta)  + \frac{\lambda^{\prime \prime }_{\eta H}} {2} [( \eta^\dagger H )^2 +  h.c ] + \lambda_{HS} (H^\dagger H) (S^\dagger S) + \lambda_{\eta S} (\eta^\dagger \eta) (S^\dagger S).
   \label{eq:V_scalar}
\end{eqnarray}

\noindent
There can be a few additional terms like 
which are allowed by gauge and Lorentz invariance, but due to the imposed \sym
symmetry these terms transform non-trivially and hence are forbidden (see e.g.,
last four terms in Table \ref{tab:fields}(b)). 
This in turn ensures that the new scalars $S,\eta$ do not acquire any induced VEV.
As usual $\mu_H^2$ can take negative values.
As stated, to generate the mass terms for the SM neutrinos,
$\mathcal{Z}_2^\prime$ symmetry can be broken explicitly by introducing a soft breaking
term
at the scalar potential,
\begin{equation}
  \mathcal{L}_{SB}= \mu^\prime \eta^\dagger H S \,.
  \label{eq:SBT}
\end{equation}
\noindent
Since $\mu^\prime$ breaks the $\mathcal{Z}_2^\prime$, it may be argued to be very small,
thus may be
helpful in fitting neutrino masses. Similarly, $\bar{L}_1 L_2$ can also accommodate a soft beraking term. The mass term can also be
generated at the two loops ($\propto \mu^\prime$) which we assume to be small for further consideration. If
the VL states would be considered to transform identically under $\mathcal{Z}_2^\prime$, then
we will have
a restricted class of the Yukawa terms and consequently accommodating
the anomalous magnetic momemts of $\mu$ and $e$ simultaneously cannot be realized in this proposed model with the given particle content. However, we may
consider a global $U(1)$ symmetry 
(the charge assignments could read as $L_1,~S,~\Psi,~\eta$=1 while $L_2$=-1 with all the SM particles including $\nu_R$
assume zero charges), then our model and its phenomenology would be completely unchanged. Infact,
it will make the dark matter stable thus $\mathcal{Z}_2$ can be assumed to be replaced.

Before discussing the phenomenology, let us briefly outline the role of
different discrete symmetries
in the present analysis. We assume $\mathcal{Z}_2$ to be an exact symmetry
which always ensures that
$(i)$
a tree level Dirac like neutrino mass term, e.g., $\bar{\ell} \tilde{H} \psi$
would be absent and
$(ii)$ $\chi_{0}$, the singlet like admixture of $L_1^0$ and $\psi$, a state odd
under $\mathcal{Z}_2$
may become stable to form the cold dark matter. On the other hand
$\mathcal{Z}_2^\prime$ forbids the usual
tree level Yukawa interaction $\bar{\ell}\tilde{H} \nu_R$, but it needs to be
broken softly to generate
neutrino masses through radiative corrections.
Additionally, there are a few other couplings among the new fields and the
SM fields which fail to
qualify as the valid interactions.
For a better understanding,
we list them in Table~\ref{tab:fields}
along with their transformations under
the proposed symmetry group.
Here $\surd$ and $\times$ refer to the occasions when a particular interaction term turns out to be
even or odd under a symmetry operation respectively.

\begin{table}[ht]
\begin{tabular}{|c|c|c|c|c|}
\hline
Fields   & Generation & $SU(2)_L \times U(1)_Y$ & $\mathcal{Z}_2$ & $\mathcal{Z}_2^\prime$ \\ 
\hline
	$\ell=(\nu_L\quad e_L)^T$  & 3&2, -1/2  & +1  &  +1 \\
		$\ell_R= (e_R,\mu_R,\tau_R)$ & 3&1, -1 & +1 & +1 \\
		$Q_L=(u_L\quad d_L)^T$  & 3&2, 1/6  & +1  &  +1 \\
		$U_R=(u_R,c_R,t_R)$  & 3&1, 2/3  & +1  &  +1 \\
		$D_R=(d_R,s_R,b_R)$  &3& 1, -1/3  & +1  &  +1 \\
		$H = \big(0 \quad \frac{1}{\sqrt{2}}(v+h)\big)^T $     &1& 2, 1/2  & +1  & +1 \\
		$\nu_R$ &3& 1, 0  & +1 & -1 \\
		$\psi$   &1& 1, 0 & -1 & +1 \\
		$L_{1}=(L_1^0 \quad L_1^-)^T$   &1& 2, -1/2 & -1 & +1\\
		$L_{2}=(L_2^0 \quad L_2^-)^T$   &1& 2, -1/2 & -1 & -1\\
		$\eta=\big(\eta^+ \quad \eta^0\big)^T$  &1& 2, 1/2 & -1 & +1 \\
 		$S$      &1&1, 0  & -1  &-1\\
\hline
\end{tabular}
\hspace{1 cm}
\begin{tabular}{|c|c|c|c|c|}
				\hline 
				Forbidden terms	& $SU(2)_L$ & $U(1)_Y$ & $\mathcal{Z}_2$ & $\mathcal{Z}_2^\prime$ \\ 	\hline 
			$\bar{\ell}  H \psi_R~(\bar{\ell}  \tilde{H} \psi_R)$	& $\surd$ &$\times~(\surd)$  &$\times$  & $\surd$ \\
			$\bar{\ell} H \nu_R~(\bar{\ell} \tilde{H} \nu_R)  $	& $\surd$ & $\times~(\surd)$  & $\surd$ & $\times$ \\
			$\bar{\ell}\eta \nu_R~(\bar{\ell}\tilde{\eta} \nu_R) $	 & $\surd$ & $\times~(\surd)$ & $\times$& $\times$ \\ 
			
			$\bar{\ell}_R S \psi$	& $\surd$ & $\times$ & $\surd$ & $\times$\\ 
			$\bar{L}_2 H\psi~(\bar{L}_2 \tilde{H}\psi)$ & $\surd$ & $\times~(\surd)$ & $\surd$ & $\times$ \\
			$\bar{\ell}L_1 S$ & $\surd$ & $\surd$ & $\surd$ & $\times$ \\
			$\bar{L}_2\eta\ell_R~(\bar{L}_2\tilde{\eta}\ell_R)$ & $\surd$ & $\surd~(\times)$ & $\surd$ & $\times$\\
			$\bar{L}_1\eta\nu_R~(\bar{L}_1\tilde{\eta}\nu_R)$ & $\surd$ & $\times~(\surd)$ & $\surd$ & $\times$\\ 
			$\bar{L}_1L_2S$ & $\surd$ & $\surd$ & $\times$ & $\surd$\\ 
			\hline \hline
			
			$\lambda_{\eta HSS}(\eta^{\dagger}H)(S^{\dagger}S)$	& $\surd$ & $\surd$ & $\times$ & $\surd$ \\ 
			
			$\lambda(\eta^{\dagger}H)$	& $\surd$ & $\surd$ & $\times$ & $\surd$ \\ 
			
			$\lambda_{SSS}(S^{\dagger}S)S$	& $\surd$ & $\surd$ & $\times$ & $\times$ \\ 
			
			$\lambda_3 S$  & $\surd$ & $\surd$ & $\times$ & $\times$ \\
				\hline 
				
			\end{tabular} \\
			\begin{center}
			\hspace{1 cm}(a)\hspace{9.6 cm}(b)
			\end{center}
			
                        \caption{ (a)~Particles and their transformations under
                          $ SU(2)_L \times U(1)_Y \times \mathcal{Z}_2 \times \mathcal{Z}_2^\prime $.
                          (b)~Forbidden interaction terms and their transformations
                          under different gauge and discrete symmetries.}
\label{tab:fields}
\end{table}
{\bf Possible completion of the model at the GUT scale :}
Here we discuss a possibility to embed our low energy model to a larger gauge group e.g., $SO(10)$. Specific gauge breaking chains may include, e.g.,
left-right (LR) symmetric phase at the intermediate scale \cite{Pati:1974yy, Mohapatra:1974gc, Senjanovic:1975rk,
  Aulakh:1997ba, Duka:1999uc},
\begin{align}
	SO(10) \underset{M_{GUT}}{\to}
	SU(3)_C \times SU(2)_L \times SU(2)_R \times U(1)_{B-L} \underset{M_{LR}}{\to}
	\text{SM}
\end{align}
with $M_{GUT}$ denoting the breaking scale of $SO(10)$ gauge group
which is subsequently broken to the SM at $M_{LR} < M_{GUT}$.
There are a few reasons for considering the LR models:
(i) the particle content contains automatically the right-handed neutrino, (ii)
a
TeV scale LR symmetric intermediate phase
may be obtained within a class of renormalizable $SO(10)$ GUTs with a perfect gauge coupling unification \cite{Arbelaez:2013nga}. Here one has to account for
a few copies of one or two types of extra fields; e.g., additioanl triplet and/or doublet scalars under $SU(2)_R$. However, for different
possibitites, we refer the reader to Ref.~\cite{Arbelaez:2013nga}. Of course, the
new scalars can effect the low energy phenomenology e.g., $(g-2)_\mu$ through
a gauge invariant interaction at the LR scale.
The matter content
of the model along with their possible transformations at each intermediate stage 
is given in table~\ref{table_gut}. Here $Q$, $Q^c$, $L$ and $L^c$ (we follow the notation in \cite{Aulakh:1997ba})
are the quark and lepton families with the addition of (three) right-handed neutrino(s)
$\nu_R$. The SM Higgs and the inert doublet can be included as bidoublets under
$SU(2)_L \times SU(2)_R$. More than a single bidoublet is required for a correct Yukawa Lagrangian at the
low scale \cite{Arbelaez:2013nga}. Similarly, transformations of the VL states
$L_{1,2}$ and the SM singlet states are noted. The
electric charges of particles are calculated through the
eigenvalues of the left $(T_{3L})$ and right $(T_{3R})$
generators of the $SU(2)_L$ and $SU(2)_R$ groups, respectively, as
$Q_{EM} = T_{3L} + T_{3R} + (B-L)/2$.
\begin{table}[ht]
\begin{tabular}{|c|c|c|c|c|}
\hline
Fields   & Generation &$ 3_c2_L2_R1_{B-L}$&  $SO(10)$ \\ 
\hline
		$Q$  & 3&(3, 2, 1,  1/3 )   &  16 \\
		$Q^c $& 3& ({$\bar 3$}, 1, 2, -1/3)& ${16}$\\
		$L $ &3& (1, 2, 1, -1)   & 16  \\
  $L^c $ &3& (1, 1, 2, 1)  & ${16}$  \\
    $\Phi$&2&(1, 2, 2, 0)&10 \\
${L}_{1,2}$   &2& (1, 2, 1, -1)& 16\\
$\hat{L}_{1,2}$   &2& (1, 2, 1, $ 1$)
& $\overline {16}$\\
$\psi$  &1 & (1, 1, 2, -1) & $\overline{16}$\\
$\hat{\psi}$   &1 & (1, 1, 2, 1) & 16 \\
 $S$      &1& (1, 1, 1, 0)  &1\\

\hline
\end{tabular}
\caption{One of the Possible completion of the particle content under $SO(10)$.}
\label{table_gut}
\end{table}
The index $c$ refers
to the equivalent SM field which transforms under $SU(2)_R$. All the interaction
terms in Eq.~\eqref{eq:L_new} can now be cast under the enlagred gauge symmetry.
For example, 
$Y_{1(1i)} \bar{L}_{1L}\eta\ell_{Ri}$ can be cast as $L^T_1 \Phi L^c$ which, under,
$SO(10)$ goes as ${\bf{16}} \times {\bf{10}} \times {\bf{16}}$. Similarly,
$Y_{6(1i)} S\bar{\psi}_L\nu_{Ri}$ can be cast as $S  \psi L^c$ which, under,
$SO(10)$ goes as ${\bf{1}} \times {\bf{\overline{16}}} \times {\bf{16}}$.
Though the particle contents can easily
be accommodated under a unified gauge group, one has to admit a minor change, e.g.,
$\psi$ in Eq.~\eqref{eq:L_new} should refer to
the neutral component of  $SU(2)_R$  doublet in Table~\ref{table_gut}.
Alternatively, one may also consider
the symmetry breaking chain as $SO(10) \rightarrow SU(5) \times U(1)_X \rightarrow SU(3)_C \times SU(2)_L \times U(1)_Y \times U(1)_X$
which was earlier considered in Ref.~\cite{Antusch:2017tud,Chen:2019nud}.

{\bf Mixings and couplings of the VL states with bosons and fermions:} As can be seen from Eq.~\eqref{eq:L_new} that lepton phenomenology is primarily governed by the
new Yukawa couplings $Y_i (i=1...6)$. 
Apparently, the first four couplings are more important for the phenomenology in the lepton sector, 
while $Y_5$ primarily controls the DM physics. The Yukawa interactions involving the singlet states
may 
contribute to neutrino masses and also the dark matter relic abundance. 
For a generic study, we
keep all the couplings with $Y_i (i=1...6)$ in the flavor space.

Let us first start our discussion with the interactions mediated by $Y_5$
in Eq. \eqref{eq:L_new}. The Yukawa interaction, $Y_5 \bar{L}_1\tilde{H}\psi$
generates a mass matrix,
\begin{equation}
\mathcal{M}=\left(\begin{array}{c c} M_{\psi} & \frac{Y_5v}{\sqrt{2}}\\
\frac{Y_5v}{\sqrt{2}} & M_{L_1}
\end{array}\right)~,
\end{equation} 
in the basis of $(\psi,\, L_1^0)$. We can rotate this to the mass basis with the help of $(2\times 2)$
orthogonal matrix, such that $\mathcal{M}_D=U^\dagger \mathcal{M}U$, where,
\begin{align}
U=\left(\begin{array}{c c} {\rm cos}\theta & -{\rm sin}\theta\\
{\rm sin}\theta & {\rm cos}\theta
\end{array}\right).
\label{eq:U_matrix}
\end{align}
The two mass eigenstates can be defined as,
\begin{align}
\chi_0&={\rm cos}\theta\,\psi+{\rm sin}\theta\, L_1^0,\\
\chi_1&=-{\rm sin}\theta\,\psi+{\rm cos}\theta\, L_1^0,
\end{align}
with the masses are given by,
\begin{align}
M_{\chi_0}&=M_{L_1}{\rm sin}^2\theta+M_\psi{\rm cos}^2\theta+\frac{Y_5v}{\sqrt{2}}{\rm sin}2\theta~,
\label{eq:Mx2}\\
  M_{\chi_1}&=M_{L_1}{\rm cos}^2\theta+M_\psi{\rm sin}^2\theta-\frac{Y_5v}{\sqrt{2}}{\rm sin}2\theta.
  \label{eq:Mx1}
\end{align}
The mixing angle is defined as, 
\begin{align}
{\rm tan}2\theta=\frac{\sqrt{2}(Y_5v)}{M_\psi-M_{L_1}}~.
\end{align}
If we assume a small mixing angle i.e., $\theta<<1$ then $\chi_1$ is
dominantly doublet-like with a small admixture of singlet $\psi$,
while $\chi_0$ is mostly singlet-like. Since the direct detection
experiments require DM to be mostly singlet dominated, we can propose $\chi_0$
as the DM candidate with the condition that $M_{\chi_0}<M_{\chi_1}$, which is further ensured by the
choice $M_\psi<M_{L_1}$. 
The Yukawa coupling $Y_5$, now being a dependent parameter,
can be expressed in terms of $M_{\chi_1}$, $M_{\chi_0}$ and $\theta$
through the following relation,
\begin{align}
Y_5=-\frac{(M_{\chi_1}-M_{\chi_0})\,{\rm sin}2\theta}{v\sqrt{2}}.
\label{eq:Y5}
\end{align}

At this point we can recast the Yukawa terms in Eq.~\eqref{eq:L_new} in this new basis of
$(\chi_0,\, \chi_1)$ as:
\begin{align}
\label{eq:massbasis1} 
  \mathcal L_{new} & \supset Y_{1(1i)}\,
  \Big[{\rm cos}\theta\bar{\chi}_1\eta^+\ell_{Ri}
    +\,{\rm sin}\theta \bar{\chi}_0\eta^+\ell_{Ri} +
    \bar{L^-_1} \eta^0 \ell_{Ri} \Big] +
  Y_{2(1i)}\,\Big[\bar{L}^0_2 \eta^0
    -\bar{L}^-_2 \eta^- \Big]{\nu}_{Ri}\\ \nonumber&+Y_{3(i1)}\,
  \Big[\bar{\nu}_{li} L^0_{2}S +
      \bar{e}_{li} L^-_{2}S\Big]+Y_{4(i1)}\,\Big[\bar{\nu_{li}}
      \eta^0 \Big({\rm cos}\theta \chi_0 - {\rm sin}\theta \chi_1  \Big) -
      \bar{e}_{li} \eta^- \Big( {\rm cos}\theta \chi_0-{\rm sin}\theta \chi_1 \Big) \Big]\\
  \nonumber&+Y_{6(1i)}\,\Big[{\rm cos}\theta S\bar{\chi}_0\nu_{Ri} - \,{\rm sin}\theta
    S\bar{\chi}_1\nu_{Ri}\Big]+\,h.c.~+\frac{Y_5}{\sqrt{2}}
  h\Big[(\bar{\chi}_0\chi_0-\bar{\chi}_1\chi_1)\,{\rm sin}2\theta+(\bar{\chi}_1\chi_0
    +\bar{\chi}_0\chi_1)\,{\rm cos}2\theta\Big]~.   
\end{align}

All the Yukawa couplings appearing above need
to satisfy a generic condition $|Y| \le 4\pi$ so to remain perturbative at the TeV scale.
Similarly, the terms appearing in the covariant derivative can be collected to write down the
couplings with the gauge bosons. Using
$D_\mu= \partial_\mu - i\frac{g}{\cos\theta_W} \Big(T^3-\sin^2\theta_W Q\Big)Z_\mu -ieQA_\mu$,
one finds that,
\begin{align}
\label{eq:massbasis2}
  \mathcal L_{new}  &\supset \frac{g}{\sqrt{2}}\Big[{\rm cos}\theta\bar{\chi}_1\gamma^\mu L_1^-
    +{\rm sin}\theta\bar{\chi}_0\gamma^\mu  L_1^-+
    \bar{L}^0_2 \gamma^\mu L_2^- \Big] W^+_\mu +h.c.\nonumber\\ &+\frac{g}{2\,{\rm cos}\theta_W}
  \Big[{\rm cos}^2\theta\, \bar{\chi}_1\gamma^\mu  \chi_1+{\rm sin}^2\theta\, \bar{\chi}_0\gamma^\mu
    \chi_0+\frac{1}{2}{\rm sin}2\theta\left(\bar{\chi}_1\gamma^\mu \chi_0+\bar{\chi}_0\gamma^\mu
    \chi_1\right) + \bar{L}^0_2\gamma^\mu L^0_2\Big]Z_\mu\nonumber\\
  &{+\frac{g}{\cos\theta_W}\left(-\frac{1}{2}+\sin^2\theta_W\right)\Big[\bar{L}^-_1\gamma^\mu L_1^-
       + \bar{L}^-_2\gamma^\mu L_2^-\Big]Z_\mu-e\Big[\bar{L}^-_1\gamma^\mu L_1^- + \bar{L}^-_2\gamma^\mu
       L_2^-\Big]A_\mu}~.
\end{align}
Note that, all the other terms in Eq.~\eqref{eq:L_new} will not be affected by this basis change.

\section{Bounds related to different experiments and theories}
\label{sec:bounds}
Here we review different bounds related to experimental searches and theories.
We will use the limits in delineating the parameter space consistent with the anomalous magnetic moments of
leptons, charged 
lepton flavor violations and the dark matter abundance.

 \subsection { Anomalous magnetic moment and different LFV decays}
{\bf Bounds on anomalous magnetic moment:} From the first
precision measurement of the magnetic
 dipole moment of the muon $a_\mu$
 at BNL (Brookhaven National Laboratory), the persistent discrepancy in its
 determination compared to its SM prediction has been undoubtedly
 one of the most promising hints towards a new physics signal
 at the TeV scale. 
 The discrepancy can be expressed through
its experimental measurements ($\equiv a^{exp}_\mu$) and the SM prediction 
($ \equiv a^{SM}_\mu$). The difference in the two values can be
seen to be driven by the BSM contributions ($\equiv \Delta a_\mu$).
For the last many years, the experimental data produced a roughly $3.7\sigma$
 deviation from the
 standard
 model (SM) value~\cite{Bennett:2006fi,Keshavarzi:2018mgv,Tanabashi:2018oca,Aoyama:2020ynm}.
For a better understanding of the known physics,
it was imperative
to resolve the tension related to the hadronic vacuum polarization (HVP)
of
$a^{SM}_\mu$~\cite{Borsanyi:2020mff,Passera:2008jk,Davier:2017zfy,Davier:2019can,Colangelo:2020lcg,Crivellin:2020zul,Keshavarzi:2020bfy,Malaescu:2020zuc}
(see also~\cite{Aoyama:2020ynm} and references therein). The tension lies in the fact that a recent lattice-QCD~\cite{Borsanyi:2020mff} estimation of the HVP may bring the SM prediction
of $a_\mu$ into agreement with experiments which
seems to be in contradiction with $e^+e^- \rightarrow $ hadrons
cross section data and global fits to electroweak precision
observables~\cite{Crivellin:2020zul,Malaescu:2020zuc}.
The Fermilab-based Muon g-2
experiment has just reported a new result~\cite{Abi:2021gix,Albahri:2021ixb}
which, if combined with
the BNL result reads $4.2\sigma$
 deviation from the SM value\footnote{Recent measurement at the Fermilab has drawn some interests to our community~\cite{Chakraborti:2021dli,Chakraborti:2021bmv,Arcadi:2021cwg,Criado:2021qpd,Wang:2021bcx, Ibe:2021cvf,Babu:2021jnu,Bai:2021bau,Keung:2021rps,Athron:2021iuf,Aboubrahim:2021rwz,Escribano:2021css, Endo:2021zal,Crivellin:2021rbq,Zhang:2021dgl,Yin:2021mls}.}.
 \begin{equation} 
  \Delta a_\mu = (25.1 \pm 5.9) \times 10^{-10}.
  \label{eq:delamu}
  \end{equation}
Thus, as stated earlier, from Eq.~\eqref{eq:delamu} it is clearly visible that one needs a positive BSM contribution to satisfy the experimental constraint on $\Delta a_\mu$. In the context of $a_e$, 
the experimental value has been updated in 2018 \cite{Aoyama:2017uqe} from
a precision
measurement of the fine-structure constant \cite{Parker_2018} that relies on the
caesium recoil
measurements. This measurement also shows a possible disagreement between the experimental observation and theory prediction, though with a less significance $\sim 3 \sigma$.
\begin{equation} 
  \Delta a_e = -(8.7 \pm 3.6) \times 10^{-13}.
   \label{eq:delae}
\end{equation}
\noindent
More importantly, here the measured value is lower than the corresponding
SM prediction. Following the improved estimates, specially in the evaluation
of $a_e$, attempts have been made to link the both discrepancies with a
common new physics origin
\cite{Giudice:2012ms,Crivellin:2018qmi,Liu:2018xkx,Dutta:2018fge,CarcamoHernandez:2019ydc, Bauer:2019gfk,Cornella:2019uxs, Hiller:2019mou,Dorsner:2020aaz,CarcamoHernandez:2020pxw,Calibbi:2020emz, Botella:2020xzf,Jana:2020pxx,Hati:2020fzp,Dutta:2020scq,Arbelaez:2020rbq,Endo:2019bcj, Badziak:2019gaf}.
Here we note that a very recent determination of the fine structure constant \cite{Morel:2020dww},
obtained from the measurement of the recoil
velocity on rubidium atoms, result into a positive discrepancy of about
$1.6\sigma$. Clearly the discrepancy in the measurement of $a_e$
can only be settled in the future.
This work focuses on caesium recoil
measurements, thus, Eq.~\eqref{eq:delae} in the subsequent sections.

{\bf Bounds on charged lepton flavor violating decays:} Charged lepton flavor violating processes, specifically $\ell_\alpha \to \ell_\beta \gamma$ or
$\ell_\alpha \to 3\ell_\beta $ through photon penguins may be influenced by the same dipole
operators which provides the BSM contributions to $a_{\mu/e}$. Non observations of
any cLFV processes so far, can potentially constrain the new physics parameters.
Currently, the radiative decay of $\ell_\alpha \rightarrow \ell_\beta \gamma$,
specifically $\mu \to e \gamma$, is the leading candidate among the
cLFV observables to put a
stringent constraint on the parameter space. In the future upgrades,
the MEG collaboration can reach a sensitivity of about $6 \times 10^{-14}$
after 3 years of
acquisition time~\cite{Baldini:2013ke}. Similarly, in the near future,
$\mu \rightarrow 3e$ can be probed by the Mu3e experiment~\cite{Blondel:2013ia, Perrevoort:2016nuv} with
a branching ratio sensitivity of $10^{-16}$. A significant improvement
is expected compared to the present limit, set by the SINDRUM experiment~\cite{Bellgardt:1987du}. An impressive improvement on most of the LFV
modes of the
rare $\tau$ decays can be expected from
searches in $B$ factories~\cite{Aushev:2010bq,Bevan:2014iga}.
Table~\ref{tableofconstraints} includes the present and future sensitivities of the
important cLFV processes which would be considered in this work.
\begin{table}[!htb]
	\centering
	\begin{tabular}{|c|c|c|}
		\hline\hline 
		LFV Process  &  Present Bound &  Future Sensitivity   \\ [0.5ex]
		\hline
		$Br(\mu \rightarrow e  \gamma)$ & $ 4.2 \times 10^{-13}$ \cite{TheMEG:2016wtm} & $6 \times 10^{-14}$ \cite{Baldini:2013ke} \\
		$Br(\tau \rightarrow e \gamma)$ & $3.3 \times 10^{-8}$ \cite{Aubert:2009ag} & $ \sim 3 \times 10^{-9}$ \cite{Aushev:2010bq} \\
		$Br(\tau \rightarrow \mu \gamma)$ & $4.4 \times 10^{-8}$ \cite{Aubert:2009ag} & $ \sim 3 \times 10^{-9}$ \cite{Aubert:2009ag}\\
		$Br(\mu \rightarrow 3e)$ & $1.0 \times 10^{-12}$ \cite{Bellgardt:1987du} & $\sim 10^{-16}$ \cite{Blondel:2013ia} \\
		$Br(\tau \rightarrow 3e)$ & $2.7 \times 10^{-8}$ \cite{Hayasaka:2010np} & $\sim 10^{-9}$ \cite{Aushev:2010bq} \\
		$Br(\tau \rightarrow 3\mu)$ & $3.3 \times 10^{-8}$ \cite{Hayasaka:2010np} & $\sim 10^{-9}$ \cite{Aushev:2010bq}\\
		$Br(\tau^- \rightarrow e^- \mu^+ \mu^-)$ & $2.7 \times 10^{-8}$ \cite{Hayasaka:2010np} & $ \sim 10^{-9}$ \cite{Aushev:2010bq}\\
		$Br(\tau^- \rightarrow \mu^- e^+ e^-)$ & $1.8 \times 10^{-8}$ \cite{Hayasaka:2010np} & $ \sim 10^{-9}$ \cite{Aushev:2010bq}\\
		$Br(\tau^- \rightarrow e^+ \mu^- \mu^-)$ & $1.7 \times 10^{-8}$ \cite{Hayasaka:2010np} & $ \sim 10^{-9}$ \cite{Aushev:2010bq}\\
                $Br(\tau^- \rightarrow \mu^+ e^- e^-)$ & $1.5 \times 10^{-8}$ \cite{Hayasaka:2010np} & $ \sim 10^{-9}$ \cite{Aushev:2010bq}\\
		\hline
	\end{tabular}
	\caption{Current Experimental bounds and future sensitivities for the LFV processes.}
	\label{tableofconstraints}
\end{table}
\subsection{Condition of Vacuum stability and the masses of scalars}
The scalar potential must be bounded from below i.e., it does not acquire
negative infinite value in any of the field directions for large field
values. This physical requirement puts certain constraints on the scalar
couplings. Considering the tree level scalar potential, these conditions
are listed below \cite{Kannike:2012pe}.
\begin{align}
\lambda_H, \lambda_\eta, \lambda_S >  0 \,, \nonumber \\ 
\lambda_{ \eta H} + 2 \sqrt{\lambda_\eta \lambda_H} > 0 \,, \nonumber \\ 
\lambda_{ \eta H} + \lambda^\prime _{ \eta H} -
\left| \lambda^{\prime \prime} _{ \eta H} \right| + 2 \sqrt{\lambda_\eta
  \lambda_H} > 0 \,, \nonumber \\
\lambda_{HS} + 2 \sqrt{\lambda_H \lambda_S} >0 \, , \nonumber \\ 
\lambda_{\eta S} + 2 \sqrt{\lambda_\eta \lambda_S} >0.
\end{align}

After the electroweak symmetry breaking only $H$ field gets a VEV, $
v \simeq 246$ GeV. Thus, scalar fields can be expressed as :
\begin{equation}
H = 
\begin{pmatrix}
0 \\
\frac{1}{\sqrt{2}} (v + h)
\end{pmatrix} \, , \hspace{0.5 cm}
\eta =
\begin{pmatrix}
\eta^+ \\
\frac{1}{\sqrt{2}} (\eta_R + i \eta_I)
\end{pmatrix} , \hspace{0.5 cm} 
S= S .
\label{eq:H_eta}
\end{equation}
Substituting $H$ and $\eta$ in
Eq.~\eqref{eq:V_scalar} one finds
\begin{eqnarray}
M_h^2= 2 \lambda_H v^2 \,, \nonumber \\
M_{\eta_R}^2 = \mu_\eta^2 + \frac{1}{2} ( \lambda_{ \eta H}
+ \lambda_{ \eta H}^\prime + \lambda_{ \eta H}
              ^{\prime \prime } ) v^2  \,, \nonumber \\
M_{\eta_I}^2 = \mu_\eta^2 + \frac{1}{2} ( \lambda_{ \eta H}
+ \lambda_{ \eta H}^\prime - \lambda_{ \eta H}
              ^{\prime \prime } ) v^2  \,, \nonumber \\
M_{\eta ^\pm} ^2 = \mu_\eta^2 + \frac{1}{2} \lambda_{\eta H} v^2 \,, \nonumber \\
M_S^2 = \mu_S^2 + \frac{1}{2} \lambda_ {H S} v^2 \,.
\label{eq:mscalar}
\end{eqnarray}
We identify $h$ as our SM like Higgs scalar with mass $M_h \simeq 125$
GeV.  Again for simplicity we assume that the new scalars are heavier to
forbid the constraints coming from the invisible $Z$ and $h$ decays. Similarly the mass splitting between the charged and the neutral
components of the doublet $\eta$ are considered to be negligible,  
i.e., $M_{\eta_I} = M_{\eta_R} = M_{\eta^\pm} \equiv M_\eta$. This is indeed possible (see Eq.~\eqref{eq:mscalar}),
if the couplings $\lambda_{ \eta H}^\prime$ and
$\lambda_{ \eta H}^{\prime \prime }$ can be assumed to be very small. In fact,
$\lambda_{ \eta H}^{\prime \prime }$
is absent under
a global or gauge $U(1)$
symmetry.
However, such a mass splitting
may play a significant role for its discovery at the LHC
(see e.g., \cite{Datta:2016nfz}). 

\subsection{Electroweak precision observables (EWPO)}
In the presence of two BSM scalars~($\eta$, $S$), two vector like lepton 
doublets~($L_1$, $L_2$) and a singlet fermion~($\psi$), 
our model may introduce 
corrections to the gauge boson vacuum polarization amplitudes or 
electroweak precision observables (EWPO). These observables were initially 
discussed by Peskin and Takeuchi as S, T and U parameters in 
Ref.~\cite{Peskin:1991sw}. 
Later Barbieri et al. introduced $\hat{S}$, $\hat{T}$, $W$, $Y$~\cite{Barbieri:2004qk} 
as the most general parameterization of the new physics effects. 
$\hat{S}$ and $\hat{T}$ are related to the original S and T parameters 
through the simple relations: $\hat{S}=\frac{\alpha S}{4s_W^2}$ and 
$\hat{T}=\alpha T$, where $\alpha$ is the fine structure constant and 
$s_W=\sin\theta_W$. 
Among the generalized Peskin-Takeuchi parameters, 
$W$ and $Y$ 
are important at LEP2 energy scale  
~\cite{Barbieri:2004qk,Cynolter:2008ea}, thus will not be considered here.  
Usually for any generic model, one can find from the global analysis
that the electroweak precision parameters are much
smaller (at the level of $10^{-3}$) and this does not depend on the 
mass of the Higgs scalar. 
Our calculations of the precision observables are based upon Refs.
~\cite{Cynolter:2008ea,Barbieri:2006dq}.   \\

The current experimental constraints are~
\cite{Barbieri:2004qk,Tanabashi:2018oca},
\begin{eqnarray}
\hat S &=& (0.0\pm 1.3)\times 10^{-3}\,, \\
\hat T &=& (0.1\pm 0.9)\times 10^{-3}\,. 
\end{eqnarray}

Inert doublet $\eta$ may particularly effect $T$ or $\hat{T}$ parameter
through $\lambda^\prime_{\eta H}$ and $\lambda^{\prime\prime}_{\eta H}$
~\cite{Barbieri:2006dq}.
But in the limit,
$M_{\eta_I} = M_{\eta_R} = M_{\eta^\pm} \equiv M_\eta$,
which we assume in the subsequent analysis, the
electroweak parameters seem to be unaffected by the presence of new scalars. 
Hence the correction is completely due to the effect of vector like 
fermions~(VLF), i.e., in our model $\Delta(\hat{S},\hat{T})=(\hat{S},\hat{T})_{VLF}$. Therefore, 
the correction in $\hat{T}$ parameter appearing due to the mixing between 
$L_1$ and $\psi$ for $q^2\rightarrow 0$ limit 
can be expressed as~\cite{Cynolter:2008ea},
\begin{align}
\hat{T}=&\frac{g^2}{16\pi^2M_W^2}\Bigg[\tilde{\Pi}(M_{L_1},M_{L_1},0)+\cos^4\theta\, \tilde{\Pi}(M_{\chi_1},M_{\chi_1},0)+\sin^4\theta\, \tilde{\Pi}(M_{\chi_0},M_{\chi_0},0)\nonumber\\
&+2\sin^2\theta \cos^2\theta\, \tilde{\Pi}(M_{\chi_0},M_{\chi_1},0)-2\cos^2\theta\, \tilde{\Pi}(M_{L_1},M_{\chi_1},0)-2\sin^2\theta\, \tilde{\Pi}(M_{L_1},M_{\chi_0},0)\Bigg],
\label{eq:T1}
\end{align}
where $M_{L_1}$ is the mass term for $L_1^-$, $g$ is the $SU(2)_L$ coupling constant, $\theta$ is the mixing angle between $L_1^0$ and $\psi$ as discussed earlier, $M_W$ stands for the mass of $W$ boson and 
\begin{align}
\tilde{\Pi}(m_a,m_b,0)=-\frac{1}{2}(m_a^2+m_b^2)\left[Div+{\rm ln}\left(\frac{\mu^2}{m_am_b}\right)\right]-\frac{1}{4}(m_a^2+m_b^2)-\frac{(m_a^4+m_b^4)}{4(m_a^2-m_b^2)}{\rm ln}\left(\frac{m_b^2}{m_a^2}\right)\nonumber\\
+m_am_b\left[Div +{\rm ln}\left(\frac{\mu^2}{m_am_b}\right)+1+\frac{(m_a^2+m_b^2)}{2(m_a^2-m_b^2)}{\rm ln}\left(\frac{m_b^2}{m_a^2}\right)\right],
\label{eq:Pi}
\end{align}
is the correction to gauge boson propagators in presence of the new VLF's. $Div=\frac{1}{\epsilon}+{\rm ln}(4\pi)-\gamma$ is the usual divergent piece appearing in the dimensional regularisation and $\mu$ denotes the renormalization scale. One can easily see that for $m_a=m_b$, Eq.~\eqref{eq:Pi} vanishes. Hence Eq.~\eqref{eq:T1} simplifies to 
\begin{align}
\hat{T}=\frac{g^2}{16\pi^2M_W^2}\Bigg[2\sin^2\theta \cos^2\theta\, \tilde{\Pi}(M_{\chi_0},M_{\chi_1},0)-2\cos^2\theta\, \tilde{\Pi}(M_{L_1},M_{\chi_1},0)-2\sin^2\theta\, \tilde{\Pi}(M_{L_1},M_{\chi_0},0)\Bigg].
\label{eq:T2}
\end{align}
It can be noted that the divergent part of the first term of Eq.~\eqref{eq:T2} is cancelled by the divergences encapsulated in the last two terms. Moreover in the  limit, when the mass splitting between $M_{L_1}$ and $M_{\chi_1}$ vanishes, (i.e.,
$\sin \theta \to 0$) one finds $\hat T \to 0$. \\
In our model, the correction in $\hat{S}$ can be parameterized as,
\begin{align}
\hat{S}=\frac{g^2}{16\pi^2}\Bigg[\tilde{\Pi}^\prime(M_{L_1},M_{L_1},0)-\cos^4\theta\, \tilde{\Pi}^\prime (M_{\chi_1},M_{\chi_1},0)-\sin^4\theta\, \tilde{\Pi}^\prime (M_{\chi_0},M_{\chi_0},0)\nonumber\\
-2\sin^2\theta \cos^2\theta\, \tilde{\Pi}^\prime (M_{\chi_0},M_{\chi_1},0)\Bigg],
\label{eq:S}
\end{align}
where the `$\prime$' signifies derivative with respect to $q^2$. The general expression for $\tilde{\Pi}^\prime (m_a,m_b,0)$ is given as~\cite{Cynolter:2008ea,Barman:2019aku},
\begin{align}
\tilde{\Pi}^\prime (m_a,m_b,0)=\frac{1}{3}\left[Div+{\rm ln}\left(\frac{\mu^2}{m_am_b}\right)\right]+\frac{m_a^4-8m_a^2m_b^2+m_b^4}{9(m_a^2-m_b^2)^2}
+\frac{(m_a^2+m_b^2)(m_a^4-4m_a^2m_b^2+m_b^4)}{6(m_a^2-m_b^2)^3}{\rm ln}\left(\frac{m_b^2}{m_a^2}\right)\nonumber\\
+m_am_b\Bigg[\frac{(m_a^2+m_b^2)}{2(m_a^2-m_b^2)^2}+\frac{m_a^2m_b^2}{(m_a^2-m_b^2)^3}{\rm ln}\left(\frac{m_b^2}{m_a^2}\right)\Bigg].
\label{eq:Pi1}
\end{align}
For $m_a=m_b$ the above expression reduces to
\begin{align}
\tilde{\Pi}^\prime (m_a,m_a,0)=\frac{1}{3}\left[Div+{\rm ln}\left(\frac{\mu^2}{m_a^2}\right)\right].
\label{eq:Pi10}
\end{align}
It can be directly verified that the divergent parts along with the scaling factor $\mu$ get cancelled when Eq.~\eqref{eq:Pi1} or Eq.~\eqref{eq:Pi10} is substituted in Eq.~\eqref{eq:S}.\\

\begin{figure}
\includegraphics[scale=0.5]{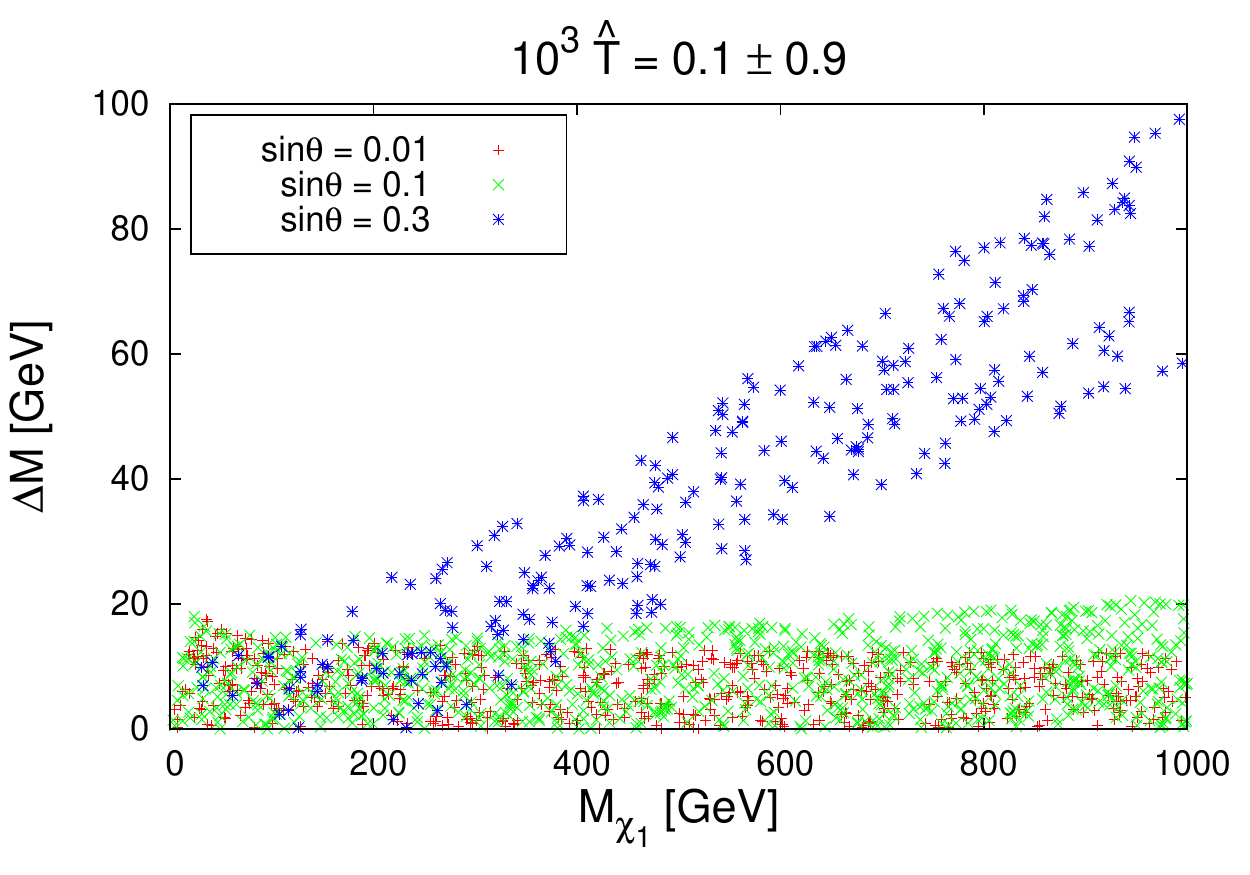}\hspace{1 cm}
\includegraphics[scale=0.5]{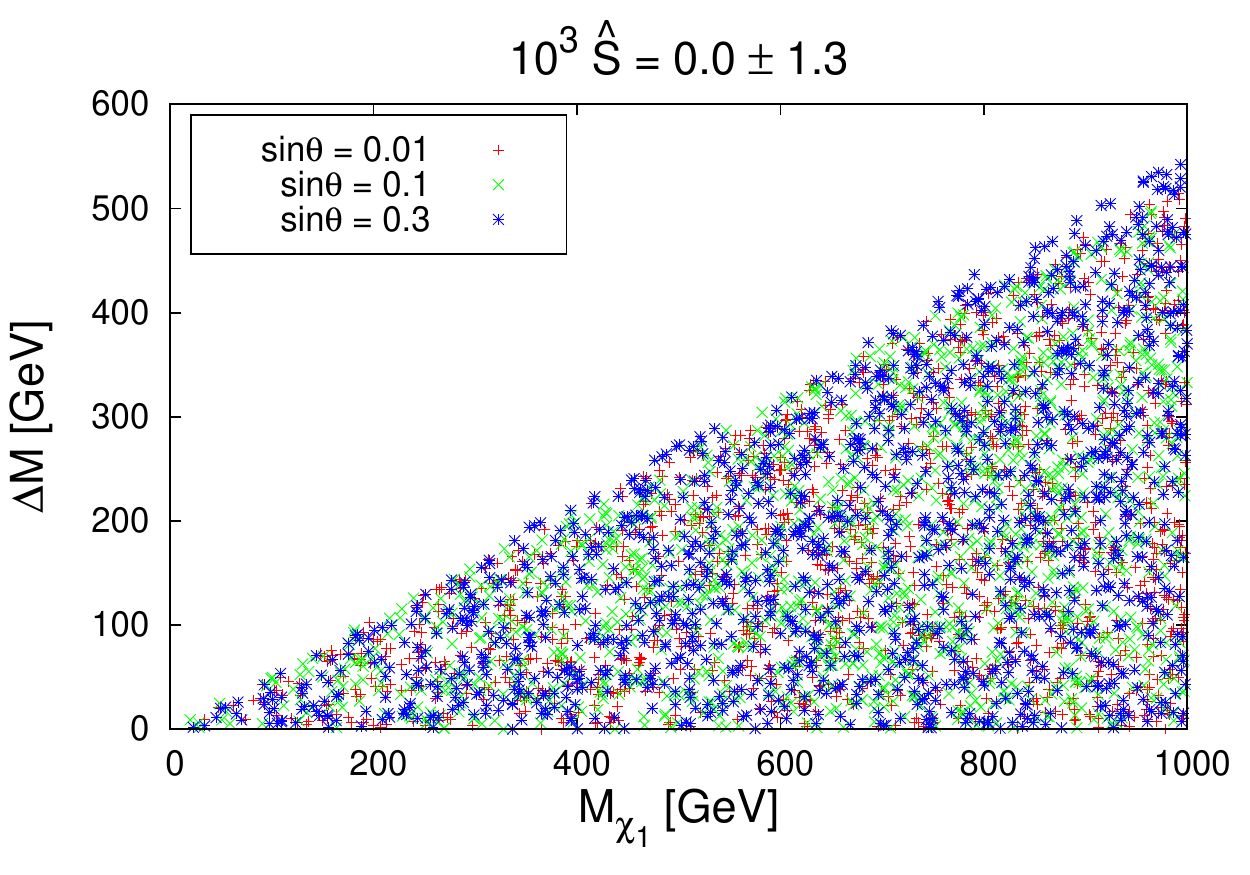}\\
\begin{center}
(a)\hspace{7.5 cm}(b)\\
\end{center}
\includegraphics[scale=0.5]{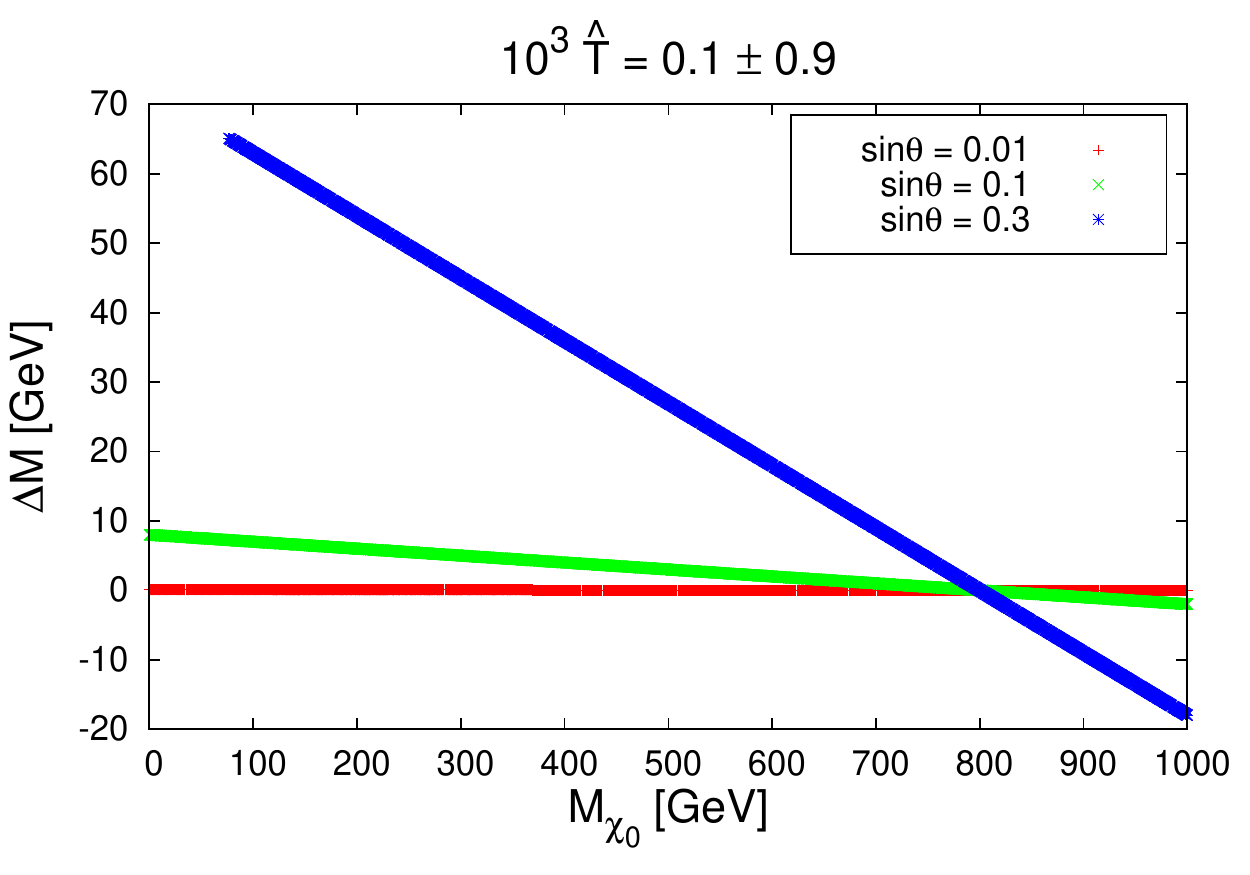}\hspace{1 cm}
\includegraphics[scale=0.5]{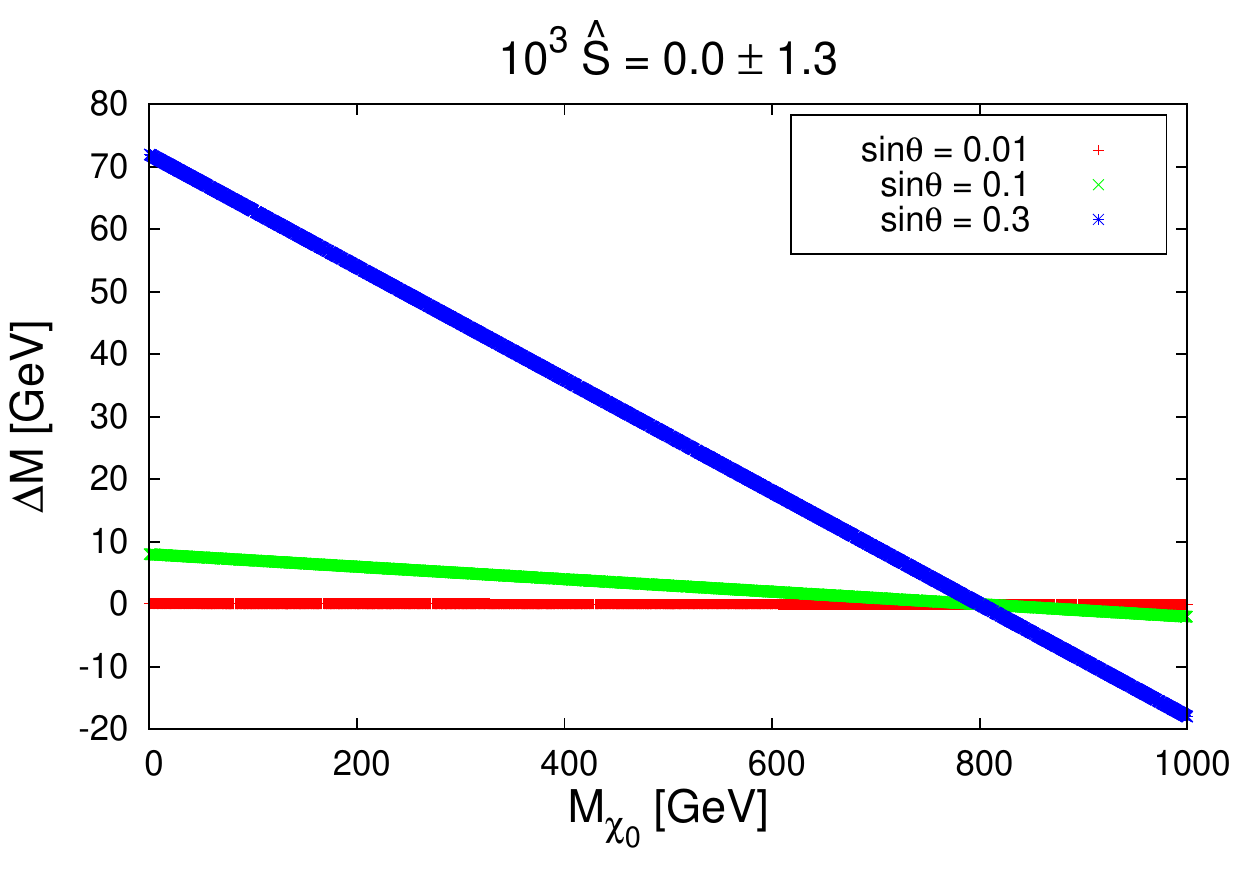}\\
\begin{center}
(c)\hspace{7.5 cm}(d)
\end{center}
\caption{Constraints on $\Delta M=(M_{\chi_1}-M_{L_1})$ coming from the EWPO (a)~$\hat{T}$ and
  (b)~$\hat{S}$ with respective to $M_{\chi_1}$ for three different values of
  $\sin\theta=0.01$, 0.1 and 0.3 when both $M_{\chi_1}$ and $M_{L_1}$ are varied randomly.
  Here $M_{\chi_0}=120$ GeV is assumed. (c) and (d) shows the variation of $\Delta M$ as a function of $M_{\chi_0}$, when $M_{\chi_1}=800$ GeV and $M_{L_1}$ is given by Eq.~\eqref{eq:ML1}.}
\label{fig:ewpo}
\end{figure}

Numerically, since the oblique parameters are sensitive to
the mass splitting
$\Delta M=(M_{\chi_1}-M_{L_1})$, we depict its variation with $M_{\chi_1}$
in
Fig.~\ref{fig:ewpo}(a) and Fig.~\ref{fig:ewpo}(b) for three different values of
$\sin\theta=0.01$, 0.1 and 0.3 keeping DM mass $M_{\chi_0}=120$ GeV. Clearly,
electroweak precision constraints on $\hat{S}$ is much relaxed
compared to the oblique parameter $\hat{T}$ to the new fermions.
For moderate or
smaller values of $\chi_1$ mass, one finds that
$\Delta M \le \mathcal O(20)$ GeV is allowed by the oblique parameter $\hat T$,
which sets an upper bound on $\sin\theta$ ($\simeq 0.1$).

The bounds can be
used to constrain the bare masses of the new fermions. For example, one may always
cast the bare masses $M_{L_1}$ and $M_\psi$ 
in terms of $M_{\chi_0}$, $M_{\chi_1}$ and mixing angle $\theta$.
\begin{align}
  M_{L_1}=M_{\chi_1}\cos^2\theta+M_{\chi_0}\sin^2\theta, \label{eq:ML1}\\ 
 M_{\psi}=M_{\chi_1}\sin^2\theta+M_{\chi_0}\cos^2\theta.\label{eq:Mpsi11}
\end{align}
Notably, the change in $\Delta M$ is negligible to the variation with
$M_{\chi_0}$ for a small mixing angle ($\sin\theta \le 0.1 $) 
(see Fig.~\ref{fig:ewpo}(c) and (d)).
In other words, the EWPOs are insensitive to the lightest neutral fermion mass
$M_{\chi_0}$ as long as the mixing angle is not much high. In the subsequent
section, we consider $\sin\theta \le 0.01 $,
thus, in this regime, 
the mass of the charged component of the VL, $M_{L_1}$ can easily be fixed
through $M_{\chi_1}$ while satisfying all the bounds coming from EWPOs. 
\subsection{Constraints from the collider observables}
\label{Constraintscollider}
For vector like quarks, the LHC pair production cross section is determined from QCD, so model independent bounds can be placed in the parameter space. However,
for the vector like leptons, the pair-production cross section is mediated
by the s-channel electroweak
vector boson exchanges, thus depends on the respective $SU(2)_L$ and $U(1)_Y$
couplings of the new states. As the cross section would reside on the lower side, much weaker bounds can be expected.  There are several searches by the LHC collaborations \cite{Aad:2015dha,CMS:2018cgi,Sirunyan:2019ofn} at $\sqrt s=8$ and 13 TeV run at the LHC. As expected, the constraint is much more stringent
for a pure $SU(2)_L$ VL pair that mixes with and decays to SM leptons.
For example, heavy lepton mass values in the range $114-176$ GeV are excluded through decay into $Z$ boson and $e,\mu$.
In some recent analysis, the CMS collaboration has published \cite{CMS:2018cgi,Sirunyan:2019ofn} the results of dedicated searches for doublet-like VLs,
based on 41.4 $fb^{-1}$ and 77.4 $fb^{-1}$  data samples at $\sqrt s = 13$ TeV.
The bounds can exclude a VL heavy $\tau^\prime$ lepton   
in  the  mass  range  of
$130-690$ GeV or $120<\tau^\prime<790$~GeV following its decays
to tau leptons. The mass of the VL is the only free parameter
both in the production cross section and in the branching fraction calculations, thus in the estimation of the bound. In a recent analysis~\cite{Bissmann:2020lge}, using a CMS search based on 77.4 $fb^{-1}$ at 13 TeV LHC a bound on doublet-like vector leptons has been presented~($\sim 800$ GeV) mainly focusing on $4\ell$ final states. Unlike most of the studies presented above, in this model,
direct couplings of $L_1,\,L_2$ with SM leptons are not allowed.
Similarly, a recent analysis~\cite{Aad:2019vnb}, using ATLAS search based on 139 $fb^{-1}$ at 13 TeV LHC presents the exclusion limits on simplified SUSY
models for a direct slepton production. Here
slepton-pair production masses up to 700 GeV are excluded assuming three generations of
mass-degenerate sleptons, considering
sleptons decaying into final states with two leptons
and missing transverse energy. However, such exclusion limits depend much
on the
masses of the lightest neutralino 
and it has been observed that even a lighter smuon mass is also allowed
depending on the value of $m_{\tilde{\chi}_1^0}$~(e.g. $m_{\tilde{\mu}}\sim 200$ GeV is allowed for $m_{\tilde{\chi}_1^0}\sim 120$ GeV). 

In the framework that we considered, we shall place $M_{L_1}(M_{\chi_1})$
at 800 GeV, but the other VL $L_2$ has to be set at a lower
value~(e.g.$\sim 200$ GeV)
in order to satisfy $(g-2)_\mu$ constraints. Here we may note a few observations
 which would be detailed in the next sections.
First of all, we will find that, the potentially
important contribution in the evaluation of $\Delta a_{\mu}$ would be driven
by the interaction involving coupling $Y_{3\mu}$ and
in the perturbative unitarity regime~(will be discussed in Sec.~\ref{sec:LFV})
$Y_{3\mu}$ can only take $\sim O(1)$ values.
We will further observe that
all other Yukawa couplings of $L_2$ would be orders
of magnitude suppressed either from the neutrino masses and mixings or from
 $(g-2)_e$ and cLFV observables. Thus, the dominant decay of $L_2$ can be considered as $L_2 \rightarrow \mu S$
followed by $S \rightarrow {\chi}_0 \nu$
($M_{L_1}(M_{\chi_1})>M_{L_2}>M_S > M_{\chi_0}$ would be followed throughout
this analysis). 
So, naturally, $PP \to (L^{\pm}_2L^{\mp}_2) \to 2\mu +\EMET$ through $Z$
boson exchange can be
considered as the most useful constraint for the present analysis. Here we may
borrow the limits from Ref.~\cite{Aad:2019vnb} as direct production of sleptons
or VL states would have same cross-section. Thus, based
upon our previous discussion, we would consider $M_{L_2}=190~$GeV
and $m_{\tilde{\chi}^0_1}=120$ GeV respectively for the calculation of different
observables in the leptonic sector.

In our model, $\eta$ couples to leptons, so can only be produced through electroweak gauge
bosons at the LHC.
Also, recall that $\eta$ does not acquire any VEV, thus do not take part in electroweak
symmetry breaking. 
In a model specific study, one would expect dilepton +$\EMET$
\cite{PhysRevD.91.115011,Datta:2016nfz}
through charged $\eta$ pair production, or mono-lepton + $\EMET$
through charged and
neutral $\eta$ productions via $Z$ boson or $W$ boson exchanges.
An observable signal
may be expected during high luminosity run of LHC through multilepton
searches
for $M_\eta \le 250$ GeV \cite{Datta:2016nfz}. Here, assuming
all the charged and neutral components of $\eta$ are of similar masses,
we consider $M_\eta >100$
GeV which is closely based on the exclusions at LEP \cite{Abbiendi:2013hk}. However, our result does not depend much on $M_\eta$.

\section{Radiative Dirac Neutrino Mass} \label{sec:nu_mass}
\begin{figure}[h]
  \includegraphics[scale=0.46]{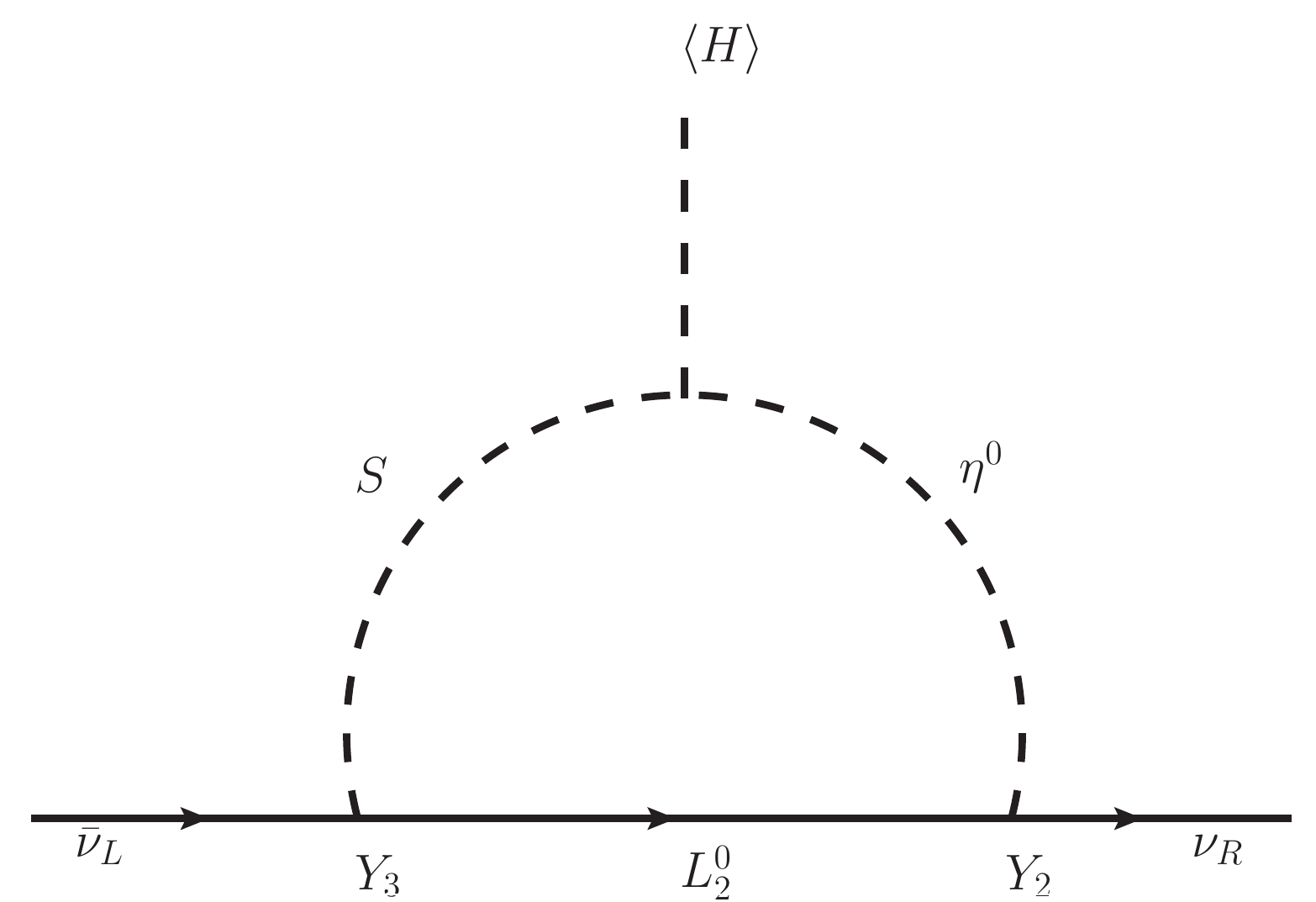}\hspace{1 cm}
 \includegraphics[scale=0.46]{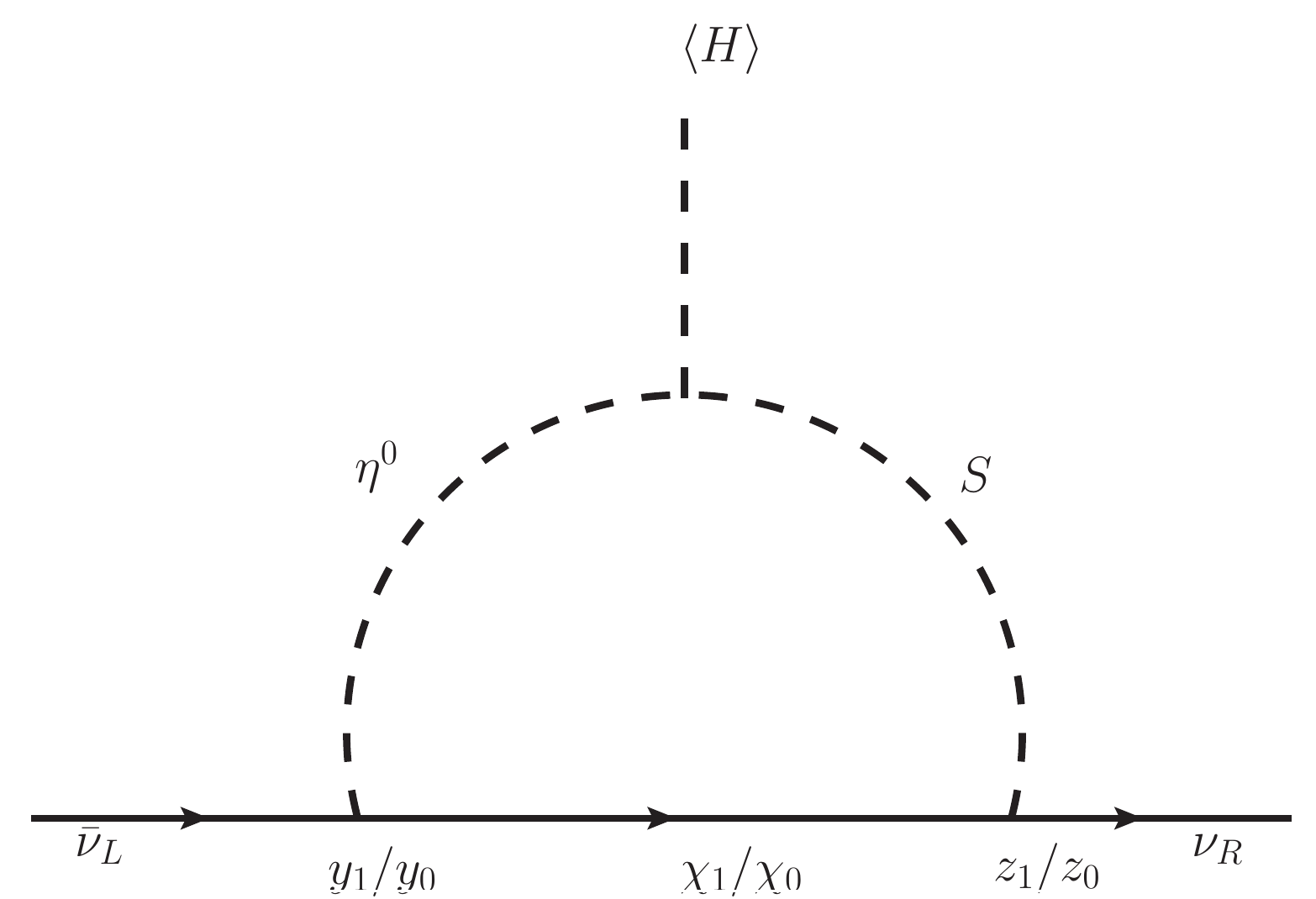}\\
 \begin{center}
 (a)\hspace{8.2 cm}(b)
 \end{center}
 \caption{Radiative mass generation for the neutrinos that adheres lepton
   number conservation. In the second diagram, neutral fermions
   are considered where ${y_1,z_1}=-(Y_{4(i1)},Y_{6(1i)}){\rm sin}\theta$,
   ${y_0,z_0}=(Y_{4(i1)},Y_{6(1i)}){\rm cos}\theta$ have been used.}.
\label{fig:nu_mass}
\end{figure}
As discussed, here neutrinos are massless at the tree level due to the
imposed \sym symmetry while they may receive appropriate radiative
corrections through the symmetry breaking term in Eq.~\eqref{eq:SBT}. Thus
one may develop a Dirac mass term for the SM neutrinos at one loop order after the Higgs
field acquires a VEV. Additionally, the neutrino loops contain a stable
particle $\chi_0$ that could be treated as the cold DM of the universe~[see Fig.~2(b)]. This intrinsically sets up a bridge between the
phenomenology of light neutrinos and the other sectors like dark matter. 
The $(3 \times 3)$ neutrino mass matrix 
        can be read as: 
	\begin{align}
 {M_\nu}_{ij}=
 \sum_{f=L_2,\chi_1,\chi_0}\frac{{y^f}_{i1}
   {M_f} z^f_{1j} r^{f}_1}{16\pi^2 M_S^2(r^{f}_1-1)} \, \epsilon \,
 \left[\frac{{\rm ln}(r^{f}_1/r_2)}{r^{f}_1-r_2}-\frac{\rm ln r_2}
   {r^{f}_1(r_2-1)}\right].
                  \label{equ:nu_mass}
	\end{align}

        Similarly, $r^f_1=\left(\frac{M_f}{M_S}\right)^2$,
        $r_2=\left(\frac{M_{\eta}}{M_S}\right)^2$ and $\epsilon =\mu^\prime
        \langle H \rangle$ is the
        symmetry breaking term, with a mass dimension of 2. 
        For each element in $f\in {(L_2,\chi_1,\chi_0)}$, the vertices $y$
        and $z$ take $(3 \times 1)$ and
   $(1 \times 3)$ elements respectively 
 which can be read as
        $y^f=(Y_{3},-Y_{4}{\rm sin}\theta,Y_{4}{\rm cos}\theta)$ and
 $z^f=(Y_{2},-Y_{6}{\rm sin}\theta,Y_{6}{\rm cos}\theta)$. Just as a
 measure of simplification, we can consider
 $M_{L_2}\sim M_{\chi_1}\sim M_{\chi_0}\equiv M_f$, so that
 Eq.~\eqref{equ:nu_mass} becomes,
\begin{equation}
  {M_\nu}_{ij}={M_{f}}\left({Y_3}_{i1} {Y_2}_{1j} +{Y_4}_{i1}
  {Y_6}_{1j}\right)\frac{r_1}{16\pi^2 M_S^2(r_1-1)}\, \epsilon \,
  \left[\frac{\rm ln(r_1/r_2)}{r_1-r_2}-\frac{\rm ln r_2}{r_1(r_2-1)}\right].
  \label{eq:nu_mass2}
\end{equation}
In the above, $\epsilon$ defines the order of the neutrino masses. Thus all
the new Yukawa couplings can be assumed to take $\mathcal O(1)$ values. The
diagonal mass terms ${\rm diag}[m_i]$ which refer to the masses for the
physical neutrino states are related to the flavor states ${M_\nu}_{ij}$
by the following equation, 
	\begin{equation}
	  {M_\nu}_{ij}=U_{PMNS}({\rm diag}[m_i])U^\dagger_{PMNS};
          \label{equ:pmns}
	\end{equation} 
        \noindent
where the PMNS matrix can be parameterized as~\cite{pdg}:
 \begin{align}
U_{\rm PMNS} & = \begin{pmatrix} 
c_{12} c_{13} & s_{12} c_{13} & s_{13} e^{-i \delta} \\
-s_{12} c_{23} - c_{12} s_{23} s_{13} e^{i \delta} & c_{12} c_{23} -s_{12} s_{23} s_{13} e^{i \delta} & s_{23} c_{13} \\
s_{12} s_{23} - c_{12} c_{23} s_{13} e^{i \delta} & -c_{12} s_{23} - s_{12} c_{23} s_{13} e^{i \delta} & c_{23} c_{13}
\label{MNS}
\end{pmatrix} \nonumber \\
& \times \text{diag}(1, e^{i \alpha_{21}/2}, e^{i\alpha_{31}/2})\,,
\end{align}
 in which  $s_{ij} \equiv \sin \theta_{ij}$, $c_{ij} \equiv \cos \theta_{ij}$; $\delta$ is the Dirac CP violating phase, and $\alpha_{21, 31}$ are  Majorana CP violating phases.  Note that, using the global fit based on the current neutrino data, one may compute $|M_{\nu ij}|$ in terms of the different mass hierarchies, namely, normal hierarchy $\Delta m^2_{32}>0$ (NH) and inverted hierarchy $\Delta m^2_{32} < 0$
 (IH) as~\cite{deSalas:2017kay}~($\Delta m^2_{ij}\equiv m^2_i - m^2_j$).
\begin{table}[ht!]\centering
  \catcode`?=\active \def?{\hphantom{0}}
   \begin{tabular}{lc}
    \hline
    parameter & best fit $\pm$ 3$\sigma$ range  \hphantom{x}
    \\
    \hline
    $\Delta m^2_{21}\: [10^{-5}\eVq]$
     & 7.05--8.14 \\[3mm]  
    $|\Delta m^2_{31}|\: [10^{-3}\eVq]$ (NH)
    &  2.41--2.60\\
     $|\Delta m^2_{31}|\: [10^{-3}\eVq]$ (IH)
     &  2.31-2.51 \\[3mm] 
    $\sin^2\theta_{12} $
     & 0.273--0.379\\[3mm] 
     $\sin^2\theta_{23} $(NH)
               & 0.445--0.599 \\[3mm]
     $\sin^2\theta_{23} $ (IH)
               & 0.453--0.598\\[3mm]
    $\sin^2\theta_{13} $ (NH)
     & 0.0196--0.0241 \\
    $\sin^2\theta_{13} $ (IH)
     & 0.0199--0.0244 \\ 
   $\delta/\pi$ (NH)
   	 & 0.87--1.94 \\
    $\delta/\pi$ (IH)	
   	 & 1.12--1.94 \\
    \hline
    \hline
     \end{tabular}
     \caption{ \label{tab:sum-2017} 
        Neutrino oscillation parameters summary determined from the 
        global analysis \cite{deSalas:2017kay}.
     }
\end{table}
Taking $m_{1(3)}=0$ for NH (IH), and zero values for the Majorana phases
($\alpha_{21(31)}=0$) and the $3\sigma$ uncertainties, the magnitudes of the neutrino mass matrix elements in units of eV for NH and IH can be estimated as:
\begin{align}
|M_{\nu ij}|\simeq \begin{pmatrix} 0.11-0.45 & 0.12-0.82 & 0.12-0.82 \\  0.12-0.82 & 2.4-3.3 & 2.0-2.2 \\ 0.12-0.82 & 2.0-2.2  & 2.2-3.1\end{pmatrix} \times 10^{-2}\,, \nonumber \\
|M_{\nu ij}|
\simeq \begin{pmatrix} 4.8-5.0 & 0.41-0.65 & 0.39-0.62 \\  0.41-0.65 & 1.9-2.8 & 2.4-2.6 \\ 0.39-0.62 & 2.4-2.6 & 2.2-3.1\end{pmatrix} \times 10^{-2}
\,. \label{eq:v_nu_mass}
\end{align}
Here, following Eqs.~\eqref{eq:v_nu_mass} and \eqref{equ:nu_mass} we may note a few observations related to the neutrino masses and mixings. In fact Eq.~\eqref{equ:nu_mass} can be cast as
${M_\nu}_{ij}=\sum_{f=L_2,\chi_1,\chi_0}{y}_{i1} {\Lambda^f} z_{1j}$ and with all the BSM particles $\sim \mathcal{O}(10^2-10^3)$ GeV,
  one may find that $\Lambda^f \simeq \mathcal O(10^{-1}-10^{-2})$ eV. Thus, the involved Yukawa
  couplings may take $\mathcal O(1)$ values to produce the correct values of the neutrino mass matrix as obtained in Eq.~\eqref{eq:v_nu_mass}. Interestingly, out of the four Yukawas, only
   ${Y_3}_{(i1)}$ and ${Y_4}_{(i1)}$ ($i \in 1...3$) appear in most of the low energy phenomenology which are of interest to us. This includes
neutrino masses and their mixings, precision observables like the 
anomalous magnetic moment of leptons or the cLFV processes and also the DM phenomenology. On the other hand, the other two Yukawa couplings ${Y_2}_{(1i)}$ and ${Y_6}_{(1i)}$~(six in total) related to the singlet state
$\nu_{Ri}$ can control the neutrino masses and mixings.
Thus, one may always use the freedom of choosing the free parameters $Y_2$ and $Y_6$ to satisfy the observed
mass square differences and mixing angles while $Y_{3}$ and $Y_{4}$ may be tuned
to satisfy the observables related to low energy lepton phenomenology.

To clarify it further numerically,
    we fix $M_S=130$ GeV, $M_f=800$ GeV and $M_\eta=300$ GeV
and use Eq.~\eqref{eq:nu_mass2} to delineate the domain 
for ${Y_2}_{(1i)}$ and ${Y_6}_{(1i)}$ that may produce correct values for
$|M_{\nu ij}|$ in Eq.~\eqref{eq:v_nu_mass} in the NH scenario.
For simplicity, we recast the parameter as
${Y_2}_{(1i)}=Y_{2i}$ and
${Y_6}_{(1i)}=Y_{6i}$ (see also the discussion in Sec.\ref{sec:LFV}).
We also fix
$\{Y_3,Y_4\}$ at the given values (see Table \ref{tab:Y2_Y6})
       which would be
       allowed by $(g-2)_\ell$ and cLFV constraints. We would further detail
       it in Sec.~\ref{sec:LFV}.
       The lower and upper limits
       in Table~\ref{tab:Y2_Y6} would refer to the minimum and maximum
       value of the
       $|{M_\nu}_{ij}|$ in Eq.~\eqref{eq:v_nu_mass}.
\begin{table}[ht]
\begin{tabular}{|c|c|}
\hline 
$\{Y_{3i},\,Y_{4i}\}$ & $\{Y_{2i},\,Y_{6i}\}$ \\ 
\hline 
$Y_{3e}=0,\,Y_{4e}=0.2$ & $0.001\leq Y_{2e}\leq 0.01$, $0.018\leq Y_{6e}\leq 0.04$ \\  
$Y_{3\mu}=2.3,\,Y_{4\mu}=0$ & $0.034\leq Y_{2\mu}\leq 0.048$, $0.11\leq Y_{6\mu}\leq 0.12$ \\  
$Y_{3\tau}=0.01,\,Y_{4\tau}=0.6$ & $0.029\leq Y_{2\tau}\leq 0.032$, $0.12\leq Y_{6\tau}\leq 0.14$ \\ 
\hline 
\end{tabular} 
\caption{Allowed range of $\{Y_{2i},\,Y_{6i}\}$ as obtained from Eq.~\eqref{eq:nu_mass2}~(for $M_S=130$ GeV, $M_f=800$ GeV and $M_\eta=300$ GeV) within which the magnitudes of the neutrino mass matrix elements for NH~[Eq.~\eqref{eq:v_nu_mass}] can be satisfied.} 
\label{tab:Y2_Y6}
\end{table}

\section{Lepton $g-2$, cLFV processes and other constraints} \label{sec:LFV}
In the lepton phenomenology, apart from tuning the $\mu$ and $e$ anomaly, 
new scalars  $\eta$, $S$, charged fermions
$L_2^\pm, L_1^\pm$ and neutral leptons $\chi_1$ and $\chi_0$
may lead to observable signatures
to lepton flavor violating processes such as
$\ell_\alpha \rightarrow \ell_\beta \gamma $, or
$\ell_\alpha \rightarrow 3 \ell_\beta$ through the
Yukawa
couplings $Y_1$, $Y_3$ and $Y_4$ that tie
the SM leptons to BSM particles. The free
parameters can be listed as:
\begin{eqnarray}
  M_{\chi_1},\ M_{\chi_0},\ M_{S},\ M_{\eta},\ M_{L_2},\ Y_{1(1i)},\ Y_{3(i1)},\ Y_{4(i1)}\ (i\in e,\mu,\tau),\ \sin\theta~.
  \label{eq:freeparam}
\end{eqnarray}
The other charged lepton mass $M_{L_1}$
can be expressed in terms of $M_{\chi_0}$, $M_{\chi_1}$ and $\theta$ via
Eq.~\eqref{eq:ML1}. Unless
otherwise stated, the
mixing parameter in the neutral lepton sector $\sin\theta = 0.01$ is being fixed in our
analysis.
For the sake of clarity, we recast the new Yukawa couplings of
Eq.~\eqref{eq:L_new}~(and hence Eq.~\eqref{eq:freeparam})
as $Y_{ij}$ where $i$ assumes different types of the couplings e.g.,
$1,2,3,4,6$ and $j$ takes the different flavors $e,\mu,\tau$. As an example,
$Y_{1(1e)}$ in the Eq.~\eqref{eq:L_new} is simply denoted as $Y_{1e}$. In this
set-up i.e., with the minimal contents of new states,
first we survey if the discrepancy between the theoretical and experimental
values of the magnetic moments of muon and electron can be
explained. Then we will consider the charged lepton flavor
violating processes. All the radiatively induced processes could be tested
in the present
and future generation of
experiments; thus a domain for
flavor specific Yukawa couplings can be derived.
\begin{figure}[ht]
\includegraphics[scale=0.46]{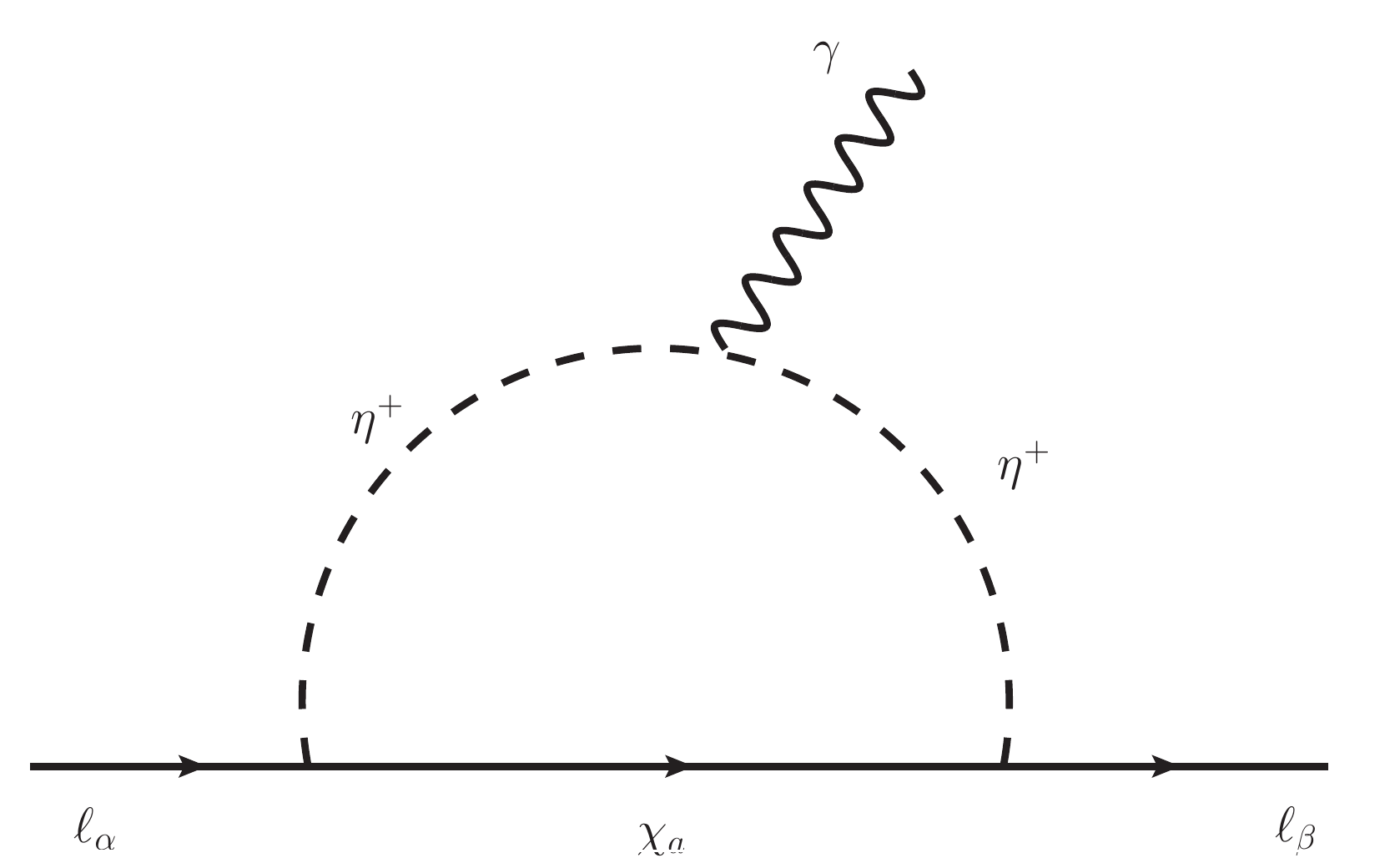}\hspace{1 cm}
\includegraphics[scale=0.46]{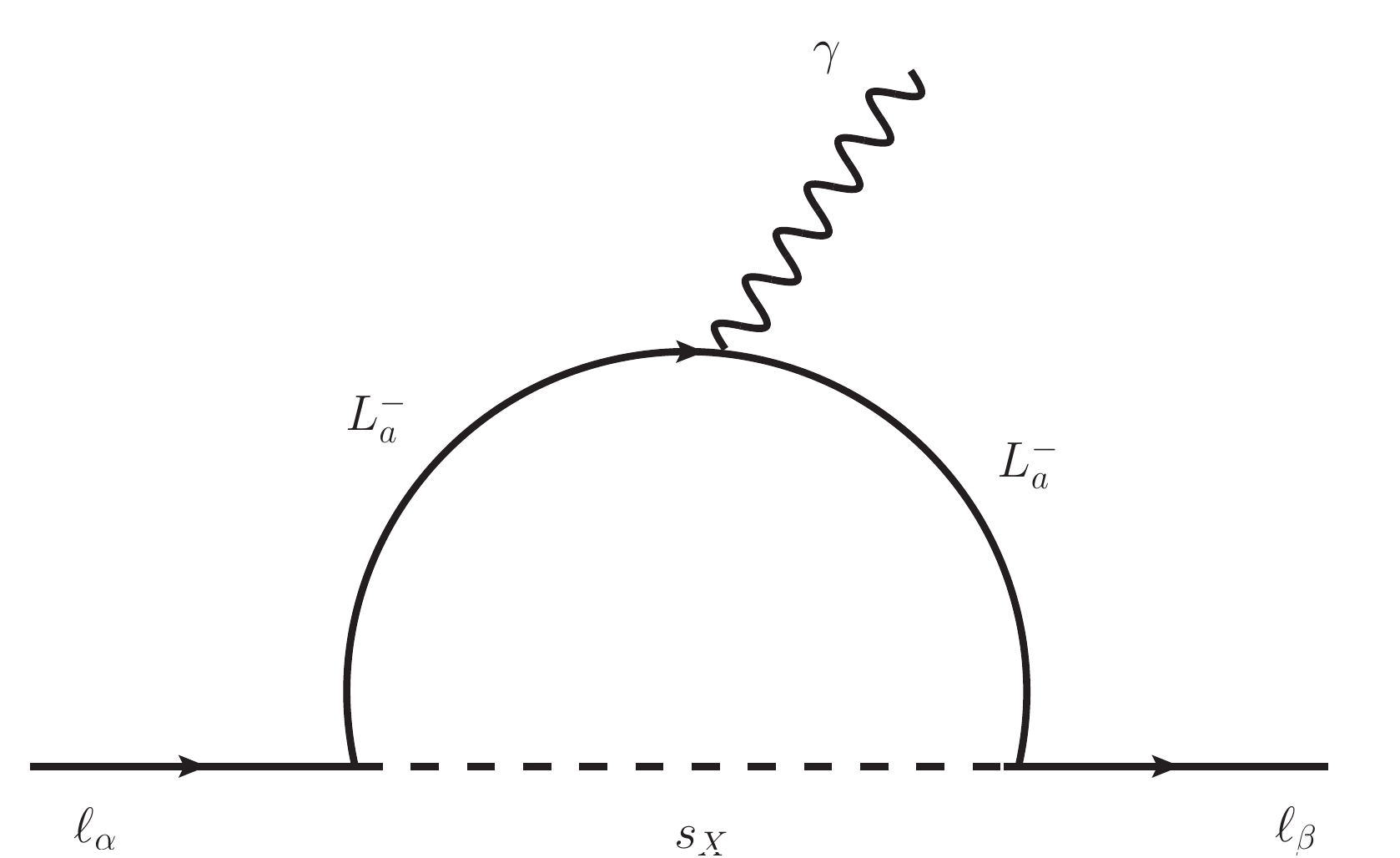}\\
\begin{center}
(a)\hspace{8.5 cm}(b)
\end{center}
\caption{Diagrams contributing to lepton $(g-2)$ and $\ell_\alpha\rightarrow\ell_\beta\gamma$ processes.}
\label{fig:lep_g}
\end{figure}
\subsection{Lepton $g-2$}

In our model, we would be able to explain $\Delta a_{\mu/e}$ simultaneously
through
  the loop diagrams, shown in Fig.~\ref{fig:lep_g}. $\ell_\alpha$ and
  $\ell_\beta$ are general
  notations for the SM leptons. 
  The total contribution for lepton $g-2$ process can be given as
  ($\ell_\alpha=\ell_\beta=\ell$):
  \begin{align}
\Delta a_\ell=\Delta a^{(c)}_\ell+\Delta a^{(n)}_{1\ell}+\Delta a^{(n)}_{2\ell},
\label{eq:g_2}
\end{align}
where, the superscripts `n' and `c' correspond to the neutral and the charged
lepton contributions in 
Fig.~\ref{fig:lep_g}(a) and Fig.~\ref{fig:lep_g}(b) respectively.
The three individual contributions of Eq.~\eqref{eq:g_2} can be expressed as,
\begin{align}
\Delta a^{(c)}_\ell=&\frac{1}{16\pi^2}\Bigg[|Y_{3\ell}|^2\left(\frac{m_\ell}{M_S}\right)^2\,F_3\left(\frac{M^2_{L_2}}{M^2_S}\right) + 
   |Y_{1\ell}|^2\left(\frac{m_\ell}{M_\eta}\right)^2\,F_3\left(\frac{M^2_{L_1}}{M^2_\eta}\right)\Bigg], \label{eq:char}\\
\Delta a^{(n)}_{1\ell}=&\frac{1}{16\pi^2}\Bigg[-|{Y_{1\ell}}|^2\,\cos^2\theta\left(\frac{m_\ell}{M_\eta}\right)^2\,F_2\left(\frac{M^2_{\chi_1}}{M^2_\eta}\right) - 
 2 (Y_{4\ell})\,(Y_{1\ell})\cos\theta\sin\theta\frac{(m_\ell\,M_{\chi_1})}{M^2_\eta}F_1\left(\frac{M^2_{\chi_1}}{M^2_\eta}\right)\nonumber \\
 &-|{Y_{4\ell}}|^2\sin^2\theta\left(\frac{m_\ell}{M_\eta}\right)^2\,F_2\left(\frac{M^2_{\chi_1}}{M^2_\eta}\right) 
  \Bigg], \label{eq:neut1}\\ 
\Delta a^{(n)}_{2\ell}=&\Delta a^{(n)}_{1\ell}\left
(\cos\theta\rightarrow -\sin\theta,\,\sin\theta \rightarrow \cos\theta,\,M_{\chi_1}\rightarrow M_{\chi_0}\right).
\label{eq:neut2}
\end{align}
The Form factors are defined in Appendix B. It is instructive to
identify the
positive and negative contributions of $\Delta a_\ell$ ($\ell\in e,\mu$) in
Eq.~\eqref{eq:char}-\eqref{eq:neut2}.
\begin{align}
  \Delta a^{(+)}_\ell=\frac{1}{16\pi^2}\Bigg[|Y_{3\ell}|^2\left(\frac{m_\ell}{M_S}
    \right)^2\,F_3\left(\frac{M^2_{L_2}}{M^2_S}\right) &+ 
    |Y_{1\ell}|^2\left(\frac{m_\ell}{M_\eta}\right)^2\,F_3\left(\frac{M^2_{L_1}}
    {M^2_\eta}\right) +\nonumber\\
    & (Y_{4\ell})\,(Y_{1\ell})\sin2\theta\frac{(m_\ell\,M_{\chi_0})}
    {M^2_\eta}F_1\left(\frac{M^2_{\chi_0}}{M^2_\eta}\right) \Bigg] \label{eq:amupl}
  .\\
\Delta a^{(-)}_{\ell}=-\frac{1}{16\pi^2}\Bigg[|{Y_{1\ell}}|^2\,
  \left(\frac{m_\ell}{M_\eta}\right)^2\,
  F_2\left(\frac{M^2_{\chi_1}}{M^2_\eta}\right) 
  &+|{Y_{4\ell}}|^2\left(\frac{m_\ell}{M_\eta}\right)^2\,F_2
  \left(\frac{M^2_{\chi_0}}{M^2_\eta}\right) +\nonumber\\
  & (Y_{4\ell})\,(Y_{1\ell})
  \sin2\theta\frac{(m_\ell\,M_{\chi_1})}{M^2_\eta}
  F_1\left(\frac{M^2_{\chi_1}}{M^2_\eta}\right)\Bigg].
\label{eq:amumi} 
\end{align}

In the above, $\sin^2\theta \rightarrow 0$ has been taken for illustration.
Additionally,
we consider that all the couplings are real and positive. 
In Eq.~\eqref{eq:amupl},
the first two terms arise from the diagram with a charged fermion and a
neutral scalar in the loop. The third term involves a neutral fermion and a
charged scalar in the loop. 
Here the DM state $\chi_0$
may provide with a positive contribution in $\Delta a_\mu$, owing to the mixing between $L_1^0$ and $\psi$.
The negative parts in $\Delta a_\ell$
(see Eq.~\eqref{eq:amumi}) involves only a 
neutral fermion and a
charged scalar in the loop which is shown in Fig. \ref{fig:lep_g}(a).  
Thus, considering the opposite signs of $\Delta a_\mu$ and $\Delta a_e$ in
mind, one
can easily
expect that $\Delta a_\mu$ should have a major contribution from
Eq.~\eqref{eq:amupl} while Eq.~\eqref{eq:amumi} may play the dominant role in
determining $\Delta a_e$. In terms of the controlling parameters, 
$\Delta a_\mu$ ($\Delta a_e$) are managed by a set of new coupling
parameters
$Y_{4\ell}$, $Y_{3\ell}$, $Y_{1\ell}$ ($\ell\in e,\mu$)
  and
  also by the masses of new scalars and mixing 
  of the neutral leptons. Electroweak precision observables
  restrict the mixing
  in the neutral leptons : $\cos\theta \sim 1$, and, thus,
  $M_{L_1} \simeq M_{\chi_1}$ may be used for illustration
  (see Eq.~\eqref{eq:ML1}).
  For the scalar mass parameters, $M_S$ is kept fixed
  at 130 GeV, while $M_\eta=300$ and $1200$ GeV are considered.
  Keeping this in mind, the variation of the
flavor dependent couplings $Y_{i\mu}$ or $Y_{ie}$ ($i \in 1,3,4$) with the mass
of new scalars or
fermions have been depicted through scattered plots where
points 
consistent with $\Delta a_i$ $(i \in e,\mu)$
within the $2\sigma$ allowance in Eq.~\eqref{eq:delamu} and
Eq.~\eqref{eq:delae} are only shown.
\begin{figure}[!ht]
  \includegraphics[scale=0.6]{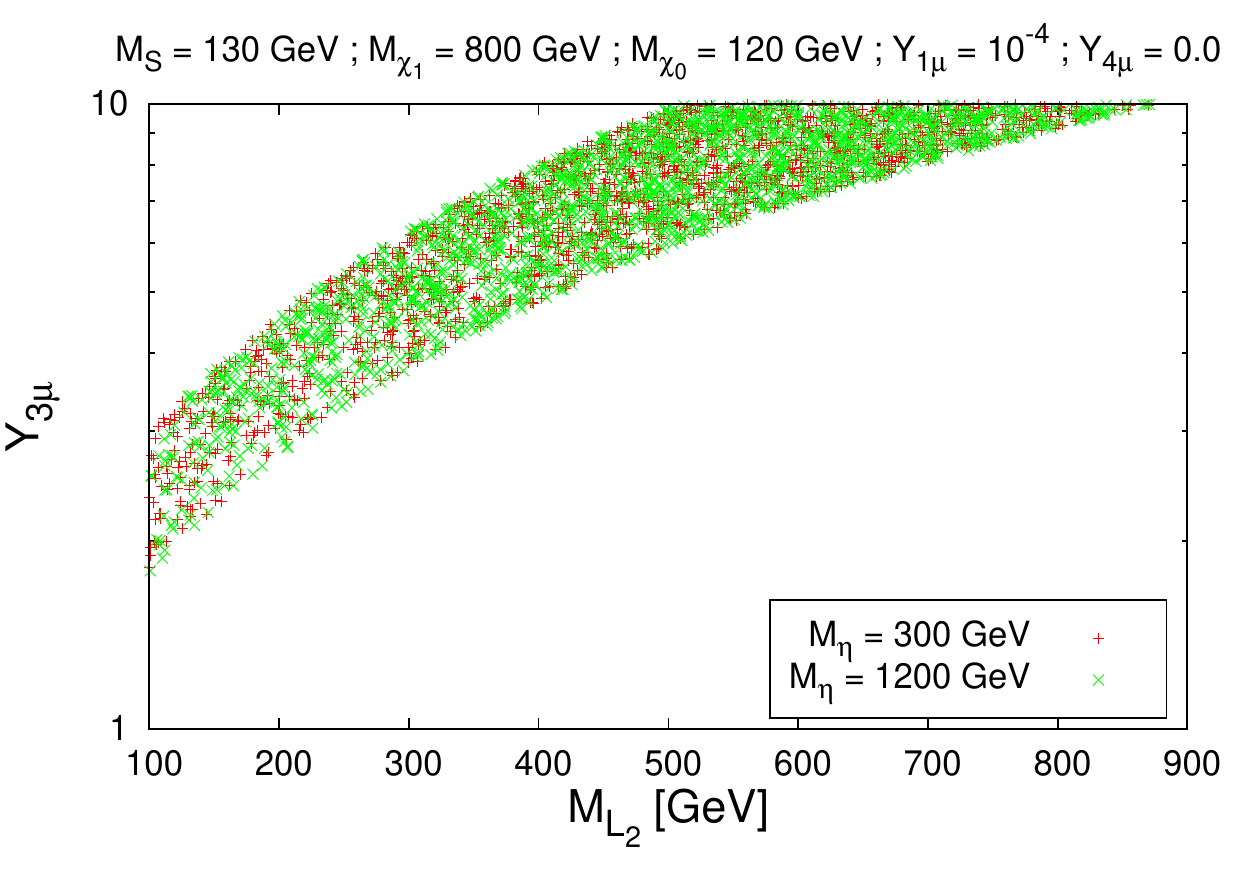}\hspace{1 cm}
  \includegraphics[scale=0.6]{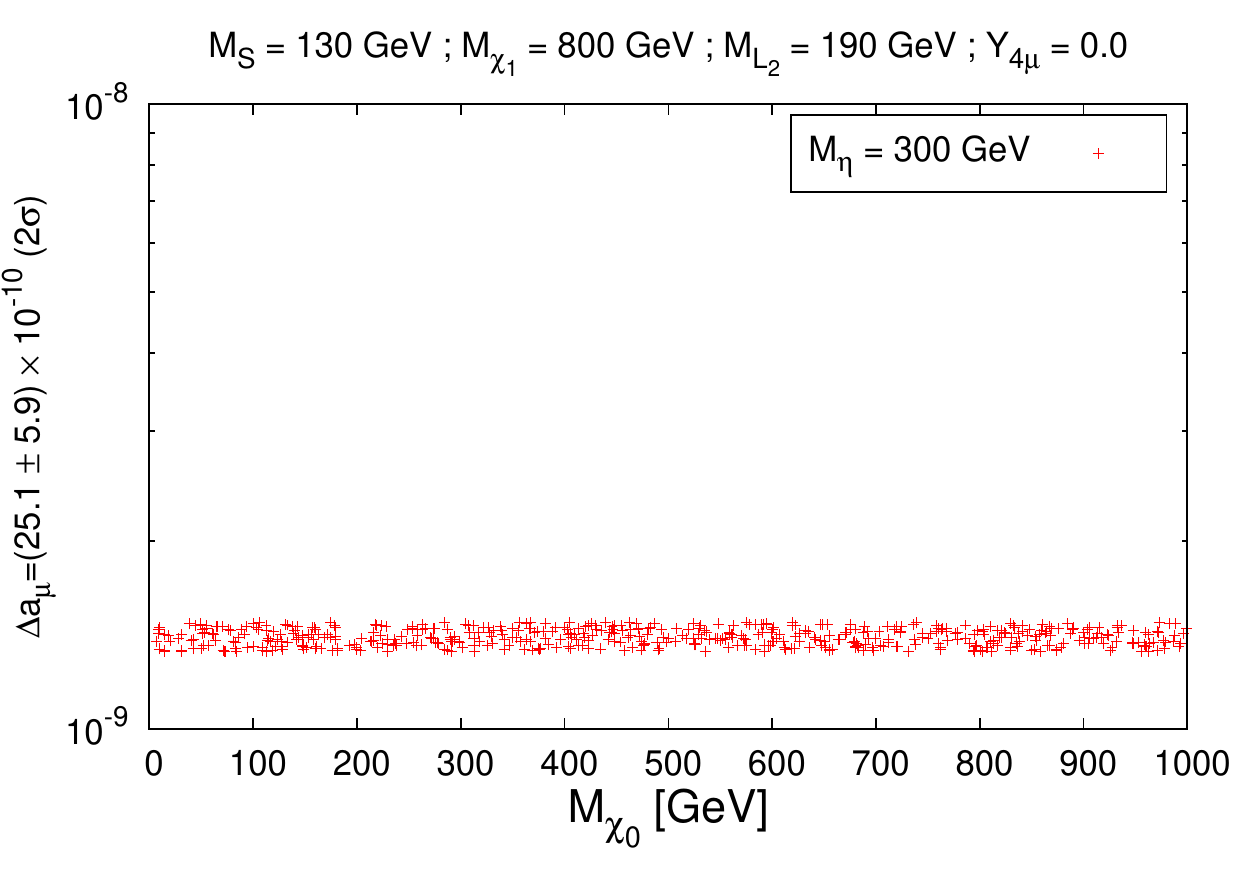}
\begin{center}
(a)\hspace{8 cm}(b)\\
\end{center}
\caption{Allowed parameter space satisfying $\Delta a_\mu$ within $2\sigma$
  bound. Here $\sin\theta=0.01$ is assumed. Here,
  red and green dots represent the scenarios corresponding
to $M_\eta=300$ GeV and $1200$ GeV respectively.}
\label{fig:mu_g2}
\end{figure}
For a better understanding of interplay of the different couplings and masses on the $\Delta a_{\mu(e)}$, we recast the
Eq.~\eqref{eq:amupl} and
Eq.~\eqref{eq:amumi} in a more convenient form:
$\Delta a_{\mu(e)}=\Delta a_{\mu(e)}^{Y_{3\mu(e)}}+\Delta a_{\mu(e)}^{Y_{1\mu(e)}}
+\Delta a_{\mu(e)}^{Y_{4\mu(e)}}
+\Delta a_{\mu(e)}^{Y_{4\mu(e)}Y_{1\mu(e)}}$, where,
\begin{align}
  \label{eq:a_mu3}
\Delta a_{\mu(e)}^{Y_{3\mu(e)}}&=\frac{1}{16\pi^2}\Bigg[|Y_{3\mu(e)}|^2
  \left(\frac{m_{\mu(e)}}{M_S}\right)^2\,F_3\left(\frac{M^2_{L_2}}{M^2_S}\right)\Bigg],
\\ \label{eq:a_mu1}
\Delta a_{\mu(e)}^{Y_{1\mu(e)}}&=\frac{1}{16\pi^2}\Bigg[|Y_{1\mu(e)}|^2
  \left(\frac{m_{\mu(e)}}{M_\eta}\right)^2\,\left\lbrace F_3\left(\frac{M^2_{L_1}}
   {M^2_\eta}\right)-F_2\left(\frac{M^2_{\chi_1}}{M^2_\eta}\right)\right\rbrace\Bigg],
\\ \label{eq:a_mu4}
\Delta a_{\mu(e)}^{Y_{4\mu(e)}}&=\frac{1}{16\pi^2}\Bigg[-|Y_{4\mu(e)}|^2
  \left(\frac{m_{\mu(e)}}{M_\eta}\right)^2\,\left\lbrace F_2\left(\frac{M^2_{\chi_0}}
   {M^2_\eta}\right)\right\rbrace\Bigg],
 \\ \label{eq:a_mu14}
\Delta a_{\mu(e)}^{Y_{4\mu(e)}Y_{1\mu(e)}}&=\frac{1}{16\pi^2}\Bigg[Y_{1\mu(e)}Y_{4\mu(e)}\sin2\theta
   \frac{m_{\mu(e)}}{M^2_\eta}
   \left\lbrace M_{\chi_0}F_1\left(\frac{M^2_{\chi_0}}{M^2_\eta}\right)-M_{\chi_1}F_1\left(\frac{M^2_{\chi_1}}{M^2_\eta}\right)\right\rbrace\Bigg].
\end{align}

We begin our discussion with $\mu$-specific couplings $Y_{i\mu}$
and the
relevant mass parameters  $M_S$, $M_\eta$, $M_{L_2}$ to probe their limits
in controlling
$\Delta a_\mu$. Here the role of $Y_{4\mu}$ is somewhat tricky and depends on
the choice of other parameters.  For example, it can provide
an unhelpful contribution through
Eq.~\eqref{eq:a_mu4}.
  Similarly, unless $Y_{1\mu} \ll Y_{4\mu}$,
Eq.~\eqref{eq:a_mu14} dominates over the $\Delta a_\mu^{Y_{4\mu}}$. However,
the contribution in Eq.~\eqref{eq:a_mu14} can take both positive and negative
values which can be controlled by the ratio ${M_{\chi_0}}/{M_{\chi_1}}$. 
Truly, a specific ratio of the neutral fermions, i.e.,
${M_{\chi_0}}/{M_{\chi_1}}$ can boost $\Delta a_\mu$
through an overall positive contribution, driven by $Y_{4\mu}$. 
However, at the same time it becomes unfriendly to obtain a correct $\Delta a_e$
(since the same bracketed term in Eq.~\eqref{eq:a_mu14} potentially contributes to $e$ magnetic moment). 
For a practical choice, we set $Y_{4\mu}=0$ as we will see that
$\Delta a_e^{Y_{4e}Y_{1e}}$ term would have to be properly tuned to fit $\Delta a_e$.
In other words, ${M_{\chi_0}}/{M_{\chi_1}}$ will be chosen to have a
negative contribution from $\Delta a_e^{Y_{4e}Y_{1e}}$ to have a consistent
$\Delta a_e$.

Thus assuming $Y_{4\mu}=0$, one finds
$\Delta a_\mu=\Delta a_\mu^{Y_{3\mu}}+\Delta a_\mu^{Y_{1\mu}}$.
A prominent cancellation between the two terms in $\Delta a_\mu^{Y_{1\mu}}$
can always be observed irrespective of the value of $Y_{1\mu}$, and, thus,
one finds  
$\Delta a_\mu \simeq \Delta a_\mu^{Y_{3\mu}}$. Thus, naturally, we may choose
$Y_{1\mu}$ at any
value within its perturbative limit while satisfying the experimental
bounds on $\Delta a_\mu$. We will see that a smaller $Y_{1\mu}$
(which will be chosen
in the subsequent analysis) would be
highly desired to
satisfy $\mu \to e \gamma$ constraint.

Fig.~\ref{fig:mu_g2}(a) shows the variation of $Y_{3\mu}$ as a function
of $M_{L_2}$ when $M_S=130$ GeV, $M_\eta=300$ GeV and $M_\eta=1200$ GeV. 
The other input parameters are $M_{\chi_1}=800$~GeV, $M_{\chi_0}=120$~GeV, $Y_{1\mu}=10^{-4}$
while $Y_{4\mu}$ is fixed at zero. Clearly, the doublet scalar does not have
any influence to the
result.
As said earlier, only $L_2-S$ loop can manage to attune
$\Delta a_\mu$, and thus, one requires somewhat larger values for $Y_{3\mu}$.
This can be further verified through Fig.~\ref{fig:mu_g2}(a). Note that, here
mass of the
singlet $M_S$ needs to be smaller to make
$Y_{3\mu}$ within the perturbative bound, and this can only be realized 
if our model
considers light dark matter (since $M_{\chi_0}<M_S$ needs to be satisfied).
However,
a heavier $\chi_0$ can also accommodate $\Delta a_\mu$ without having any
difficulties. Recall that setting $Y_{4\mu}=0$ will automatically make vanishing contributions from Eqs.~\eqref{eq:a_mu4} and \eqref{eq:a_mu14}, which include $M_{\chi_0}$. Thus, because of the choice of our parameters, $\chi_0$ can affect $\Delta a_\mu$ only through
Eq.~\eqref{eq:a_mu1}, which can only lead to insignificant contribution.
A further confirmation can be made through Fig.~\ref{fig:mu_g2}(b),
where we show variations of $\Delta a_\mu$ as a function of
$M_{\chi_0}$ for $M_\eta=300$ GeV and $M_S=130$ GeV. Neutral
and charged vector leptons are fixed at masses $M_{\chi_1}=800$ and
$M_{L_2}=190$ GeV.
Here we varied the couplings~($Y_{1\mu}:[0.0001-1], Y_{3\mu}:[0.0001-2]$) and $M_{\chi_0}$
randomly. The resultant $\Delta a_\mu$ can be seen to be consistent
over the entire $\chi_0$ range. We note here that, in 
Fig.~\ref{fig:mu_g2}(a) and Fig.~\ref{fig:mu_g2}(b), we refrain from considering
LHC bounds based on with two leptons and missing transverse energy
(see Sec.\ref{Constraintscollider}) on the parameter space. This helps us to
study the dependence of different parameters on the $\Delta a_\mu$ numerically
and
to choose a valid parameter space which is consistent with
the LHC searches. For instance, a light $L_2$ accompanied with a light scalar
$S$ may easily accommodate $\Delta a_\mu$ with a perturbative value of
$Y_{3\mu} \sim 2$. We have checked that $Y_{3\mu}$ remains perturbative upto TeV scale even when one includes dominant radiative corrections while at and above
TeV scale beta function of $Y_{3\mu}$ may include new gauge interactions. This
is in particular true if our low energy model is embedded in a TeV scale LR model.
However, as discussed earlier,
the LHC limits can be managed if one assumes a light $\chi_0$ as well. Thus
in the following sections, particularly, in the computation of cLFV
and DM observables,
we would fix a few parameters at values
$M_{L_2}=190$~GeV, $M_{S}=130$~GeV and $M_{\chi_0}=120$~GeV.

\begin{figure}[!ht]
\includegraphics[scale=0.6]{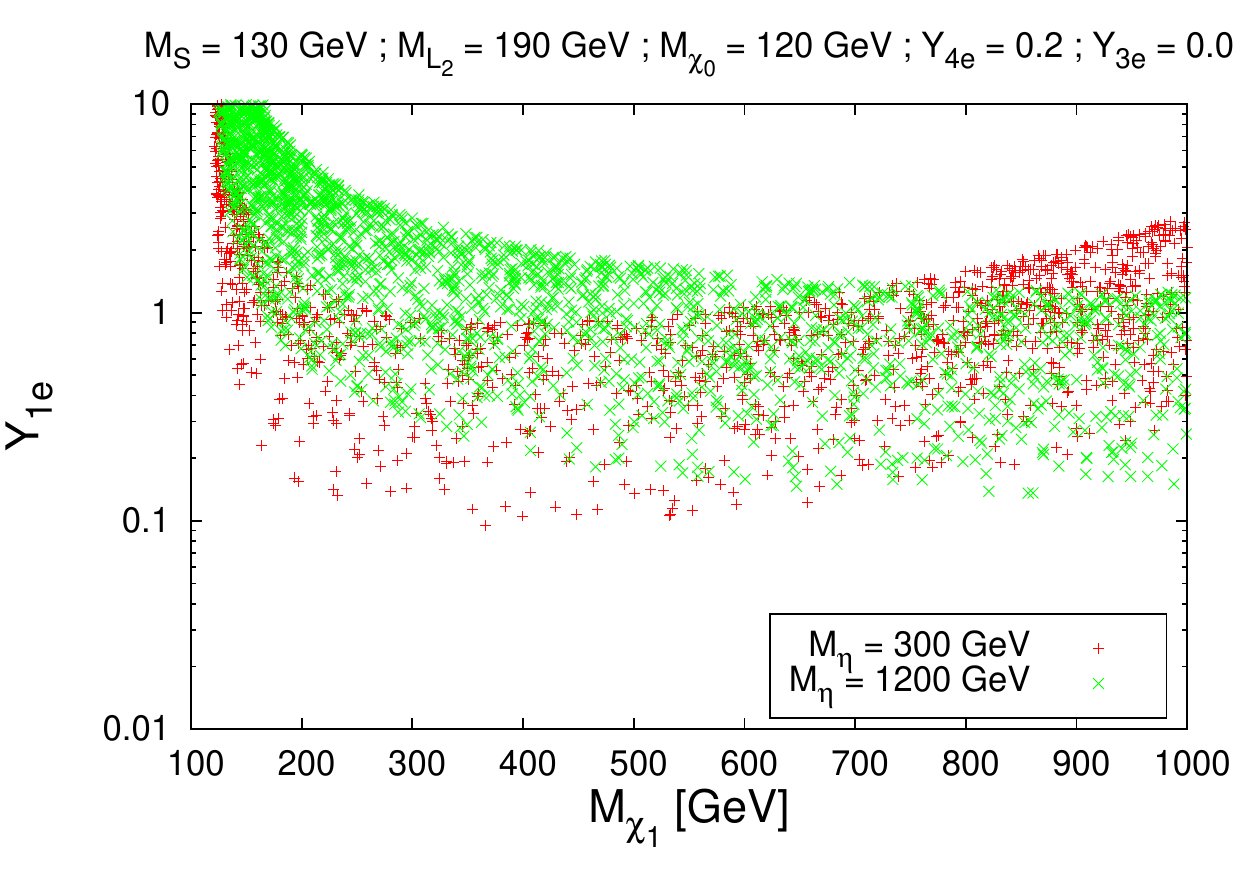}\hspace{1 cm}
\includegraphics[scale=0.6]{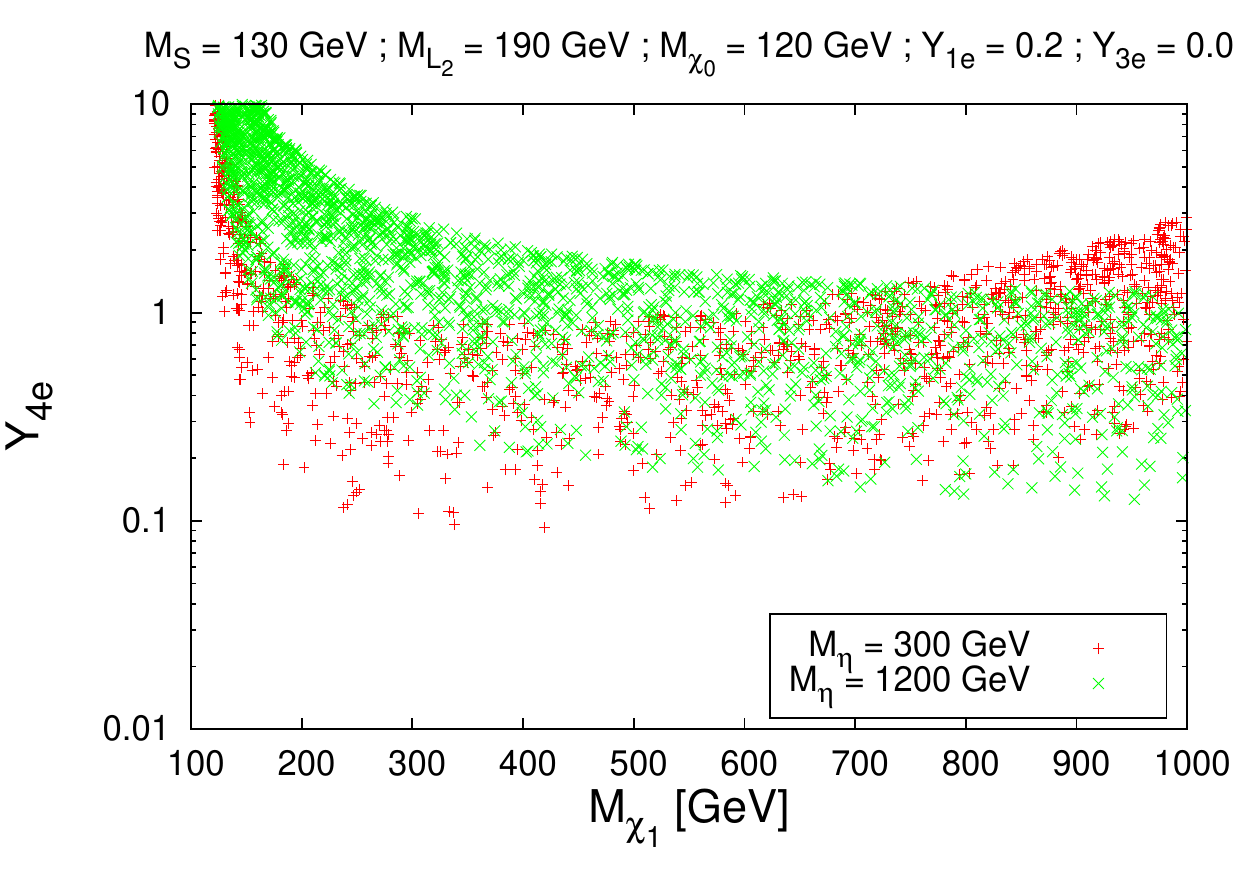}
\begin{center}
(a)\hspace{8 cm}(b)\\
\includegraphics[scale=0.6]{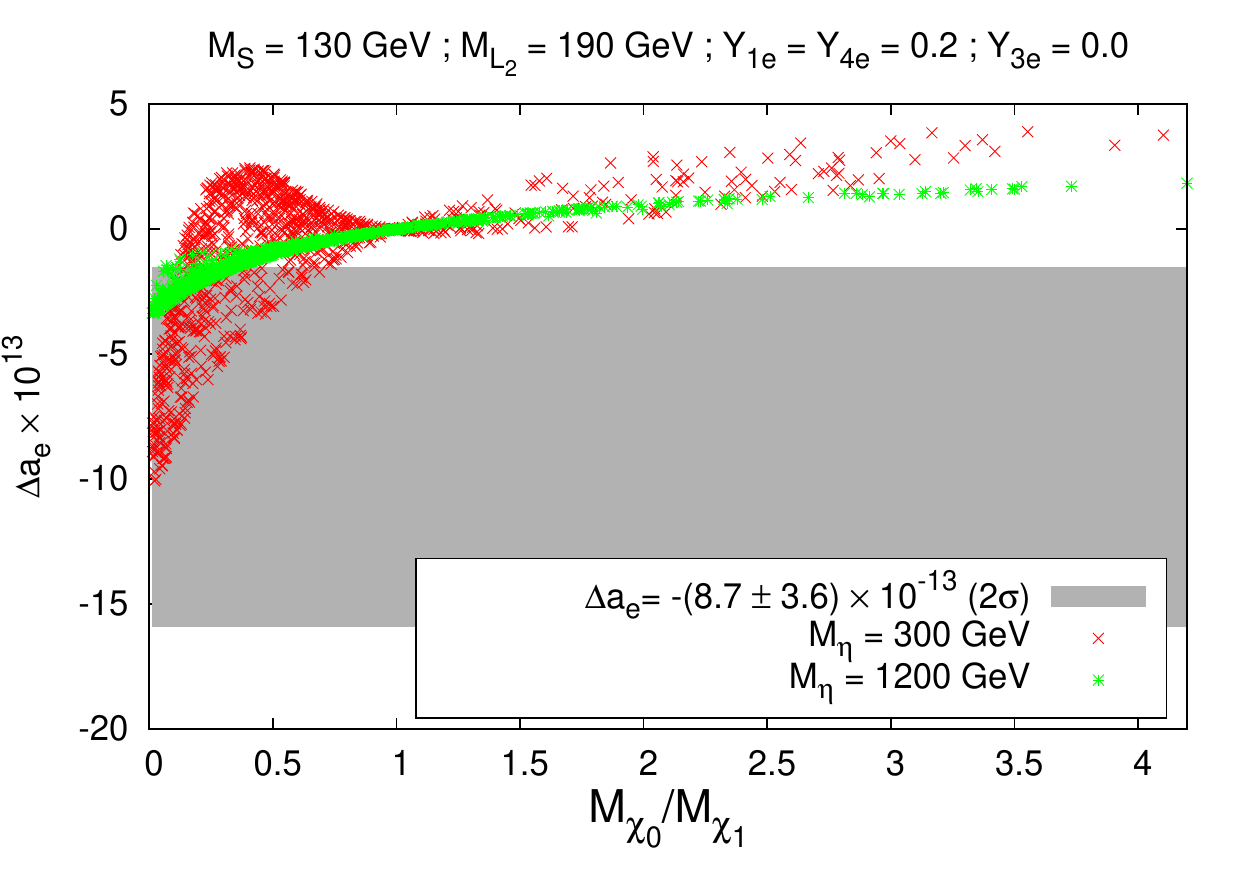}\\
(c)
\end{center}
\caption{Allowed parameter space satisfying $\Delta a_e$ within $2\sigma$
  bound. As before $\sin\theta=0.01$ is taken. Here, red and green dots
  represent the scenarios corresponding
to $M_\eta=300$ GeV and $1200$ GeV respectively.}
\label{fig:g_2e}
\end{figure}

In our next precision calculation, we will now see the role 
of different parameters in obtaining a correct value for $\Delta a_e$.
Note that, here, for practical purposes, one finds
$\Delta a_{e} \simeq \Delta a_{e}^{Y_{4(e)}Y_{1(e)}}$. The reasons are as follows.
$\Delta a_{e}^{Y_{1(e)}}$ becomes insignificant due to cancellation between different
terms. Moreover,
$\Delta a_{e}^{Y_{1(e)}}$ and $\Delta a_{e}^{Y_{4(e)}}$ are $\propto m_e^2$, thus,
are much suppressed and 
can be neglected
for the parameter space, we are interested in. Additionally,
we choose $Y_{3e}=0$ to forbid the
positive part in Eq.~\eqref{eq:a_mu3}. So we may re-express $\Delta a_e$ as
follows:\\
\begin{align}
  \Delta a_e\simeq-\frac{(Y_{1e})(Y_{4e})\sin 2\theta \,m_e}{16\pi^2 M_\eta^2}
  \Bigg[M_{\chi_1}F_1\left(\frac{M_{\chi_1}^2}{M_\eta^2}\right)-M_{\chi_0}
    F_1\left(\frac{M_{\chi_0}^2}{M_\eta^2}\right)\Bigg].
\label{eq:ae_can}
\end{align}

As before, in the numerical analysis, we fixed $M_S=130$ GeV, $M_\eta=300$
and $1200$ GeV,
$M_{L_2}=190$ GeV and $M_{\chi_0}=120$ GeV. Fig.~\ref{fig:g_2e}(a) depicts
the variation of
$Y_{1e}$ as a function of $M_{\chi_1}$, when $Y_{4e}$ is fixed at $0.2$.
And similarly for the
Fig.~\ref{fig:g_2e}(b), where $Y_{4e}$ appears as the variable and $Y_{1e}$
is fixed at
$0.2$. In both of these plots red and green dots represent the scenarios
corresponding
to $M_\eta=300$ GeV and $1200$ GeV respectively. Note that, for the smaller
values of
$M_{\chi_1}$, there is a difference between the allowed regions corresponding
to $M_\eta=300$
GeV and $1200$ GeV, while at the higher values both the red and green dots
merge~
[see Fig.~\ref{fig:g_2e}(a) and (b)]. In the lighter $\chi_1$ regime,
where $M_{\chi_1}\sim M_{\chi_0}$, a partial cancellation in
the bracketed part of Eq.~\eqref{eq:ae_can} can be observed. The suppression
is more for a heavier $\eta$, thus, a larger coupling can be helpful to tune
$\Delta a_e$.
On the other hand, for larger values of $M_{\chi_1}$, the term
$M_{\chi_1}F_1\left(\frac{M_{\chi_1}^2}{M_\eta^2}\right)$ may appear to have the
leading
contribution. At this large $M_{\chi_1}$ region, for a fixed value of $M_{\chi_1}$, $F_1\left(\frac{M_{\chi_1}^2}{M_\eta^2}\right)$ increases with the increasing value of $M_\eta$. However, the overall term $\frac{1}{M_\eta^2} F_1\left(\frac{M_{\chi_1}^2}{M_\eta^2}\right)$ becomes somewhat insensitive to the variation in $M_\eta$ and hence only a slight increment in Yukawa coupling can be observed for the lighter $M_\eta$ value.

Fig.~\ref{fig:g_2e}(c) shows the scaled variation of $\Delta a_e$
as a function of $M_{\chi_0}/M_{\chi_1}$ where we have again relaxed the potential
constraints coming from LHC. All the other masses and couplings
are fixed as before. The grey patch represents the $2\sigma$
range of the
$\Delta a_e$. One can easily see that, a
small mass ratio ($<0.25$, $<0.75$) for $M_\eta=1200,\,300$ GeV
respectively, can lead to the desired
negative contribution. For larger
$M_\eta$, to compensate the suppression,
a lighter ${\chi_0}$ is desired to produce the correct value for $\Delta a_e$. 
On the other hand, with increasing $M_{\chi_0}/M_{\chi_1}$ the positive
contribution starts to increase for a fixed $M_\eta$
[see Eq.~\eqref{eq:ae_can}] and hence a correct value of $\Delta a_e$
would be difficult to obtain. 

As a final remark, it is now evident that the
presence of the two VL states $L_1$ and $L_2$
are necessary
to accommodate the both $\Delta a_\mu$ and $\Delta a_e$.
The second doublet $L_2$ may provide the sole contribution to muon magnetic
moment, while the other one can be used to tune the magnitude and
sign of the $e$ magnetic moment. Moreover, we will find that, satisfying
different cLFV processes may become much easier in this scenario.

\subsection{cLFV constraints}
\label{subsec:cLFV}
In this model framework, in computing the cLFV observables we closely
follow Refs.~\cite{Hisano_1996,Arganda:2005ji}.  One-loop effective
vertices, relevant for the different two and three body processes 
$\ell_\alpha\rightarrow \ell_\beta\gamma$ or $\ell_\alpha\rightarrow 3\ell_\beta$ are generated through the interactions among BSM fermions~($\chi_a$, $L_a^\pm$), scalars $\eta$ and $S$  
and the SM leptons.

\subsubsection{$\ell_\alpha\rightarrow \ell_\beta\gamma$}
We start with the form factors for $\ell_\alpha
\rightarrow \ell_\beta \gamma$, where the relevant diagrams have been depicted
in Fig.~\ref{fig:lep_g}. The details
of the calculation are presented in Appendix C. Here we recast the form
factors $A^{(n)L,R}_{2}$ and $A^{(c)L,R}_{2}$
related to neutral and charged fermions in terms of our model
parameters respectively.
\begin{align} 
  A^{(n)L}_{2}=&\frac{1}{32\pi^2M^2_\eta}\Bigg[Y^\dagger_{1\beta}Y^\dagger_{4\alpha}\sin\theta \cos\theta
    \left\lbrace\frac{2M_{\chi_1}}{m_{\ell_\alpha}}\,F_1\left(\frac{M^2_{\chi_1}}{M^2_\eta}\right)
-\frac{2M_{\chi_0}}{m_{\ell_\alpha}}\,F_1\left(\frac{M^2_{\chi_0}}{M^2_\eta}\right)\right\rbrace\nonumber\\
+& Y^\dagger_{1\beta} Y_{1\alpha} \cos^2\theta \,F_2\left(\frac{M^2_{\chi_1}}{M^2_\eta}\right)+
Y_{4\beta} Y^\dagger_{4\alpha} \cos^2\theta \, \frac{m_{\ell_\beta}}{m_{\ell_\alpha}}F_2
\left(\frac{M^2_{\chi_0}}{M^2_\eta}\right)\Bigg]~,
\label{eq:AnL2M} \\
&A^{(c)L}_{2}=\frac{1}{32\pi^2M^2_S}Y_{3\beta}Y^\dagger_{3\alpha}\frac{m_{\ell_\beta}}{m_{\ell_\alpha}} F_3\left(\frac{M^2_{L_2}}{M^2_S}\right)                      
  +\frac{1}{32\pi^2M^2_\eta}Y^\dagger_{1\beta}Y_{1\alpha} F_3\left(\frac{M^2_{L_1}}{M^2_\eta}\right)~,\nonumber \\
  &A^{(n)R}_{2}=A^{(n)L}_{2}|_{Y_{4}\leftrightarrow Y^\dagger_{1}~,~
    F_2\left(\frac{M^2_{\chi_1}}{M^2_\eta}\right)\leftrightarrow
    F_2\left(\frac{M^2_{\chi_0}}{M^2_\eta}\right)}~,~~~~~
  A^{(c)R}_{2}=A^{(c)L}_{2}|_{Y_{3}\leftrightarrow Y^\dagger_{1}}~.
  \label{eq:AcL2M}
\end{align}

Finally, the coefficients in the above can be clubbed to get the total contributions.
\begin{equation} 
  A^{L,R}_2 = A^{(n)L,R}_{2}+A^{(c)L,R}_{2} ~,
  \label{eq:L-llgab}
  \end{equation}
The decay width is given by~\cite{Hisano_1996,Arganda:2005ji}
\begin{align}
&\Gamma \left( \ell_\alpha \to \ell_\beta \gamma \right) = \frac{\alpha_{em} m_{\ell_\alpha}^5}{4} \left( |A_2^L|^2 + |A_2^R|^2 \right)~.\nonumber\\
&{\rm Br} \left( \ell_\alpha \to \ell_\beta \gamma \right) = \tau_{\alpha}\frac{\alpha_{em} m_{\ell_\alpha}^5}{4} \left( |A_2^L|^2 + |A_2^R|^2 \right)~.
\label{eq:Br}
\end{align}
where $\alpha_{em}$ is the electromagnetic fine structure constant and $\tau_{\alpha}$ is the lifetime of $\ell_\alpha$.

\subsubsection{$\ell_\alpha\rightarrow 3\ell_\beta$}
Here we calculate the decay width for the processes where a heavier
SM lepton decays into three lighter leptons of the same flavor,
i.e., $\ell^-_\alpha\rightarrow\ell^-_\beta
\ell^-_\beta\ell^+_\beta$. 
We present the relevant $\gamma$-penguin, $Z$-penguin and Box diagrams contributions to get the complete decay width and hence
the branching ratio for 
$\ell_\alpha\rightarrow 3\ell_\beta$ processes. The details
of the calculation can be found in Appendix C.
\begin{figure}[H]
\includegraphics[scale=0.5]{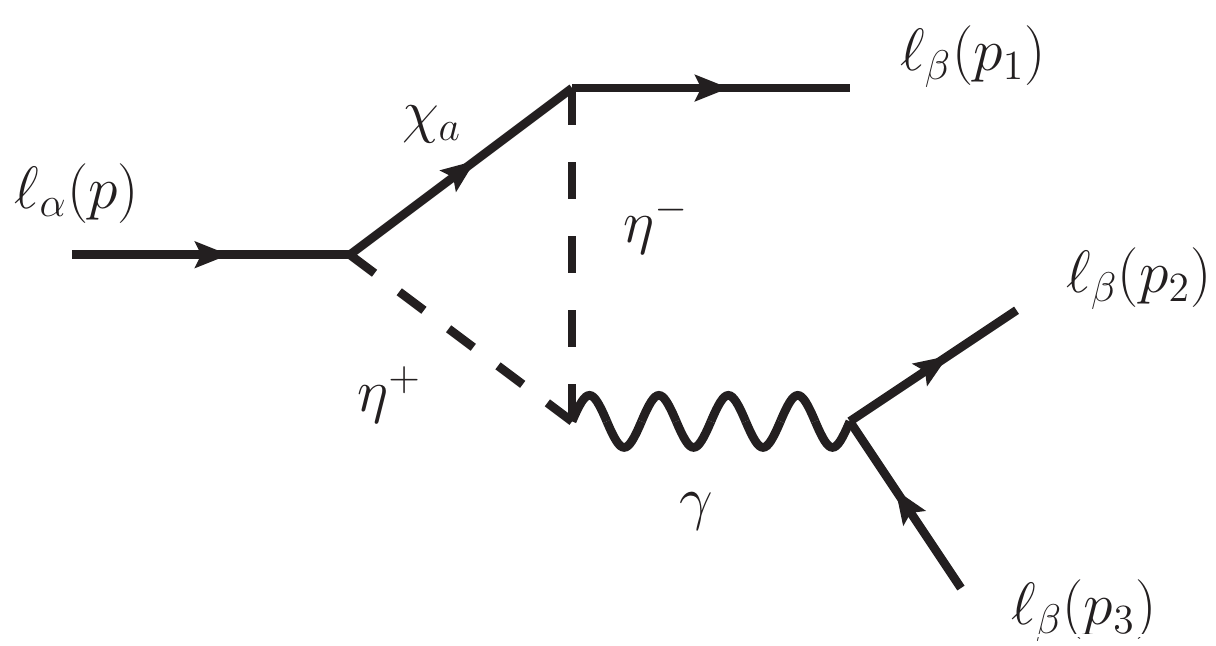}\hspace{3 cm}
\includegraphics[scale=0.5]{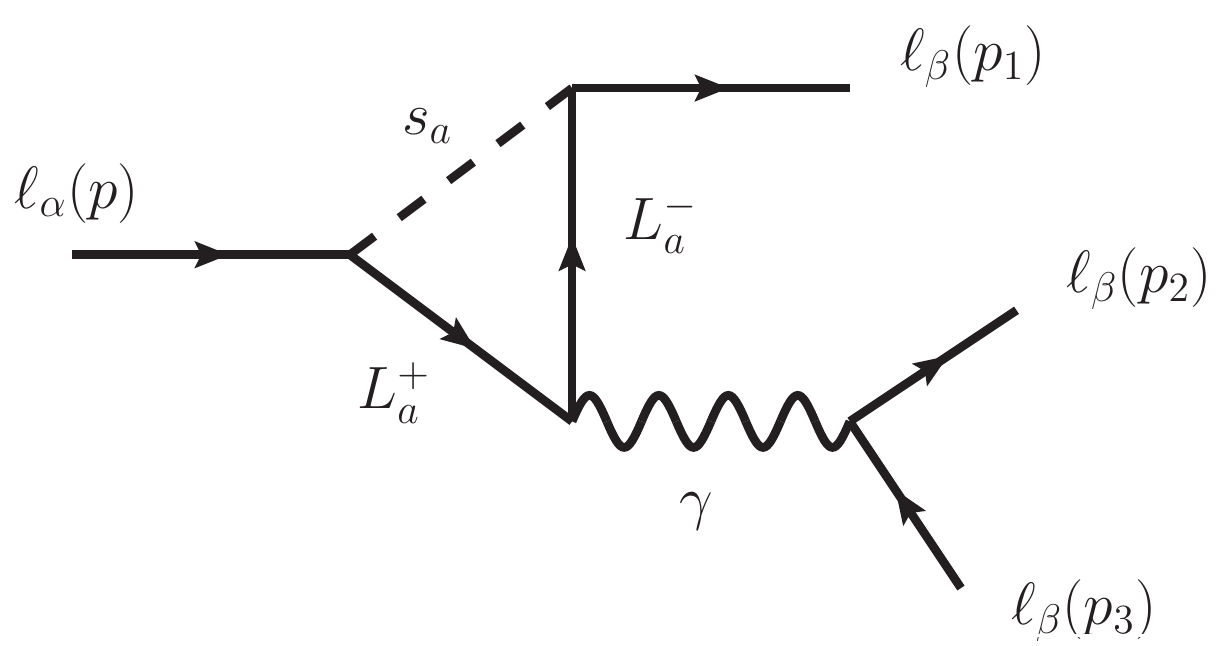}\\
\caption{$\gamma$-penguin diagrams contributing to the
  $\ell_\alpha^- \rightarrow \ell_\beta^- \ell_\beta^-\ell_\beta^+$ decay.
  The index $a$ reads $0,1$ for neutral and $1,2$ for charged fermions 
and $s_1=\eta^0$, $s_2=S$ for the charged lepton loops. The 
corresponding leg-corrections (not shown) are also taken into account.}
\label{fig:gamma_peng}
\end{figure}
\begin{itemize}
\item
  Photon penguin contribution:  As shown in Fig.~\ref{fig:gamma_peng}, the monopole contributions can be recast in terms of
  our model parameters,
\begin{align}
  &A_1^{(n)L}=\frac{1}{576\pi^2M_\eta^2}\Bigg[Y_{4\beta} Y^\dagger_{4\alpha} \cos^2\theta \,
    F_4\left(\frac{M^2_{\chi_0}}{M_\eta^2}\right)\Bigg]~
  ,~A_1^{(n)R}=A_1^{(n)L}|_{{Y_{4}\rightarrow Y^\dagger_{1}},M_{\chi_0}\rightarrow M_{\chi_1}}~, \\
&A_1^{(c)L}=-\frac{1}{576\pi^2M_S^2}\Bigg[ Y_{3\beta} Y^\dagger_{3\alpha}\,F_5\left(\frac{M^2_{L_2}}{M_S^2}\right)\Bigg]~,~
  A_1^{(c)R}= A_1^{(c)L}|_{{Y_{3}\rightarrow Y^\dagger_{1}},M_{L_2}\rightarrow M_{L_1},M_S\rightarrow M_\eta}~.
\end{align}
The dipole contributions can be read from  Eq.~\eqref{eq:AnL2M} and 
Eq.~\eqref{eq:AcL2M}.
\item
  $Z$ penguin contribution:  Dominant Feynman diagrams are shown in Fig.~\ref{fig:Z_peng}.
  We have calculated the coefficients as follows:
  \begin{figure}[H]
\includegraphics[scale=0.5]{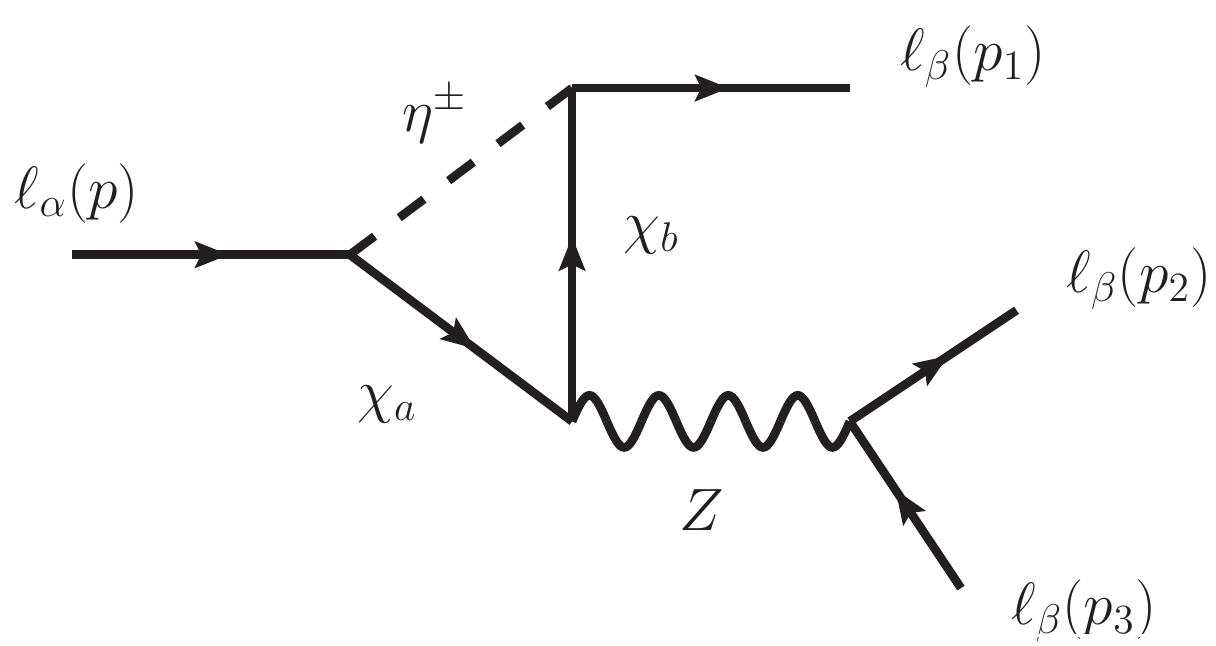}\hspace{3 cm}
\includegraphics[scale=0.5]{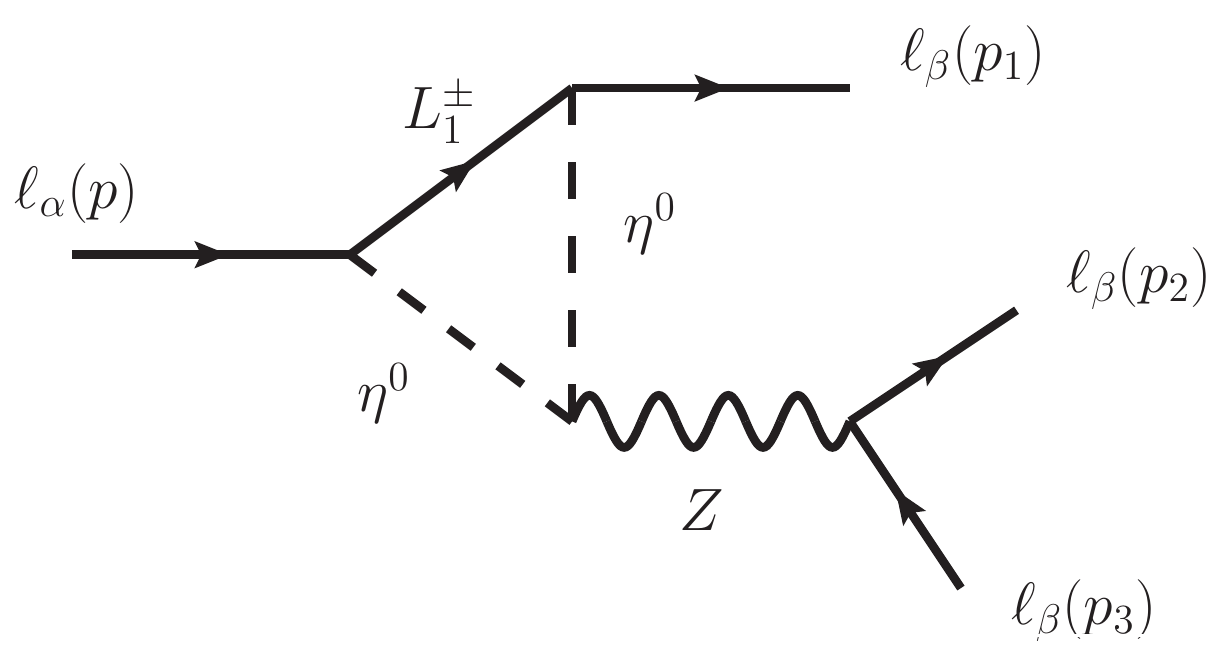}\\
\includegraphics[scale=0.5]{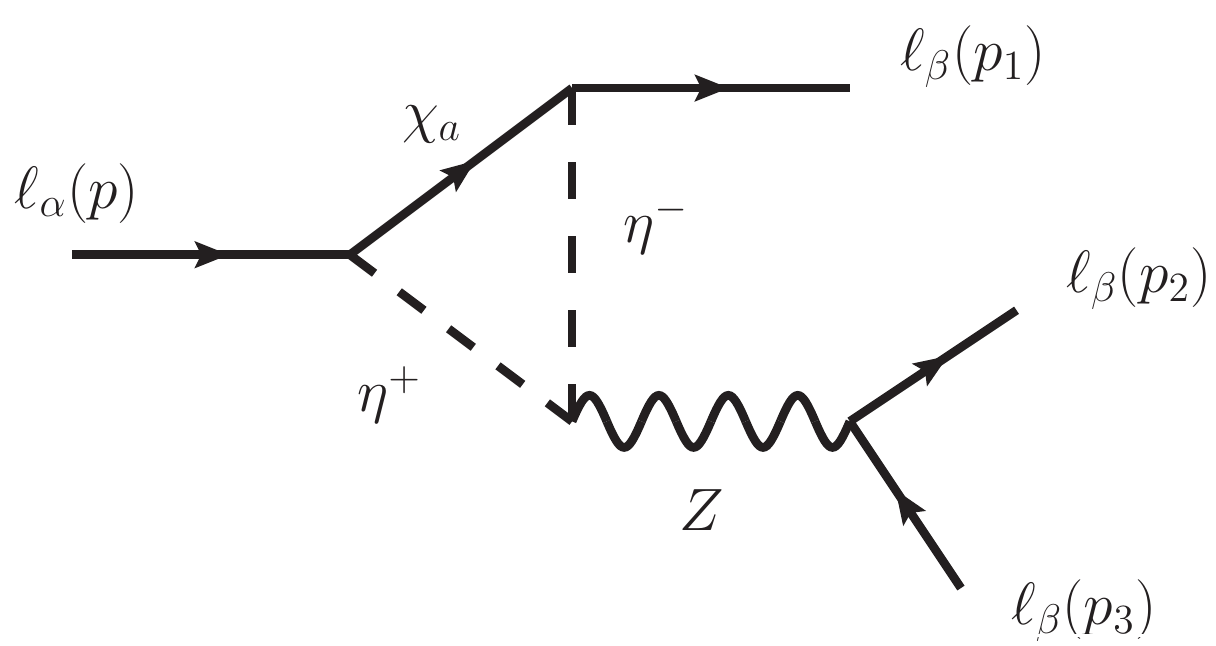}\hspace{3 cm}
\includegraphics[scale=0.5]{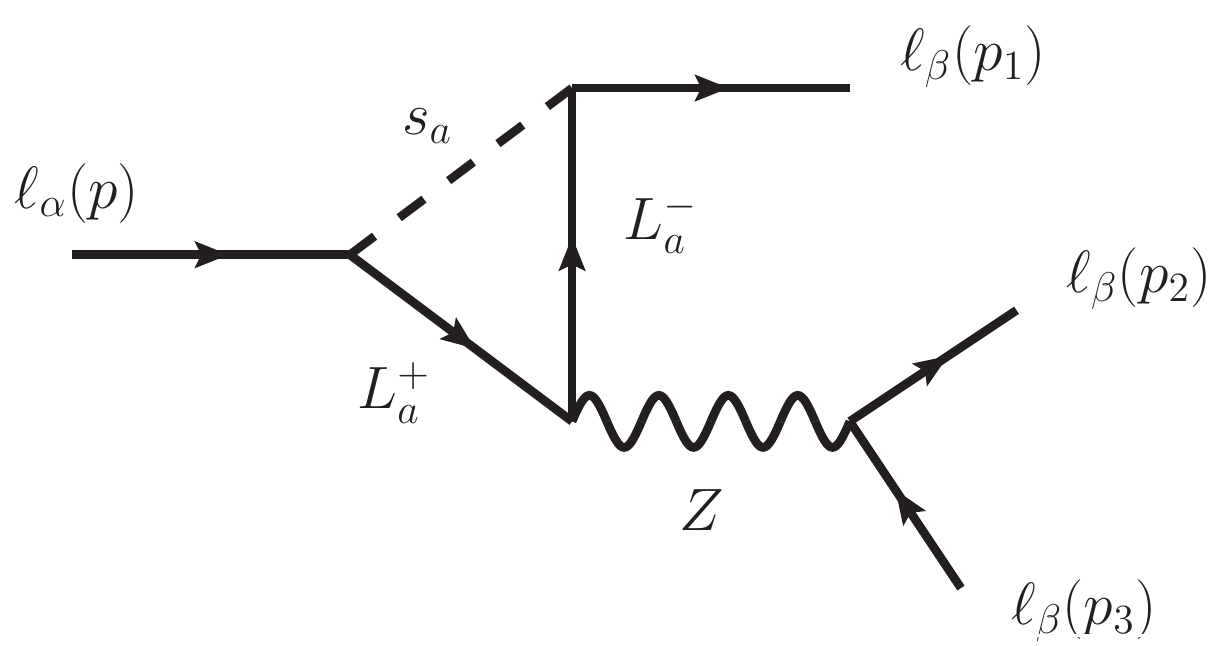}\\
\caption{Leading $Z$ penguin diagrams contributing to the $\ell_\alpha^- \rightarrow \ell_\beta^- \ell_\beta^-\ell_\beta^+$ decay.
  Leg corrections are also considered (not shown). 
Indices $a,b=1,2$~(for $L^\pm_{a,b}$), $0,1$~(for $\chi_{a,b}$) and $s_1=\eta^0$, $s_2=S$ as before.}
\label{fig:Z_peng}
\end{figure}

The expressions for the form factors are given below~\cite{Krauss:2013gya,PhysRevD.91.059902,Arganda:2014lya}:
\begin{align}
\label{eq:FL}
  F^{(n)}_{L}=&-\frac{1}{16\pi^2}\sum_{a,b=0,1}\Bigg[Y_{4\beta}Y^\dagger_{4\alpha}U_bU_a \Bigg\{ E^{R(n)}_{ba}\left(2C
    _{24}(M_\eta^2,M_{\chi_a}^2,M_{\chi_b}^2)-\frac{1}{2}\right)-\nonumber\\ &\qquad\qquad\qquad\qquad\qquad\qquad\qquad E^{L(n)}_{ba}M_{\chi_a}M_{\chi_b}C_0(M_\eta^2,M_{\chi_a}^2,M_{\chi_b}^2)\Bigg\} \nonumber\\
    &+Y_{4\beta}Y^\dagger_{4\alpha}U^2_a \left\lbrace 2Q_{\eta\eta}C_{24}(M_{\chi_a}^2,M_\eta^2,M_\eta^2)\right\rbrace +
    Y_{4\beta}Y^\dagger_{4\alpha}U^2_a \left\lbrace g_L^{(\ell)}B_1(M_{\chi_a}^2,M_\eta^2)\right\rbrace\Bigg],\\
F^{(c)}_{L}=&-\frac{1}{16\pi^2}\Bigg[ Y_{3\beta}Y^\dagger_{3\alpha}\left\lbrace E^{R(c)}_{22}\left(2C_{24}(M_S^2,M_{L_2}^2,M_{L_2}^2)-\frac{1}{2}\right)-E^{L(c)}_{22}M_{L_2}M_{L_2}C_0(M_S^2,M_{L_2}^2,M_{L_2}^2)\right\rbrace \nonumber\\
  &+Y_{3\beta}Y^\dagger_{3\alpha}\left\lbrace 2Q_{22}C_{24}(M_{L_2}^2,M_S^2,M_S^2)\right\rbrace + Y_{3\beta}Y^\dagger_{3\alpha}\left\lbrace
  g_L^{(\ell)}B_1(M_{L_2}^2,M_S^2)\right\rbrace\Bigg]~,\\
F^{(n)}_{R}=&{F^{(n)}_{L}}|_{{Y_{4}\rightarrow Y^\dagger_{1}},U\rightarrow U^\prime, g_L^{(\ell)}\rightarrow g_R^{(\ell)}}~,F^{(c)}_{R}=
{F^{(c)}_{L}}|_{{Y_{3}\rightarrow Y^\dagger_{1}},M_{L_2}\rightarrow M_{L_1},M_S\rightarrow M_\eta, Q_{22}\rightarrow Q_{11},g_L^{(\ell)}\rightarrow g_R^{(\ell)}}.
\label{eq:FR}
\end{align}
As before, $F_{L,R}=F^{(n)}_{L,R}+F^{(c)}_{L,R}$. The generic forms of $C_{24}$, $C_0$ and $B_1$ functions are listed in
Appendix B.

\item
  Box diagram contributions: Leading contributions are shown in
  Fig.~\ref{fig:Box}. 
  \begin{figure}[H]
\includegraphics[scale=0.5]{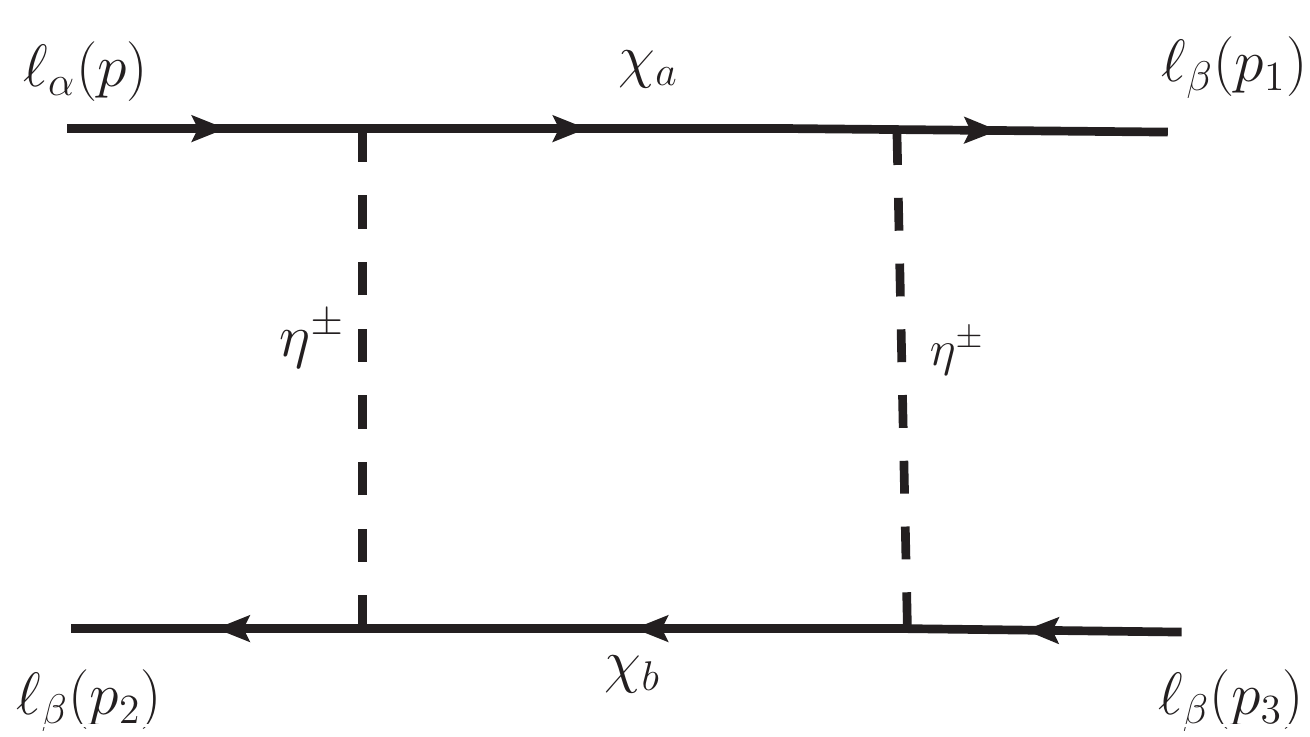}\hspace{3 cm}
\includegraphics[scale=0.5]{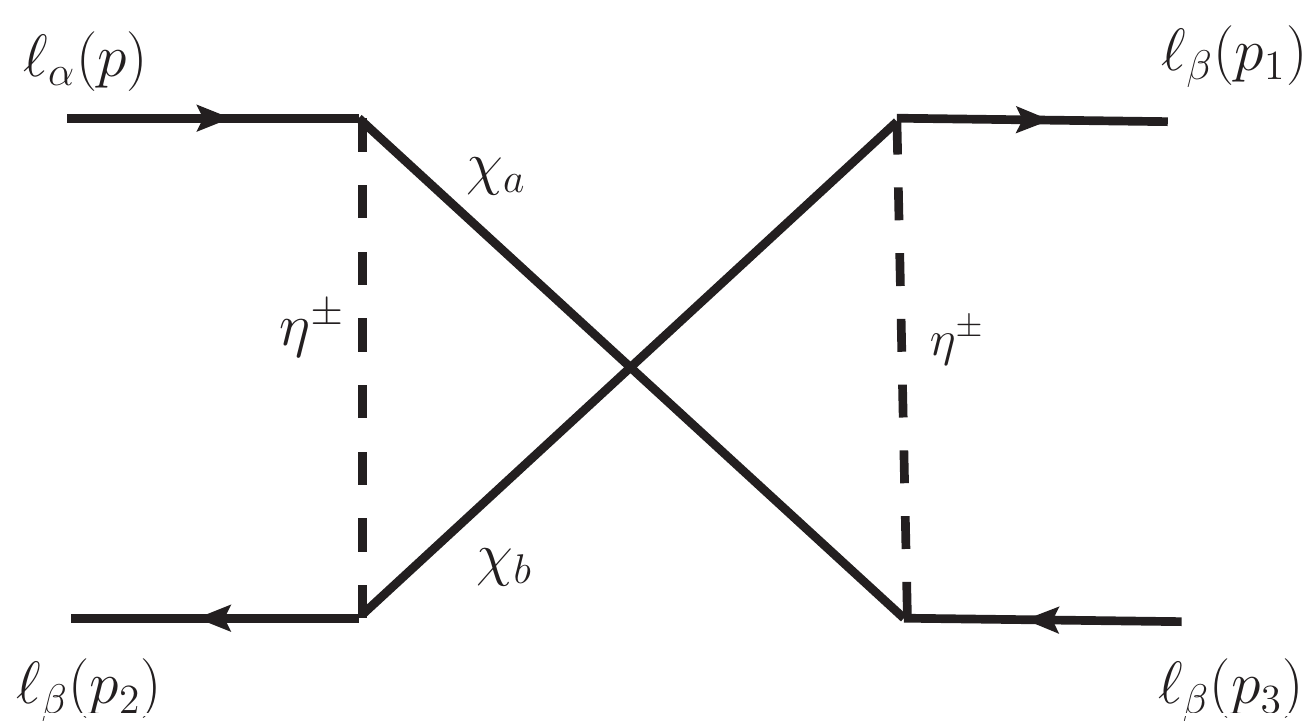}\\
\begin{center}
\includegraphics[scale=0.5]{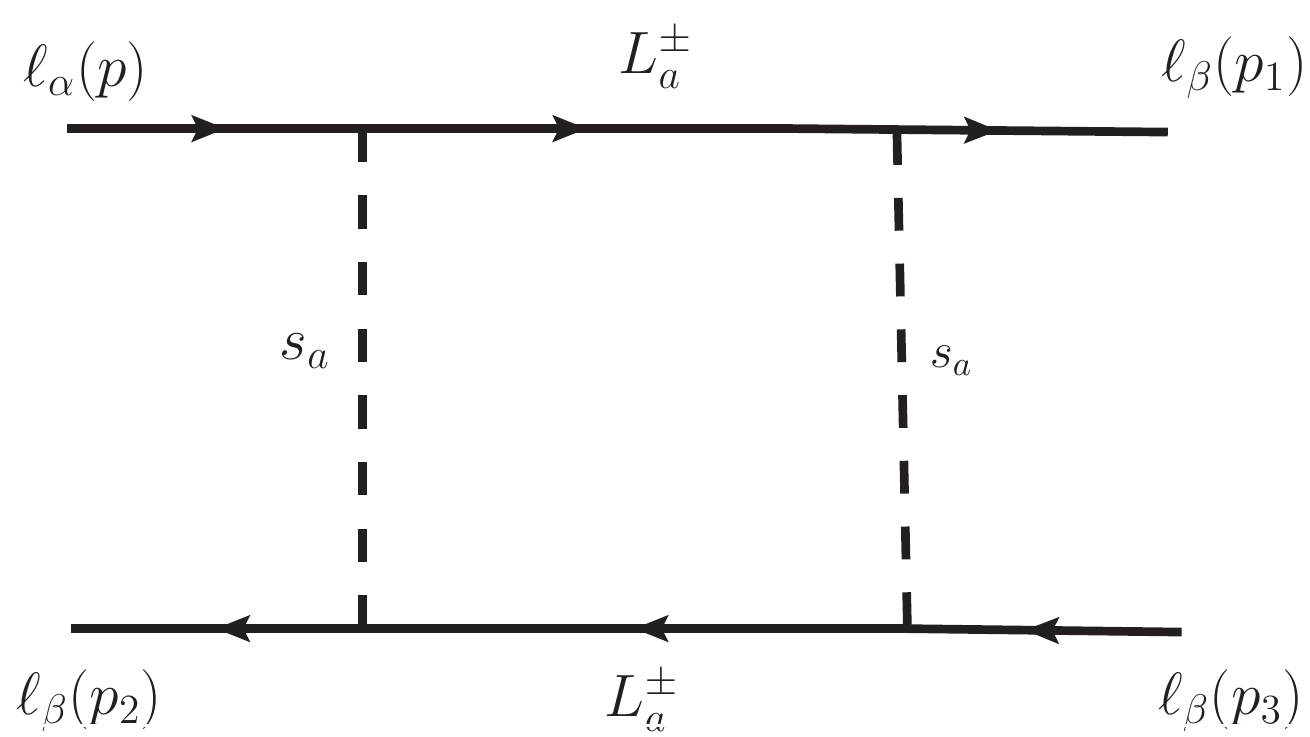}
\end{center}
\caption{Box diagrams contributing to the $\ell_\alpha^- \rightarrow \ell_\beta^- \ell_\beta^-\ell_\beta^+$ decay.
  As before, $a,b=1,2$~(for $L^\pm_{a,b}$), $0,1$~(for $\chi_{a,b}$) and $s_1=\eta^0$, $s_2=S$.}
\label{fig:Box}
\end{figure}
  The dominant $B$-factors can be calculated as,
\begin{align}
  e^2B_1^{(n)L}=&\frac{1}{16\pi^2}\Bigg[\frac{\tilde{D}_0}{2}Y_{4\beta}{Y^\dagger_{4\alpha}}Y_{4\beta}{Y^\dagger_{4\beta}}|U_a|^2|U_b|^2 + D_0\,M_{\chi_a}M_{\chi_b}Y_{4\beta}Y_{4\beta}Y^\dagger_{4\alpha}{Y^\dagger_{4\beta}}U_b^2 U^{\dagger2}_a \Bigg]\label{eq:box1}~,
\end{align}

\vskip -0.3cm
\begin{align}
  e^2B_2^{(n)L}=\frac{1}{16\pi^2}
\Bigg[\frac{\tilde{D}_0}{4} Y_{4\beta}{Y^\dagger_{4\alpha}}{Y^\dagger_{1\beta}Y_{1\beta}}
  |U_a|^2|U^\prime_b|^2 - \frac{D_0}{2}\,M_{\chi_a}M_{\chi_b}
  Y^\dagger_{1\beta}Y^\dagger_{4\alpha}Y_{4\beta}Y_{1\beta}U^\prime_a
    U^{\dagger}_a U_b U^{\prime\dagger}_b \nonumber\\
    -\frac{\tilde{D}_0}{4}
    Y^\dagger_{1\beta}Y_{4\beta} Y^\dagger_{4\alpha}Y_{1\beta}U^\prime_b U_bU^\dagger_a U^{\prime\dagger}_a
    +\frac{\tilde{D}_0}{4}Y_{4\beta} Y^\dagger_{1\beta}
    Y^\dagger_{4\alpha}Y_{1\beta}U_b U^\prime_bU^{\dagger}_aU^{\prime\dagger}_a \Bigg]~,\label{eq:box2} 
\end{align}
\begin{align}
  e^2B_1^{(n)R}&=e^2B_1^{(n)L}|_{Y_{4}\rightarrow Y^\dagger_{1}, U \rightarrow U^\prime}~,~~~~~
  e^2B_2^{(n)R}=e^2B_2^{(n)L}|_{Y_{4}\leftrightarrow Y^\dagger_{1}, U \leftrightarrow U^\prime}~.
\end{align}
\begin{align}
  &e^2B_1^{(c)L}=\frac{1}{16\pi^2}\Bigg[\frac{\tilde{D}_0}{2}\,
    Y_{3\beta} Y^\dagger_{3\alpha} Y_{3\beta} Y^\dagger_{3\beta} \Bigg]~,\\
  &e^2B_2^{(c)L}=\frac{1}{16\pi^2}\Bigg[\frac{\tilde{D}_0}{4}\, Y_{3\beta} Y^\dagger_{3\alpha}
    Y^\dagger_{1\beta} Y_{1\beta} -\frac{D_0}{2}\,
    M_{L_a}M_{L_a} Y^\dagger_{1\beta} Y^\dagger_{3\alpha}Y_{3\beta}Y_{1\beta}\Bigg]~,\\
  &e^2B_1^{(c)R}=e^2B_1^{(c)L}|_{Y_{3}\rightarrow Y^\dagger_{1}}~,~~
  e^2B_2^{(c)R}=e^2B_2^{(c)L}|_{Y_{3}\leftrightarrow Y^\dagger_{1}}~.
\end{align}
The generic functional forms for these $D_0$ and $\tilde{D}_0$ are again
available
at Appendix B. 
Though only the dominant terms are mentioned, for numerical purposes, we
calculated
all $B_i^{L,R}\quad[i=1,2,3,4]$. Finally, there may be Higgs penguin
    diagrams as well, but the Higgs couplings to the SM leptons are much
    suppressed~($\sim\mathcal{O}(\leq 10^{-2})$) compared to that of
    $\gamma$ and $Z$, and hence we can ignore them\footnote{In some specific models, Higgs penguin may lead to  significant contributions~\cite{Abada:2011hm,Babu:2002et,Dedes:2002rh}.}.
  \\
\end{itemize}

\subsubsection{Numerical Results}

\begin{figure}[!ht]
\begin{center}
\includegraphics[scale=0.44]{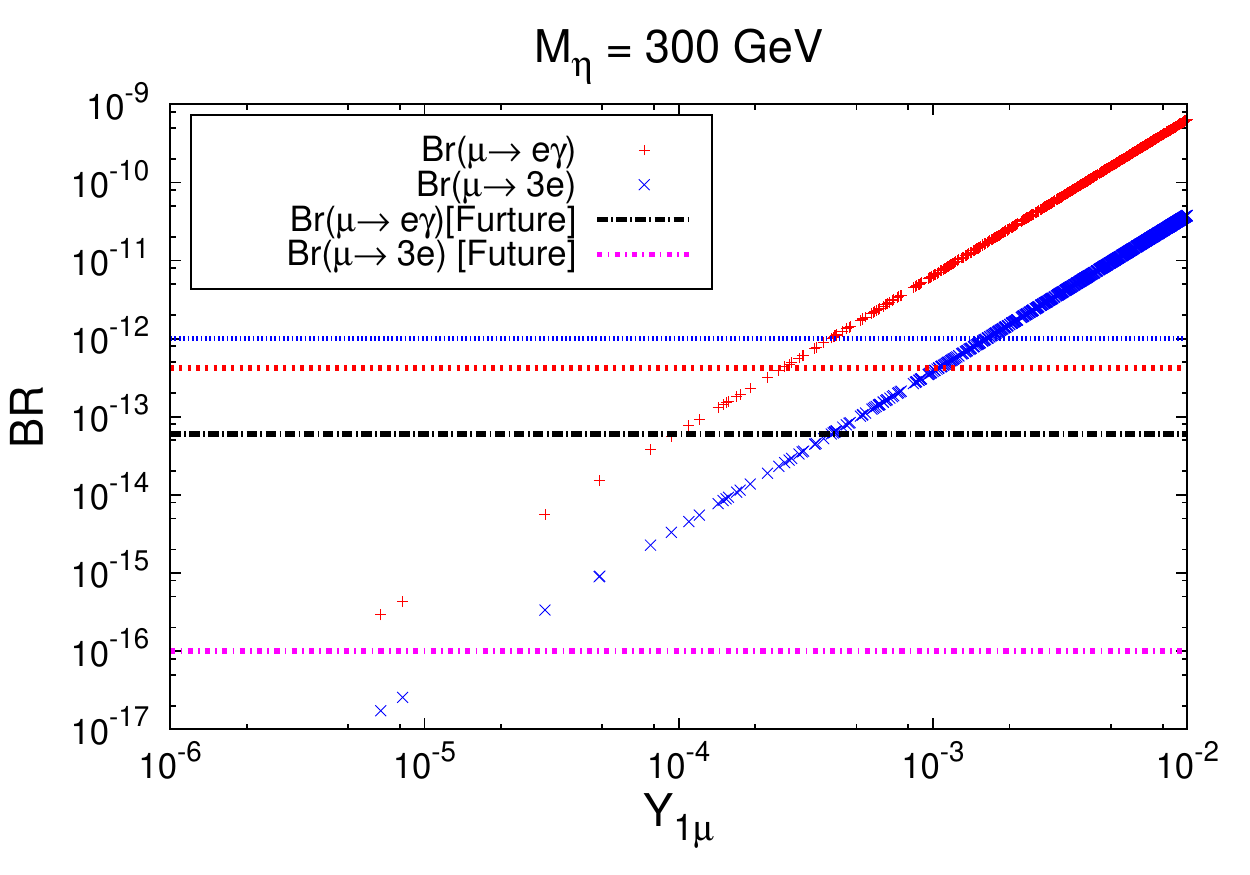}\\
(a)\\
\end{center}
\includegraphics[scale=0.44]{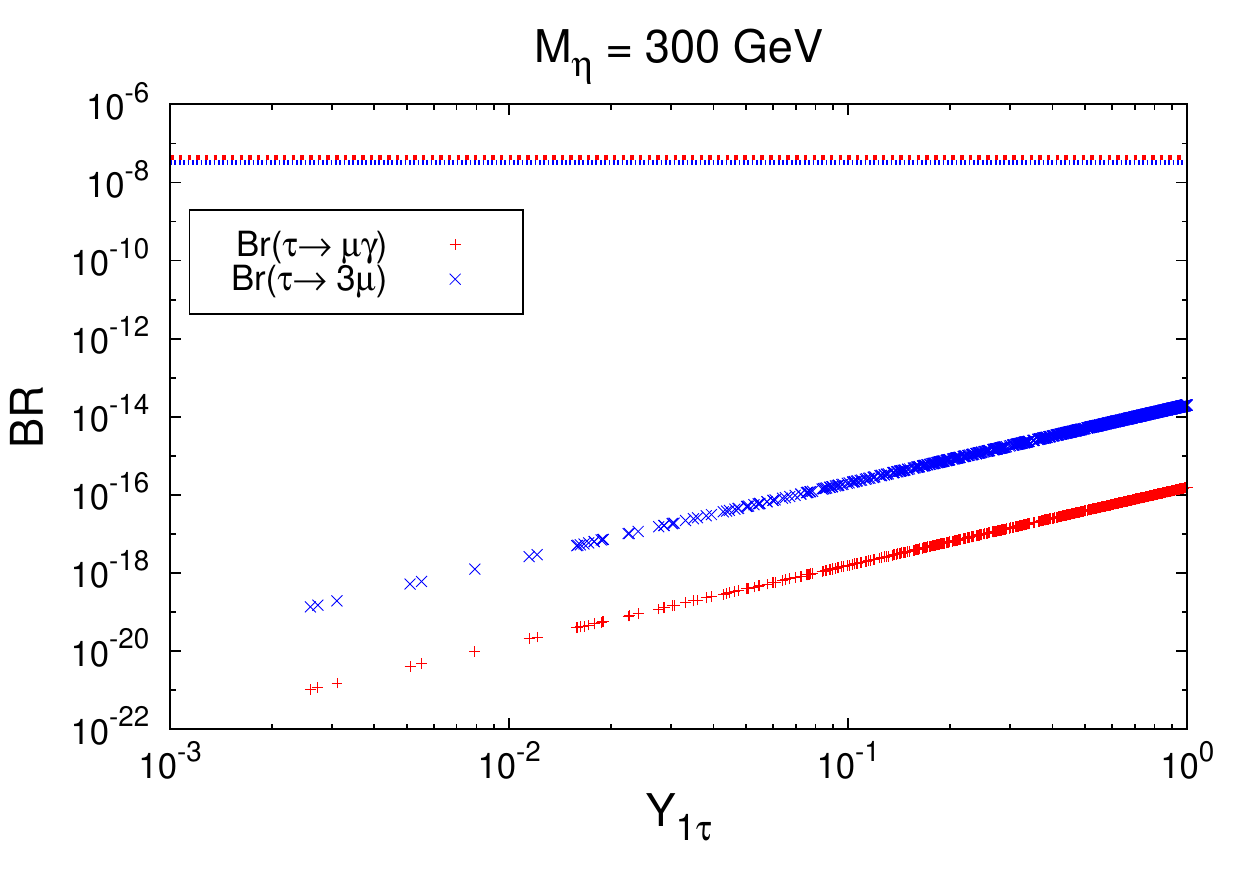}
\includegraphics[scale=0.44]{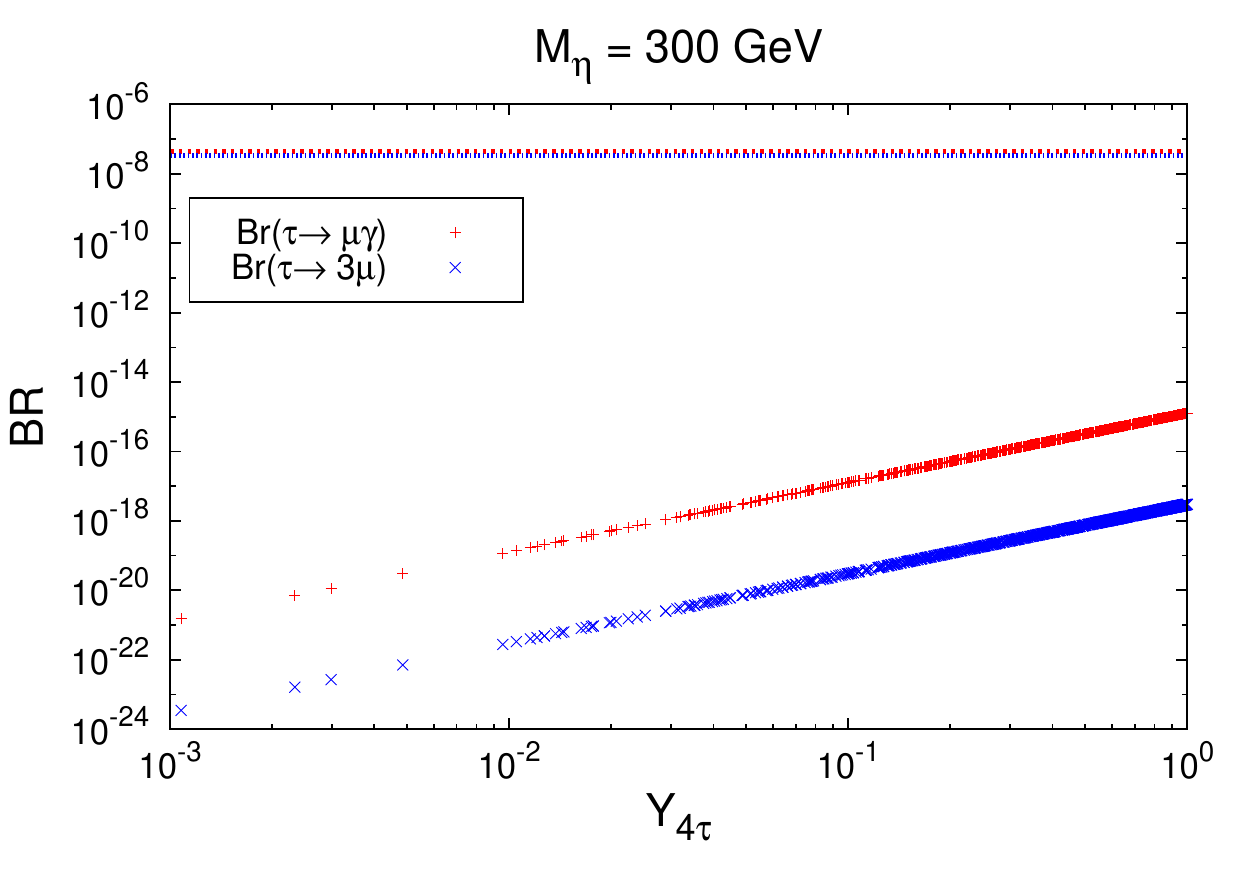}
\includegraphics[scale=0.44]{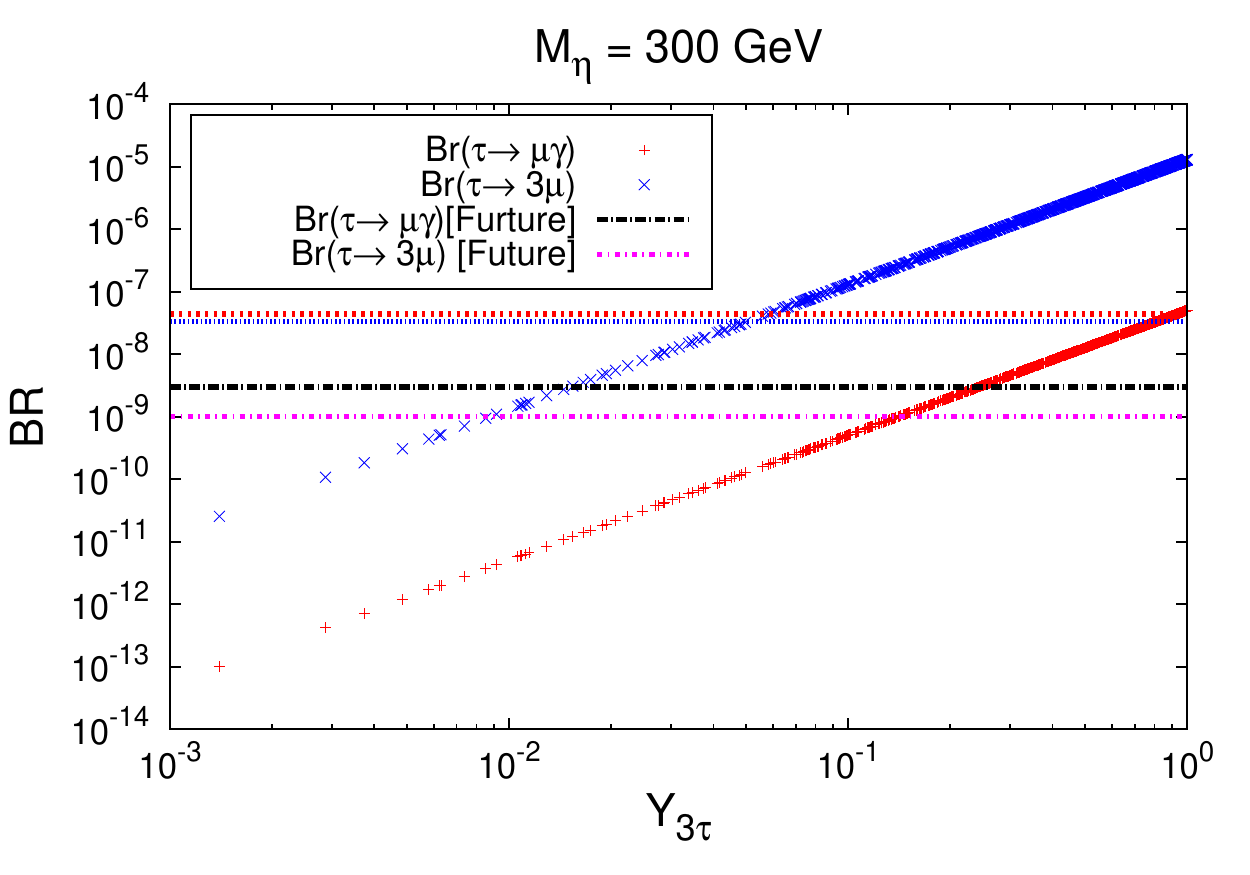}\\
\begin{center}
(b)\hspace{5 cm}(c)\hspace{5.2 cm}(d)\\
\end{center}
\includegraphics[scale=0.44]{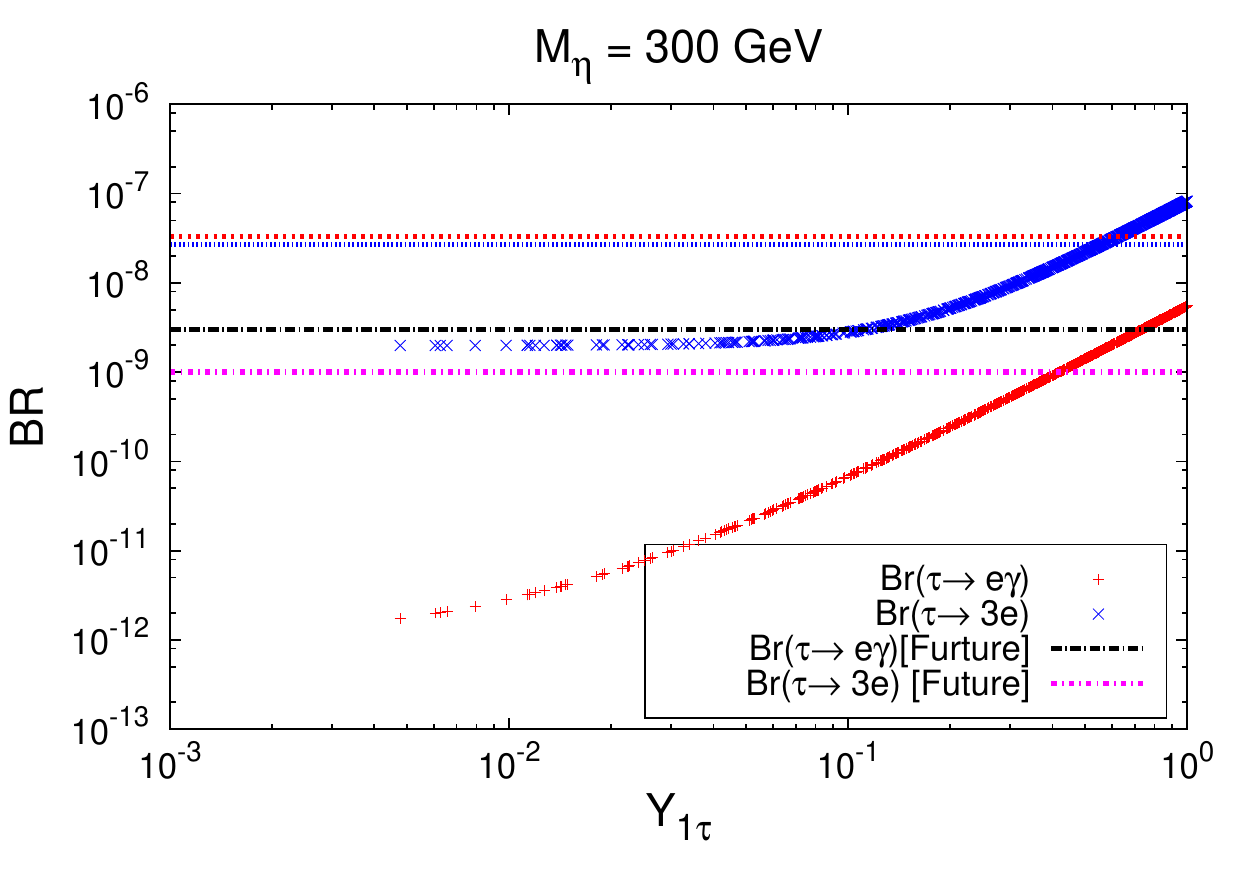}
\includegraphics[scale=0.44]{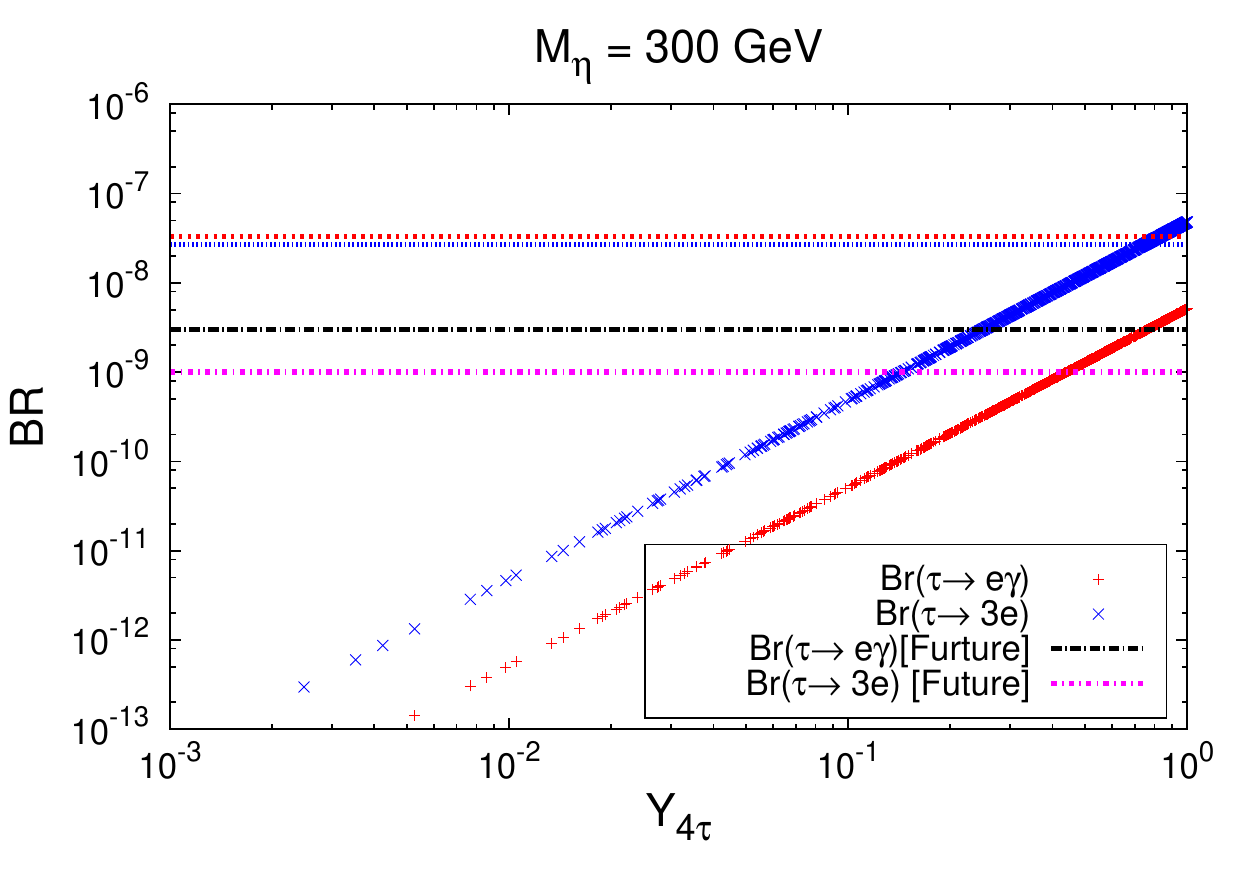}
\includegraphics[scale=0.44]{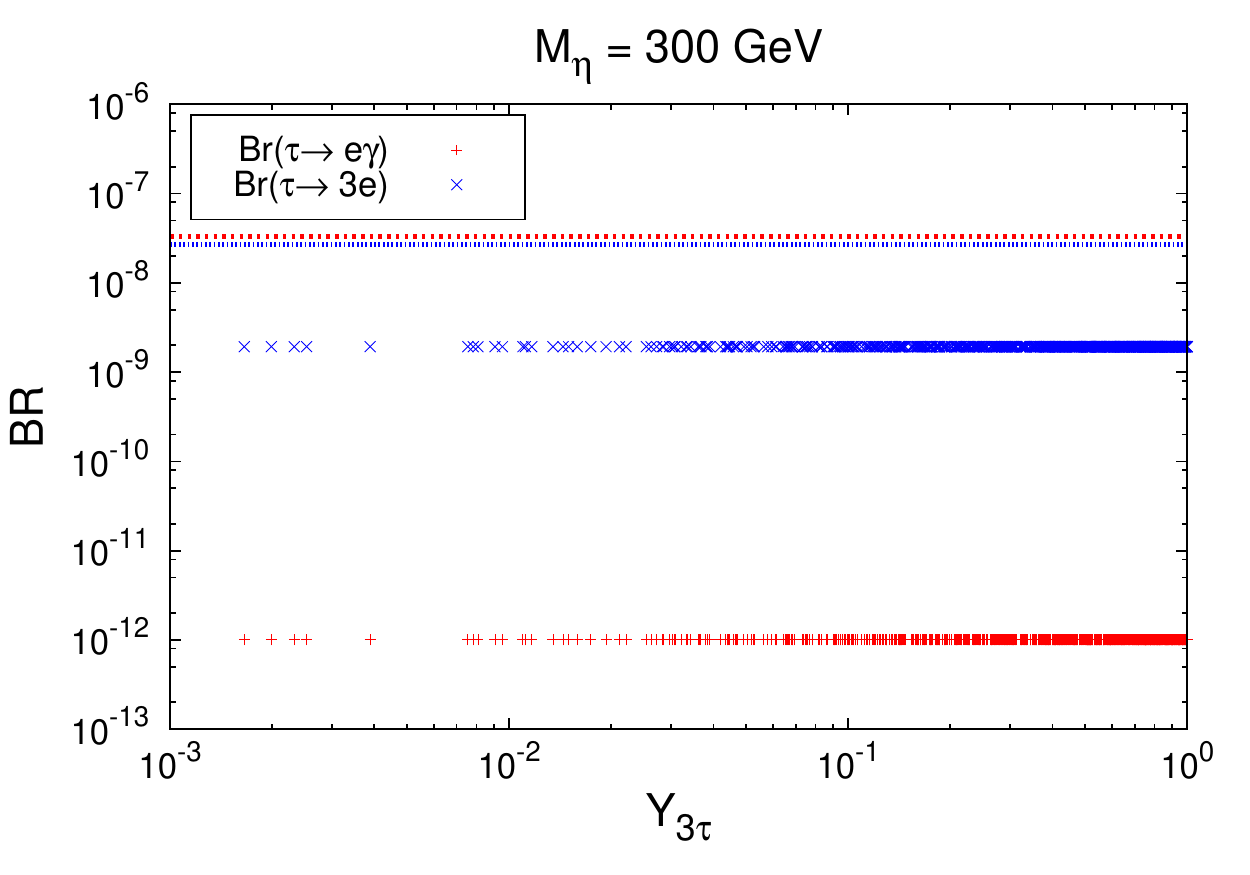}\\
\begin{center}
(e)\hspace{5 cm}(f)\hspace{5.2 cm}(g)\\
\end{center}
\caption{Variation of ${\rm Br}(\ell_\alpha\rightarrow\ell_\beta\gamma)$~(Red) and ${\rm Br}
  (\ell_\alpha\rightarrow 3\ell_\beta)$~(Blue) as a function of $Y_{i\ell}$~[$i=1,3,4$]. Input parameters are set as $M_\eta=300$ GeV, $M_S=130$ GeV, $M_{\chi_1}=800$ GeV, $M_{L_2}=190$ GeV and $M_{\chi_0}=120$ GeV. In the $\mu$-sector, i.e., for plot (a) the $e$ and $\mu$-specific couplings are fixed at those values which are mentioned in the text. In the $\tau$-sector, we choose for the plot (b) $Y_{4\tau}=Y_{3\tau}=0$, (c) $Y_{1\tau}=Y_{3\tau}=0$, (d) $Y_{1\tau}=0$ \& $Y_{4\tau}=0.01$, (e) $Y_{4\tau}=0.01$ \& $Y_{3\tau}=0$, (f) $Y_{1\tau}=Y_{3\tau}=0$ and (g) $Y_{1\tau}=0$ \& $Y_{4\tau}=0.01$. In the plots (a), (d), (e), (f) the projected future bounds corresponding to Br$(\ell_\alpha\rightarrow \ell_\beta \gamma)$ and Br$(\ell_\alpha\rightarrow 3\ell_\beta)$ have been marked with the black and magenta horizontal lines respectively.}
\label{fig:Br_LFV_400}
\end{figure}

\begin{figure}[!ht]
\begin{center}
\includegraphics[scale=0.44]{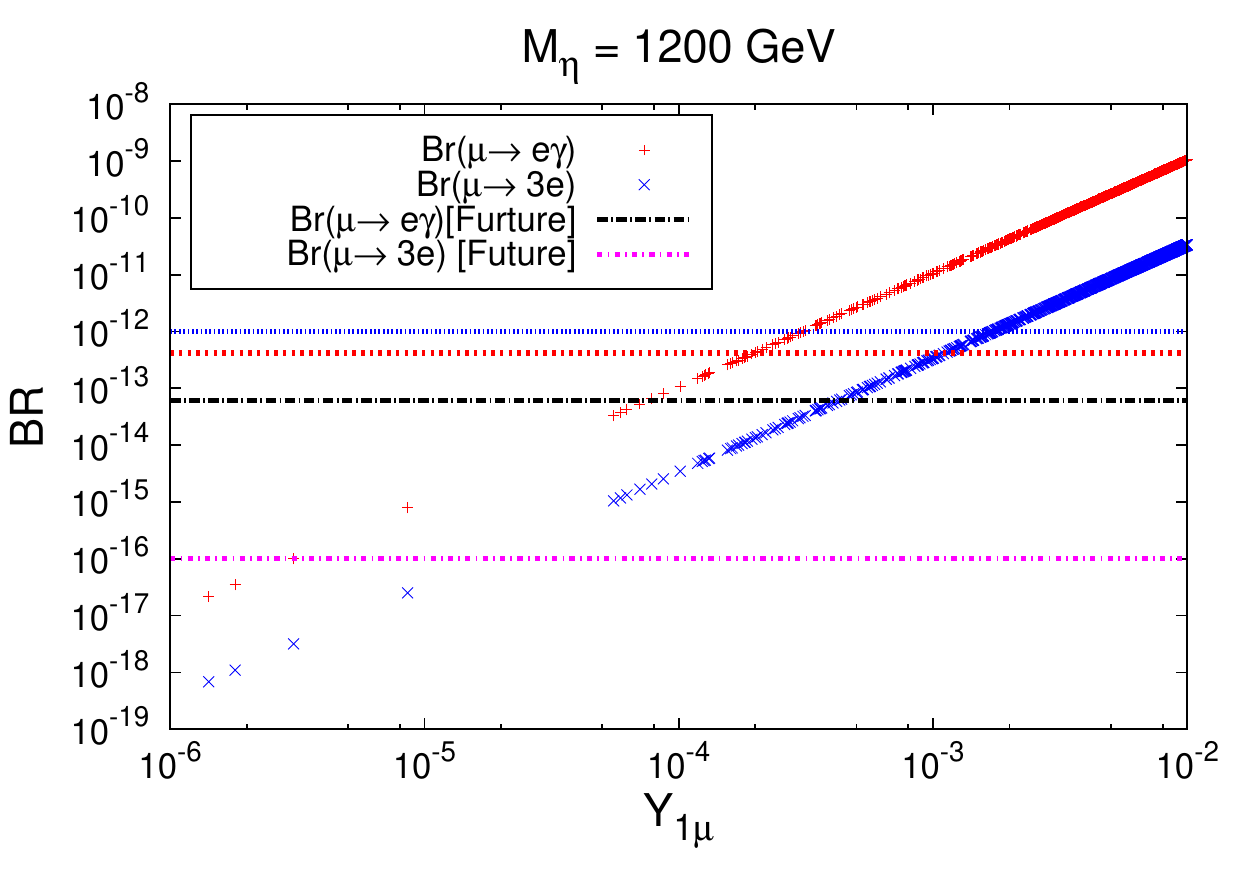}\\
(a)\\
\end{center}
\includegraphics[scale=0.44]{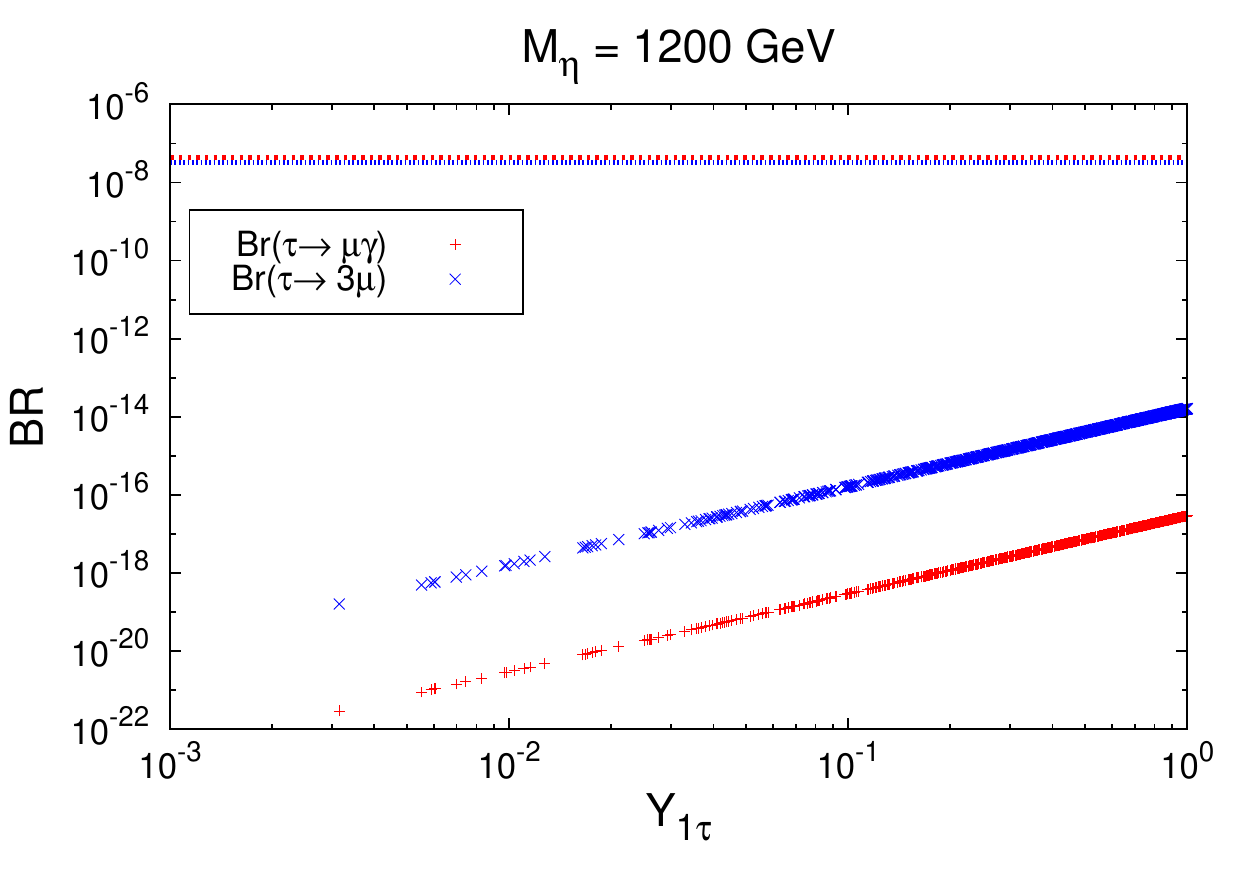}
\includegraphics[scale=0.44]{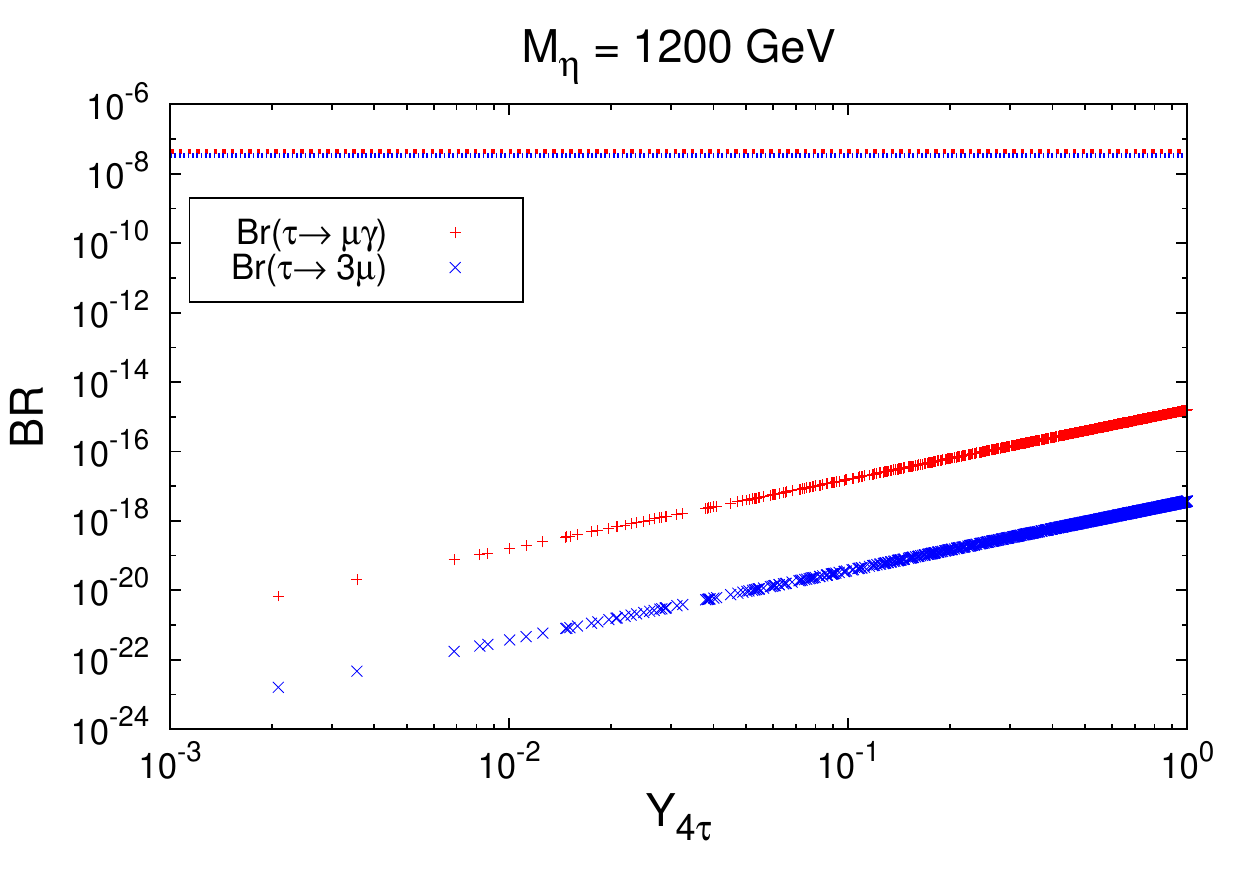}
\includegraphics[scale=0.44]{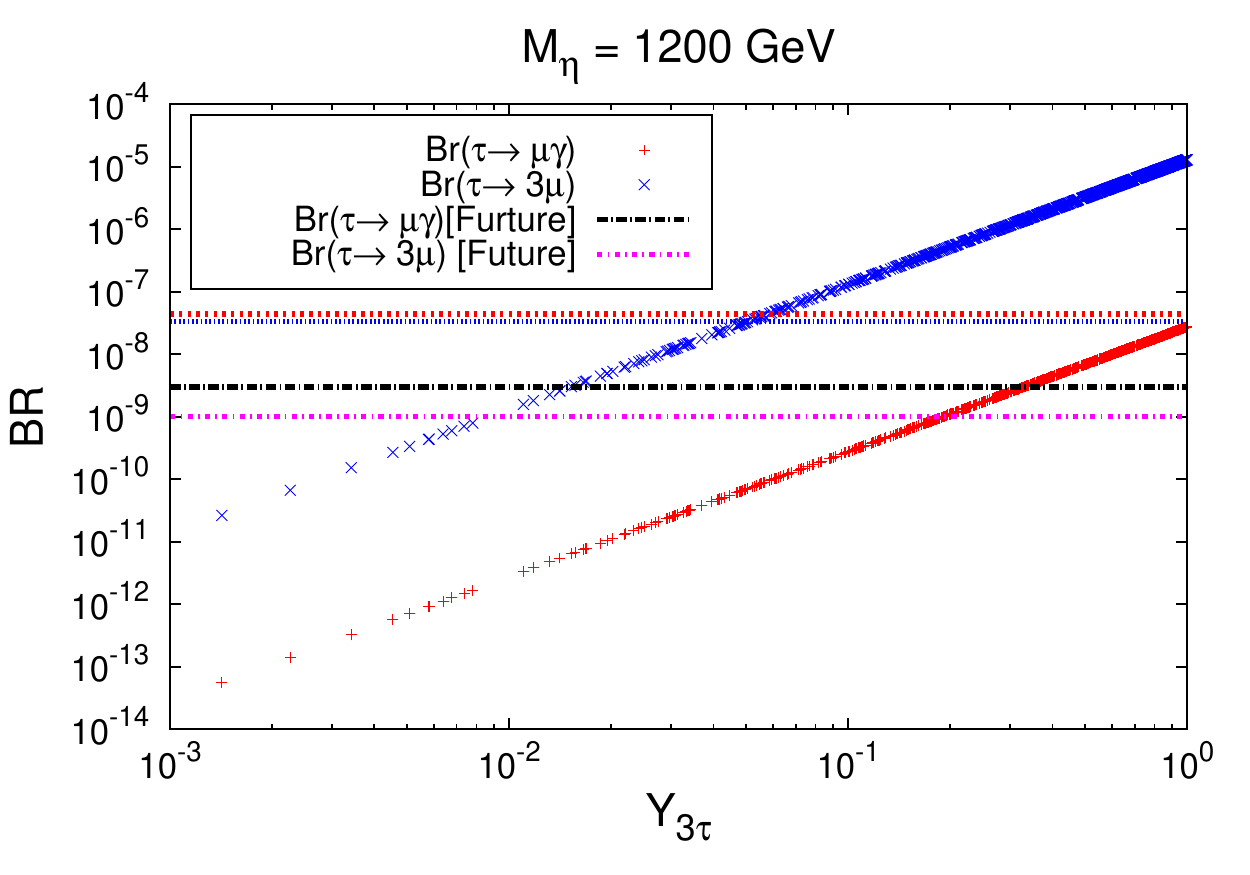}\\
\begin{center}
(b)\hspace{5 cm}(c)\hspace{5.2 cm}(d)\\
\end{center}
\includegraphics[scale=0.44]{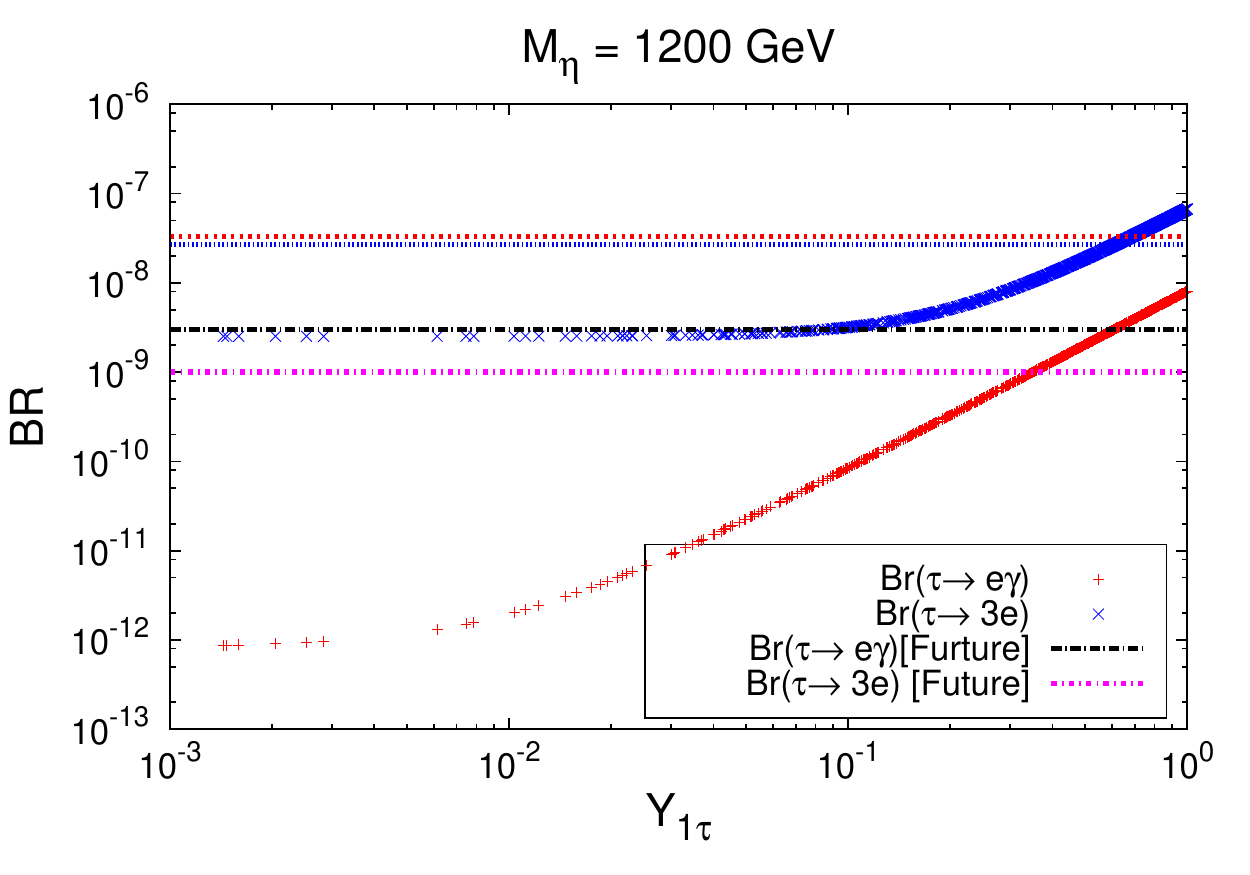}
\includegraphics[scale=0.44]{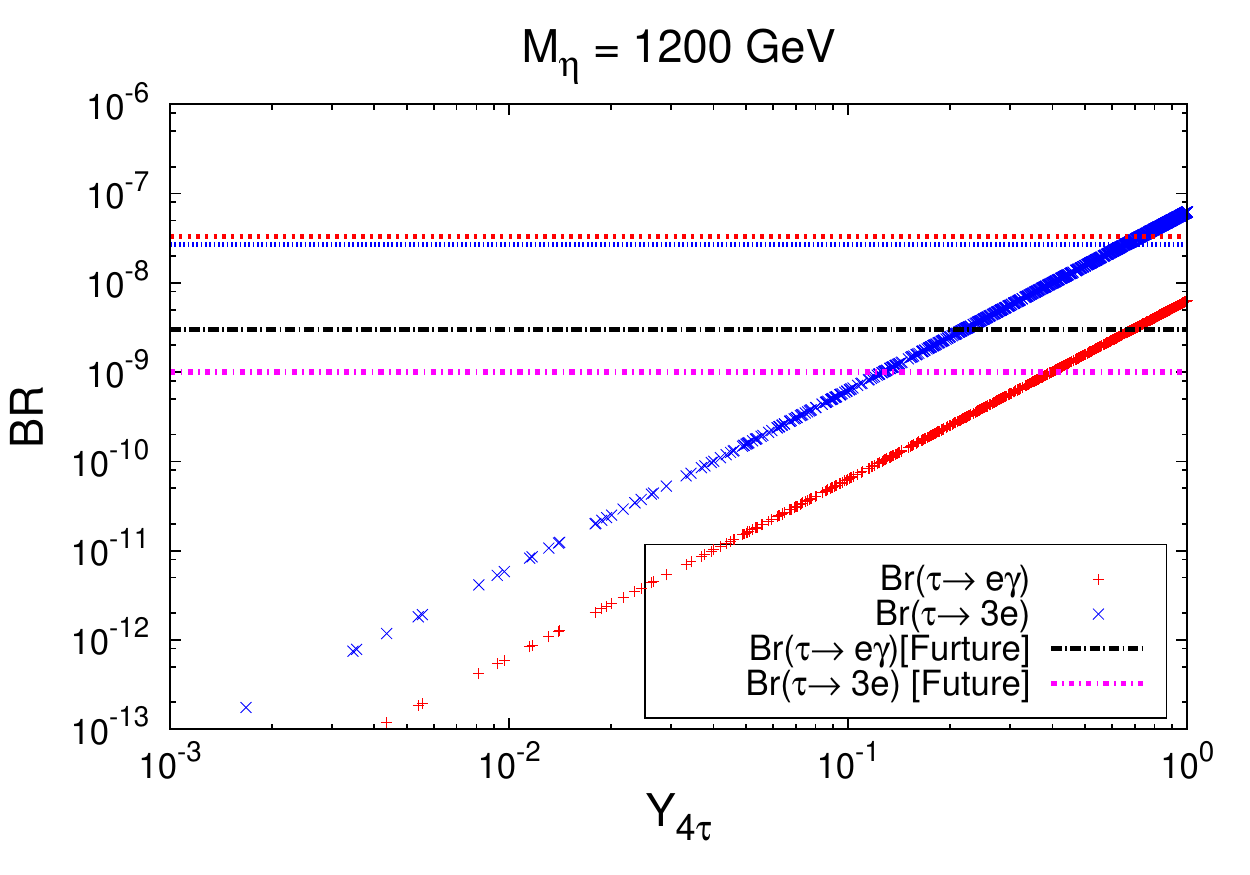}
\includegraphics[scale=0.44]{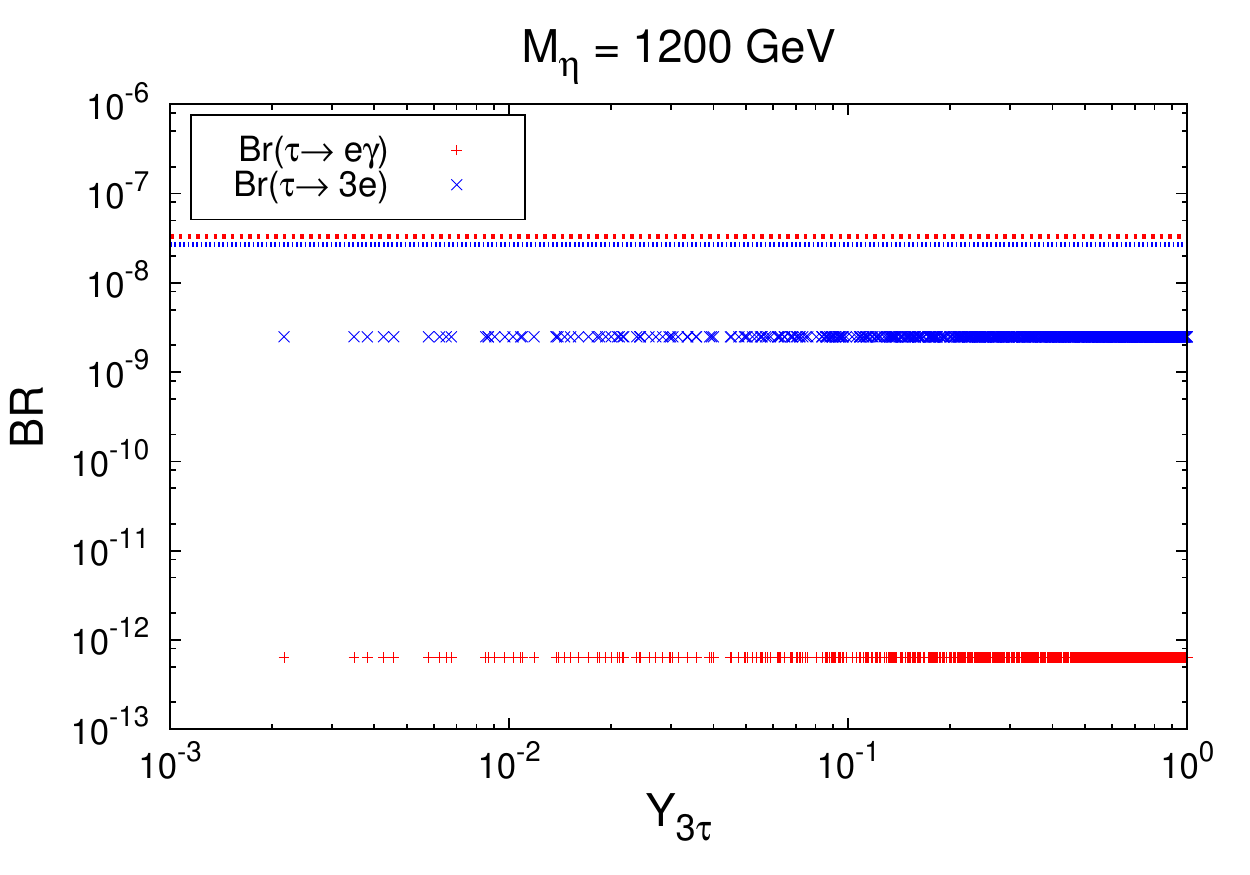}\\
\begin{center}
(e)\hspace{5 cm}(f)\hspace{5.2 cm}(g)\\
\end{center}
\caption{Variation of ${\rm Br}(\ell_\alpha\rightarrow\ell_\beta\gamma)$~(Red) and ${\rm Br}
  (\ell_\alpha\rightarrow 3\ell_\beta)$~(Blue) as a function of $Y_{i\ell}$~[$i=1,3,4$]. Input parameters are set as $M_\eta=1200$ GeV, $M_S=130$ GeV, $M_{\chi_1}=800$ GeV, $M_{L_2}=190$ GeV and $M_{\chi_0}=120$ GeV. In the $\mu$-sector, i.e., for plot (a) the $e$ and $\mu$-specific couplings are fixed at those values which are mentioned in the text. In the $\tau$-sector, we chose for the plot (b) $Y_{4\tau}=Y_{3\tau}=0$, (c) $Y_{1\tau}=Y_{3\tau}=0$, (d) $Y_{1\tau}=0$ \& $Y_{4\tau}=0.01$, (e) $Y_{4\tau}=0.01$ \& $Y_{3\tau}=0$, (f) $Y_{1\tau}=Y_{3\tau}=0$ and (g) $Y_{1\tau}=0$ \& $Y_{4\tau}=0.01$. In the plots (a), (d), (e), (f) the projected future bounds corresponding to Br$(\ell_\alpha\rightarrow \ell_\beta \gamma)$ and Br$(\ell_\alpha\rightarrow 3\ell_\beta)$ have been marked with the black and magenta horizontal lines respectively.}
\label{fig:Br_LFV_1200}
\end{figure}

Here, we will particularly identify the allowed regions
of parameter space associated with free parameters and masses
as introduced in Eq.~\eqref{eq:freeparam}, in regard to different cLFV decays.
Some of the free parameters, as
already tuned by $\Delta a_i$~($i \in e , \mu$), collider or the electroweak precision searches would be set within their allowed domains. 
In Figs.~\ref{fig:Br_LFV_400} and \ref{fig:Br_LFV_1200},
the variation of branching ratios for the
different cLFV processes with respect to the relevant
couplings have been shown for $M_\eta=300$ GeV and $1200$ GeV respectively.
We have followed a particular color code for all these plots, i.e., the red
signifies ${\rm Br}(\ell_\alpha\rightarrow\ell_\beta\gamma)$ while blue stands for
${\rm Br}(\ell_\alpha\rightarrow 3\ell_\beta)$. The horizontal lines specify the present experimental bounds~[see Table~\ref{tableofconstraints}]
on the respective cLFV processes as indicated by the color code. Moreover, to have an idea of the future prospects of our results, in the plots (a), (d), (e), (f)~[of Figs.~\ref{fig:Br_LFV_400} and \ref{fig:Br_LFV_1200}], the projected future bounds corresponding to Br$(\ell_\alpha\rightarrow \ell_\beta \gamma)$ and Br$(\ell_\alpha\rightarrow 3\ell_\beta)$ have been marked with the black and magenta horizontal lines respectively.

For the numerical set-up we have fixed,
\begin{itemize}
\item Scalar masses: $M_\eta=300$ GeV and $1200$ GeV, $M_S=130$ GeV. Here,
  Fig.~\ref{fig:Br_LFV_400} considers $M_\eta=300$ GeV and
  Fig.~\ref{fig:Br_LFV_1200} assumes
  $M_\eta=1200$ GeV.
\item Vector lepton masses and mixings:  $M_{\chi_1}=800$ GeV, $M_{L_2}=190$ GeV,
  $M_{\chi_0}=120$ GeV, and $\sin\theta=0.01$.
\item $\mu$-specific flavor dependent couplings: $Y_{4\mu}= 0.0$ and
  $Y_{3\mu}=2.3$. 
\item $e$-specific flavor dependent couplings: $Y_{1e}= 0.2,~Y_{4e}= 0.2$
  and $Y_{3e}=0.0$. 
\end{itemize}
So, at this point, 
we are left with only four flavor specific free parameters,
i.e., $Y_{1\mu},Y_{1\tau}$, 
$Y_{3\tau}$ and $Y_{4\tau}$. Our aim would be to constrain these free couplings
using the present and future limits of the cLFV branching ratios for
$\ell_{\alpha}\rightarrow \ell_{\beta}\gamma$ and $\ell_{\alpha}\rightarrow 3\ell_{\beta}$ processes (where $\alpha,\beta = e,\mu,\tau$).
Thus, we have varied the free couplings randomly, 
and calculated the corresponding values for
${\rm Br}(\ell_{\alpha}\rightarrow \ell_{\beta}\gamma)$ and ${\rm Br}
(\ell_{\alpha}\rightarrow 3\ell_{\beta})$. Focusing on a particular flavor
at a time, in the following, we present the possible 2-body and 3-body decays.

\begin{itemize}
\item
  ${\rm Br}(\mu\rightarrow e\gamma)$ and ${\rm Br}(\mu\rightarrow 3e)$ :
  The first rows of the Figs.~\ref{fig:Br_LFV_400} and \ref{fig:Br_LFV_1200}
  depict the variation of $\mu\rightarrow e$ branching fractions.
 Here the relevant couplings can be read as 
 $Y_{(1,3,4)i}\sim (i=e,\,\mu)$. However, only $Y_{1\mu}$ can be regarded as
 the free parameter since all the other couplings have
 already been fixed by the precision measurements of $\mu$ and $e$
 anomalous magnetic moments. As can be evident from the plot,
 for $Y_{1\mu} \le 10^{-4}$ both the ${\rm Br}(\mu\rightarrow e\gamma)$
 and ${\rm Br}(\mu\rightarrow 3e)$ can be made satisfied. This explains
 our choice for $Y_{1\mu}$ in the earlier $(g-2)_\mu$ analysis. Thus, to have a simultaneous validation of the $(g-2)_\mu$ and cLFV
 constraints~(i.e. ${\rm Br}(\mu\rightarrow e\gamma)$ and ${\rm Br}
 (\mu\rightarrow 3e)$) one certainly needs a much smaller value of
 $Y_{1\mu}$~($\sim 10^{-4}$).

    \item
      ${\rm Br}(\tau\rightarrow \mu\gamma)$ and ${\rm Br}(\tau\rightarrow 3\mu)$ :
      The second rows of Figs.~\ref{fig:Br_LFV_400} and \ref{fig:Br_LFV_1200} correspond
      to these processes. All the $\mu$ specific couplings are already fixed:
      $Y_{3\mu} $ and $Y_{4\mu}$ have been set to their earlier values and $Y_{1\mu} = 10^{-4}$
      is considered (in accordance with Figs.~\ref{fig:Br_LFV_400}(a) and
      \ref{fig:Br_LFV_1200}(a)). Thus we have varied the $\tau$ specific free
      parameters
      $Y_{j\tau}$~($j=1,4,3$) and calculated the branching ratios. The
      allowed ranges of
      these couplings where ${\rm Br}(\tau\rightarrow \mu\gamma)$ and
      ${\rm Br}(\tau\rightarrow 3\mu)$ are satisfied, can be seen from
      Figs.~\ref{fig:Br_LFV_400}~(b), (c), (d)~and \ref{fig:Br_LFV_1200}~(b),
      (c), (d)
      respectively. Clearly, only meaningful constraint can be derived for $Y_{3\tau}$ which
      reads as $Y_{3\tau} \le  0.04$.  
      The bound can be placed using ${\rm Br}(\tau\rightarrow 3\mu)$ which seems to be
      much stringent compared to ${\rm Br}(\tau\rightarrow \mu\gamma)$. This is a result
      of the $Z$-penguin dominance in that region of the parameter space.

      To illustrate it further, we focus on the dominant parts of $\gamma$
      penguin contributions. In case of photon initiated
      2-body ${\rm Br}(\ell_\alpha\rightarrow \ell_\beta\gamma)$, or 3-body
      ${\rm Br}(\ell_\alpha\rightarrow 3\ell_\beta)$ decays, dipole terms become more
      important, and specially the most significant parts read as:
      \begin{align}
      A^{(n)}_{2} \supset &\,\sin\theta \cos\theta\Bigg[Y^\dagger_{1\beta}Y^\dagger_{4\alpha}
    \left\lbrace\frac{2M_{\chi_1}}{m_{\ell_\alpha}}\,F_1\left(\frac{M^2_{\chi_1}}{M^2_\eta}\right)
-\frac{2M_{\chi_0}}{m_{\ell_\alpha}}\,F_1\left(\frac{M^2_{\chi_0}}{M^2_\eta}\right)\right\rbrace \nonumber \\ &+ Y_{4\beta}Y_{1\alpha}
    \left\lbrace\frac{2M_{\chi_1}}{m_{\ell_\alpha}}\,F_1\left(\frac{M^2_{\chi_1}}{M^2_\eta}\right)
    -\frac{2M_{\chi_0}}{m_{\ell_\alpha}}\,F_1\left(\frac{M^2_{\chi_0}}{M^2_\eta}\right)\right\rbrace\Bigg].
    \end{align}
      The other terms related to dipole or monopole terms
      are proportional to the products of the other flavor specific
      couplings~$Y_{1\beta}Y_{1\alpha},Y_{4\beta}Y_{4\alpha},Y_{3\beta}Y_{3\alpha}$.
      However, generically, considering the couplings for any $\alpha$, $\beta$ are of the same size, these terms are
      few orders of magnitude smaller compared to
      $A^{(n)}_{2}$. 
      For $\tau-\mu$ cLFV processes, $A^{(n)}_{2} \propto \sin\theta \cos\theta
      (Y_{1\mu}Y_{4\tau}+Y_{1\tau}Y_{4\mu})$, thus extremely
      suppressed, unless $Y_{4\tau}$ is reasonably large. This suppression can be attributed to the tinyness of $\sin\theta$ and our choice of Yukawa couplings. In fact $A^{(n)}_2$ may become large if $Y_{4\tau}$ is reasonably large or moderate. This can be
      verified from Figs.~\ref{fig:Br_LFV_400}~(c) ~and
      \ref{fig:Br_LFV_1200}~(c) where due to the choice of $Y_{1\tau}=Y_{3\tau}=0$, the 2-body process dominates over the entire range of $Y_{4\tau}$. Similarly, based on the relative choice of Yukawa couplings the $Z$-penguin diagrams may become more important or comparable
      to the $\gamma$ initiated ones in some cases. For illustration, we choose a particular set of $\tau$-specific couplings as mentioned in the captions of Figs.~\ref{fig:Br_LFV_400} and
      \ref{fig:Br_LFV_1200}. For example, in Figs.~\ref{fig:Br_LFV_400}~(b), (d)
      ~and
      \ref{fig:Br_LFV_1200}~(b), (d),
      $Y_{4\tau}=0, 0.01$ have been chosen respectively, thus, $\gamma$ penguin is always suppressed which results in the dominance of ${\rm Br}(\tau \rightarrow 3\mu)$
      over ${\rm Br}(\tau \rightarrow \mu\gamma)$.  
      We may note
      here that, in the $\mu-e$ processes, the choice of
      parameters~(particularly $Y_{4\mu}=Y_{3e}=0$ and $Y_{1\mu}=10^{-4}$)
      makes the 3-body BR always suppressed in comparison to that of
      2-body~(see the first row of the
      Figs.~\ref{fig:Br_LFV_400}~and \ref{fig:Br_LFV_1200}).

    \item
      ${\rm Br}(\tau\rightarrow e\gamma)$ and ${\rm Br}(\tau\rightarrow 3e)$ :
      Third rows of
    Figs.~\ref{fig:Br_LFV_400} and \ref{fig:Br_LFV_1200} show the plots for these two
    processes. Here the only free parameters are $Y_{j\tau}$~($j=1,4,3$), as the
    electronic couplings are fixed by the $(g-2)_e$ results. Indeed,
    the $\tau$ specific parameters are same as in the $\tau\rightarrow \mu$ analysis. 
     The ranges of $Y_{j\tau}$
    couplings where ${\rm Br}(\tau\rightarrow e\gamma)$ and ${\rm Br}(\tau\rightarrow 3e)$ can be simultaneously satisfied, have been shown in
    Figs.~\ref{fig:Br_LFV_400}~(e), (f) and (g) and \ref{fig:Br_LFV_1200}~(e), (f) and (g) respectively. We may observe
    that $Z$-penguin diagrams become dominant over photon penguins in
    Figs.~\ref{fig:Br_LFV_400}, \ref{fig:Br_LFV_1200}~(f) 
    since $Y_{4\tau} Y_{4e}$ can now contributes
    significantly. 
        From these plots~(Figs.~\ref{fig:Br_LFV_400}~(e), (f)
    and \ref{fig:Br_LFV_1200}~(e), (f)), we are able to
    constrain the two $\tau$-specific couplings as: $Y_{1\tau}\leq 0.5$
    and $Y_{4\tau}\leq 0.7$.
    Note that, the variation of BRs with respect to $Y_{3\tau}$
    has been appearing as
    two horizontal lines, implying that the BRs are apparently
    independent of this
    coupling. This result is a sole outcome of the choice $Y_{3e}=0$.
    Since in both
    ${\rm Br}(\tau\rightarrow e\gamma)$ and ${\rm Br}(\tau\rightarrow 3e)$,
    the
    coupling structure appears as $Y_{3e}Y_{3\tau}$, putting $Y_{3e}=0$
    automatically
    ensures the invariance of the BRs with respect to $Y_{3\tau}$. 
\end{itemize}

So finally, collecting all the constraints, i.e., from the anomalous
magnetic moment data
and non observation of the cLFV processes, we find that all the flavor
specific couplings
$Y_{1,3,4}$ may assume $\sim \mathcal{O}(1-10^{-4})$ values, some of
which may be tested in the near future. 

\subsection{$Z$ and $h$ observables}
\begin{itemize}
  \item[1)] Invisible decays $Z,h\to\chi_0\chi_0$ :
In this model, a light DM is natural and the parameter space
associated with it can be observed to be consistent with
the all low energy data. It is well known that for a light DM,
invisible decays of $Z$ and $h$ which lead to $Z,h\rightarrow \chi_0\chi_0$ can
be substantial to constrain the parameter space.
The corresponding decay widths are given by,
\begin{align}
\Gamma(Z\rightarrow\chi_0\chi_0)&=\frac{1}{48\pi}M_Z\left(\frac{g^2\sin^4\theta}{\cos^2\theta_W}\right)\left(1+\frac{2M_{\chi_0}^2}{M_Z^2}\right)\left(1-\frac{4M_{\chi_0}^2}{M_Z^2}\right)^{1/2} ,\\ \nonumber
\Gamma(h\rightarrow\chi_0\chi_0)&=\frac{(Y_5\sin 2\theta)^2}{16\pi}m_h\left(1-\frac{4M_{\chi_0}^2}{m_h^2}\right)^{3/2},
\end{align}
where, $Y_5=-\frac{(M_{\chi_1}-M_{\chi_0})}{v\sqrt{2}}\sin 2\theta$, with $M_{\chi_1}$ fixed at 800 GeV. We also plot the valid regions in 
$\sin\theta-M_{\chi_0}$ plane. For depicting our results, we use
(i) the observed invisible partial width of $Z$ boson,
$\Gamma^{inv}_Z=499\pm 1.5$ MeV which is below the
SM prediction $\Gamma^{inv}_{SM}=501.44\pm 0.04$ MeV at 1.5$\sigma$ C.L.
~\cite{PhysRevD.98.030001} and (ii)
the experimental bound on invisible $h$ decay reads as ${\rm Br}^{\rm inv}<0.26$~
\cite{ATLAS:2019cid}. Note also that, $\Gamma^{\rm SM}_h=4.07$ MeV,
has been taken~\cite{PhysRevD.98.030001}.
\begin{figure}[!ht]
  \includegraphics[scale=0.6]{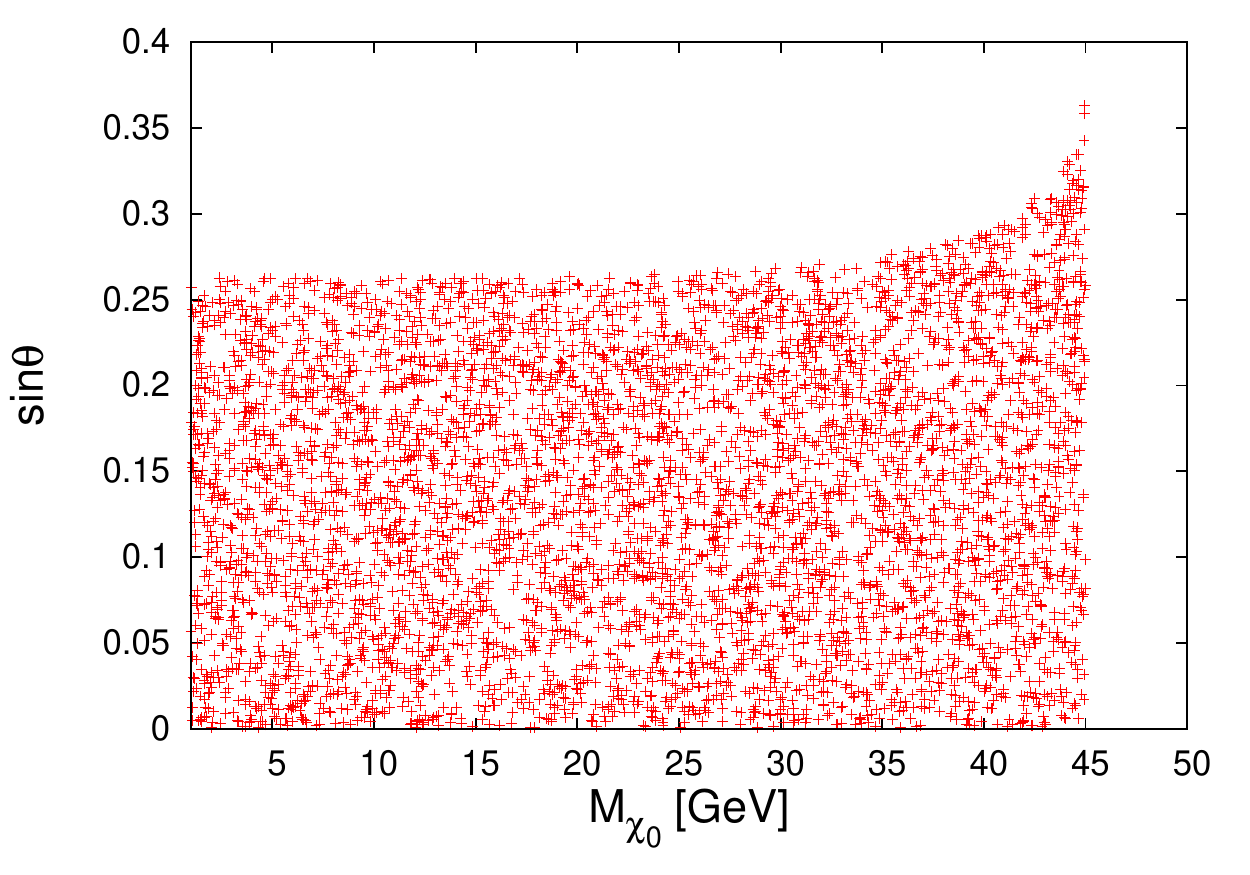}\hspace{1 cm}
  \includegraphics[scale=0.6]{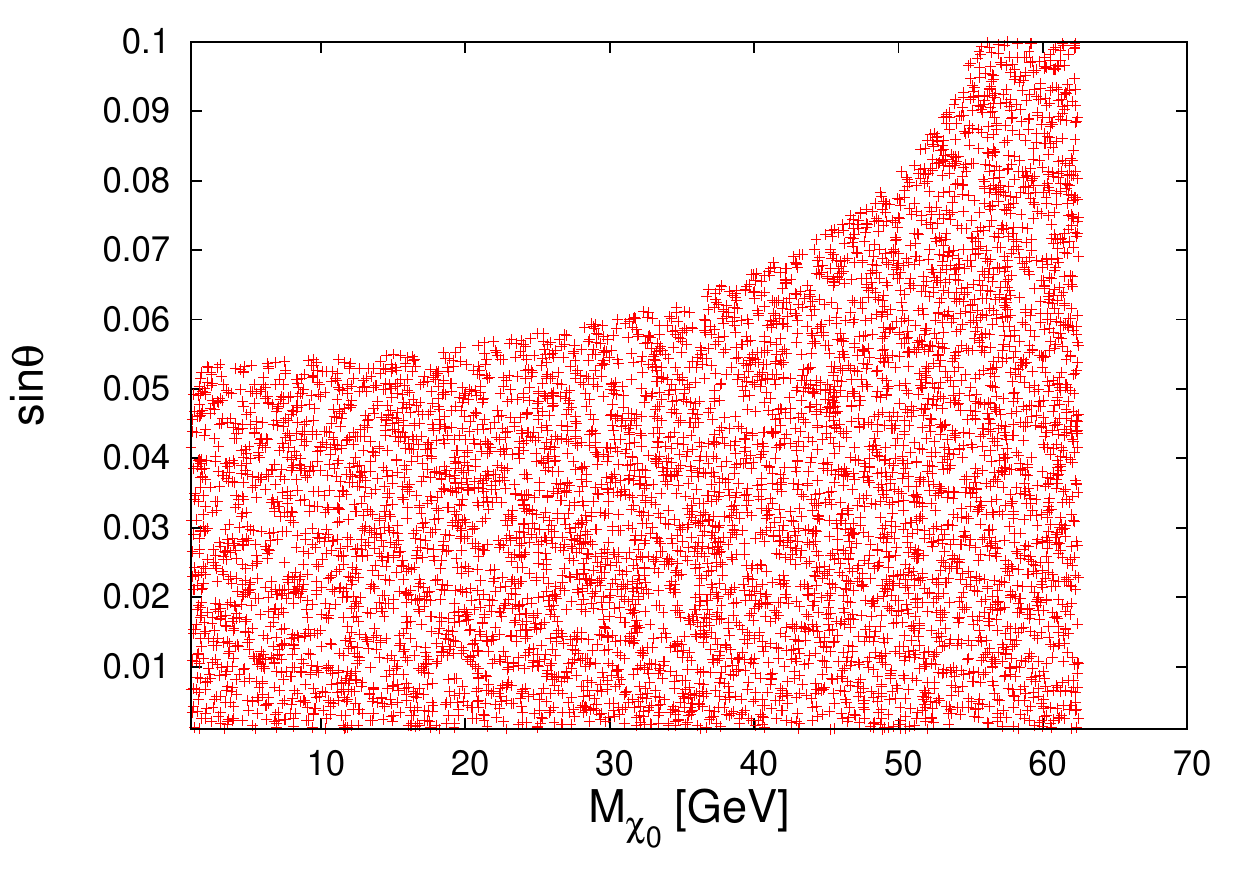}
\begin{center}
(a)\hspace{8 cm}(b)\\
\end{center}
\caption{(a)~The allowed values of $\sin\theta$ for different DM
  masses~($\leq M_Z/2\simeq45$ GeV) from the invisible $Z$ decay constraints
  and (b) from the invisible $h$ ($\leq M_h/2\simeq 62$ GeV) decay constraints.}
\label{fig:inv}
\end{figure}

Clearly, a more stringent bound on the model parameters comes from the
invisible $h$ decay, compared to that of the $Z$ decay, but for
$\sin\theta\simeq 0.01$ the entire parameter space is allowed.
\item[2)]  $Z \to \ell^\pm_i\ell^\mp_i, \ell^\pm_i\ell^\mp_j$ :
\begin{figure}[ht]
\begin{center}
\includegraphics[scale=0.5]{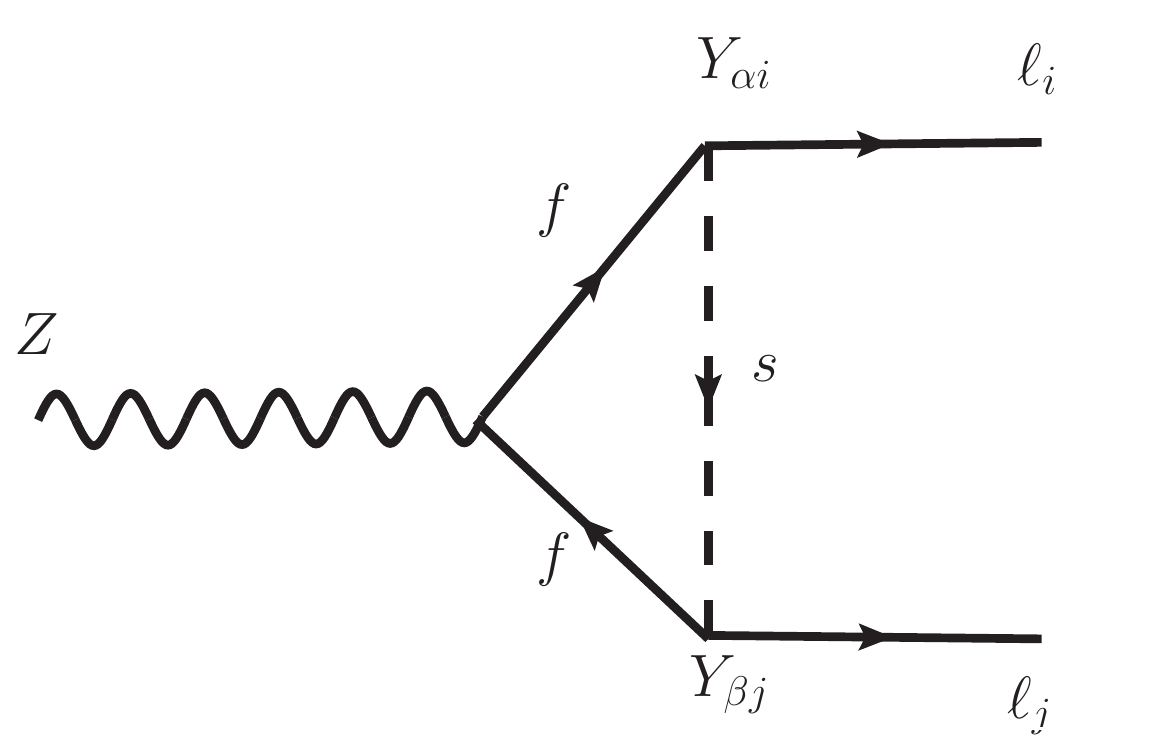}\hspace{1 cm}\includegraphics[scale=0.5]{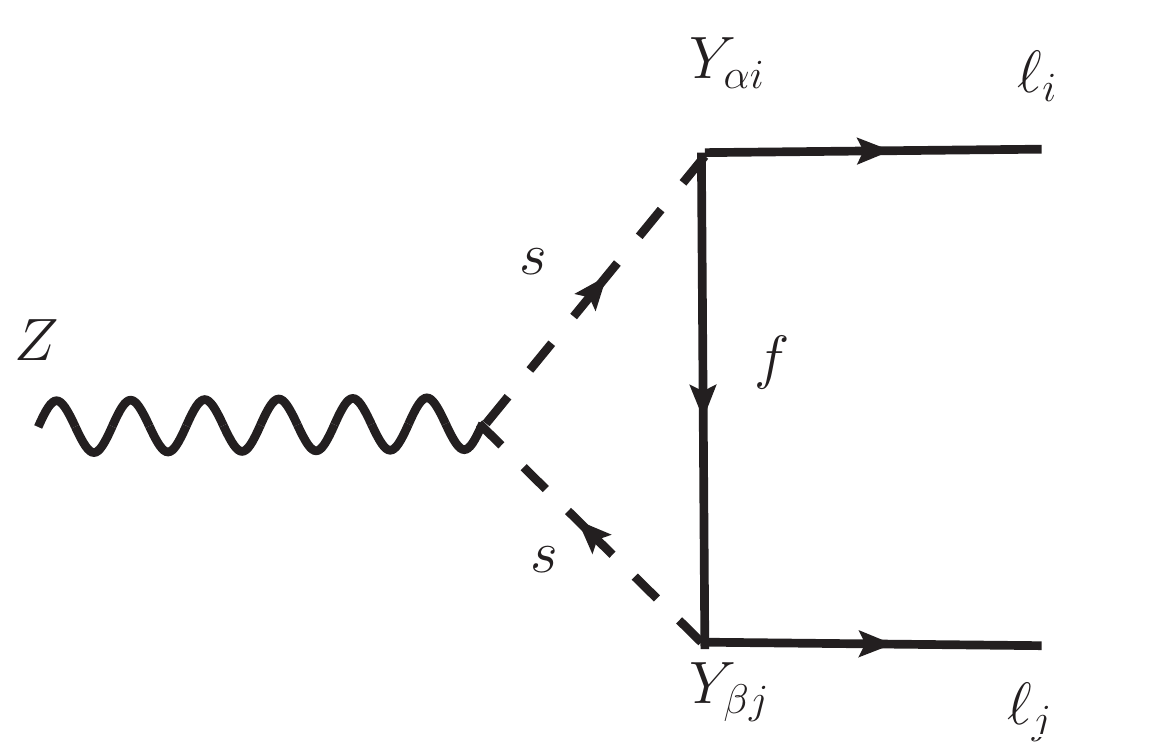}
\end{center}
\caption{Representative diagram for $Z\rightarrow \ell_i\ell_j$ processes. Here $f=\chi_0,\,\chi_1,\,L_1,\,L_2$ and $s \in \eta,\, S$. The indices stand for $i,j=e,\mu,\tau$ and $\alpha,\beta=1,3,4$ (see Eq.~\eqref{eq:massbasis1}).}
\label{fig:2}
\end{figure}
The new fermions $f=\chi_0,\,\chi_1,\,L_1,\,L_2$ and the scalars
$s=\eta,\, S$ can lead to $Z\rightarrow \ell_i\ell_j$ decays. Rare
charged lepton flavour violating (cLFV)
$Z$ decays also inherit a possible complementarity test with
low-energy cLFV searches. The current LHC limits put stringent bounds
compared to the old limits obtained by the LEP experiments on the 
three flavor violating decay modes of $Z$ boson.
Similarly, future sensitivity can be
estimated from \cite{Dam:2018rfz} which considers the future $e^+e^-$ colliders CEPC/FCC-ee \cite{CEPCStudyGroup:2018ghi,FCC:2018evy} experiments assuming $3\times10^{12}$
visible $Z$ decays. 
The present limits and the future
bounds can be read as,
\begin{itemize}
\item[a)] Br$(Z\rightarrow e^\pm\mu^\mp)\leq 7.5\times 10^{-7}$~\cite{ATLAS:2014vur}~~;~~
  $10^{-8}-10^{-10}$~\cite{Dam:2018rfz}
\item[b)] Br$(Z\rightarrow e^\pm\tau^\mp)\leq 5\times 10^{-6}$~\cite{ATLAS:2020zlz,ATLAS:2021bdj}~~;~~$10^{-9}$~\cite{Dam:2018rfz}
\item[c)] Br$(Z\rightarrow\mu^\pm\tau^\mp)\leq 6.5\times 10^{-6}$~\cite{ATLAS:2020zlz,ATLAS:2021bdj}~~;~~$10^{-9}$~\cite{Dam:2018rfz}
\end{itemize}

The branching ratio can be expressed as~\cite{Delepine:2001di,Flores-Tlalpa:2001vbz},
\begin{align}
{\rm Br}(Z\rightarrow \ell_i^\pm\ell_j^\mp)=\frac{\alpha}{3\sin^22\theta_W}\left(\frac{M_Z}{\Gamma_Z}\right)\left(|F_L|^2+|F_R|^2\right),
\label{eq:Zll1}
\end{align}
where, $\sin2\theta_W= 2 \sin \theta_W \cos \theta_W$, $F_L$ and $F_R$ are defined
via Eqs.~\eqref{eq:FL}$-$\eqref{eq:FR} and \eqref{eq:appenZpen}. Here, considering the on-shell decay
of $Z$, $M_Z$ dependence has been incorporated in the
definitions of $F_L$ and $F_R$. 
The form factors $F_L$ and $F_R$ control the loop induced couplings for
$Z \ell_i^\pm\ell_j^\mp$; its numerical values ($|F_L| = |F_R| \sim 10^{-5}$) can be found to be orders of magnitudes suppressed compared to the tree level couplings, specially in the parameter space where
cLFV contraints are satisfied.
The total width $\Gamma_Z$ includes the contributions from all the new BSM modes in addition to the contributions from SM. For numerical evaluations of the
branching fractions, we consider the parts of the parameter space where all
$(g-2)_\ell$, cLFV, and 
DM abundance are simultaneously satisfied. Thus
we fix $M_\eta=300$ GeV, $M_S=130$ GeV, $M_{\chi_1}=800$ GeV, $M_{\chi_0}=120$ GeV and $M_{L_2}=190$ GeV, as chosen in the previous sections. All the Yukawa couplings are fixed at values, as given in following Table \ref{tab:Y} which will subsequently be helpful to obtain a correct relic density for the DM.
\begin{table}[ht]
\begin{center}
\begin{tabular}{|c|c|c|c|c|c|c|c|c|}
\hline 
$Y_{1e}$ & $Y_{3e}$ & $Y_{4e}$ & $Y_{1\mu}$ & $Y_{3\mu}$ & $Y_{4\mu}$ & $Y_{1\tau}$ & $Y_{3\tau}$ & $Y_{4\tau}$ \\ 
\hline 
0.2 & 0.0 & 0.2 & $10^{-4}$ & 2.3 & 0.0 & 0.0 & 0.01 & 0.6 \\
\hline 
\end{tabular}
\end{center}
\caption{ Values of the Yukawa couplings for the evaluation of
$Z\rightarrow \ell_i\ell_j$. }
\label{tab:Y} 
\end{table}
Substituting these values in Eq.~\eqref{eq:Zll1}, we get the following branching ratios:\\
$\bullet$ Br$(Z\rightarrow e^\pm\mu^\mp)=4.16\times 10^{-16}$\\
$\bullet$ Br$(Z\rightarrow e^\pm\tau^\mp)=7.48\times 10^{-10}$\\
$\bullet$ Br$(Z\rightarrow \mu^\pm\tau^\mp)=5.67\times 10^{-11}$.\\
The first branching fraction is much supressed due to the choice of the Yukawa
couplings. Thus, the chances of observing the LFV decays of $Z$ bosons even in the future
are not quite attractive. Similarly, we have observed that
BSM loop contributions to Br$(Z\rightarrow \ell^\pm \ell^\mp)$
($\ell \in e,\mu,\tau$), arising in our framework are lying below the
present limits~\cite{PhysRevD.98.030001}.\\

\item[3)]
{$h\rightarrow \ell^\pm\ell^\mp$} :
\begin{figure}[ht]
\begin{center}
\includegraphics[scale=0.5]{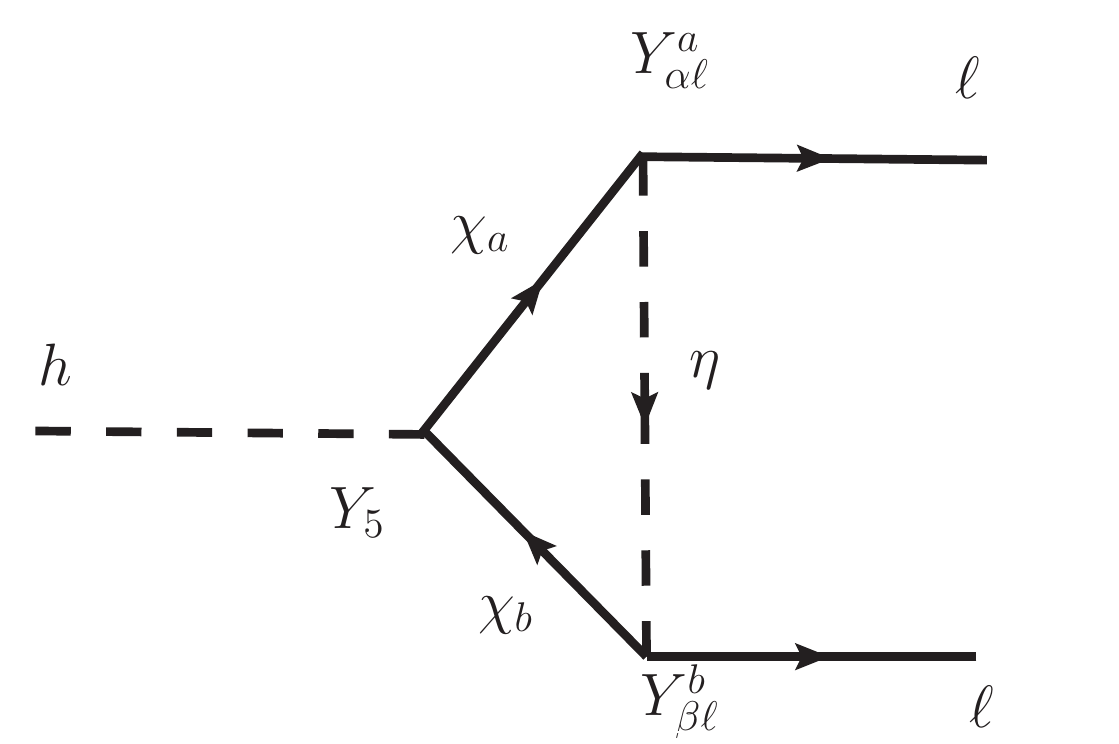}
\end{center}
\caption{Representative diagram for $h\rightarrow \ell\ell$ processes.
  Here $a,b=0,1$ and $\alpha,\beta$ stand for the flavor specific new couplings
  $=1,4$ (see Eq.~\eqref{eq:massbasis1}). Leg corrections are also considered (not shown). }
\label{fig:3}
\end{figure}
The radiative corrections to Yukawa couplings of SM leptons~($y_\ell$) can also be
generated through the new neutral fermions $\chi_0,\chi_1$ in the loop
(see Fig.~\ref{fig:3}). The new physics contributions at one loop can be calculated as,
\begin{align}
{\tilde Y_\ell}^h \equiv \frac{Y_5}{16\pi^2}\sum_{a,b=0,1}&\Bigg[Y^a_{\alpha \ell}Y^b_{\beta \ell}\Big\{M_\eta^2 C_0(0,0,m_h^2,M_{\chi_a}^2,M_\eta^2,M_{\chi_b}^2)+B_0(m_h^2, M_{\chi_a}^2,M_{\chi_b}^2)\nonumber\\
&+M_{\chi_a}M_{\chi_b}C_0(0,0,m_h^2,M_{\chi_a}^2,M_\eta^2,M_{\chi_b}^2)\Big\}-\left(Y^a_{\alpha \ell}\right)^2 B_0(0,M_{\chi_a}^2,M_\eta^2)\Bigg],
\end{align}
where, in terms of our definitions of Yukawa couplings, we define
$Y^0_{\alpha \ell}= Y_{4\ell}$ and $Y^1_{\alpha \ell}= Y_{1\ell}$ for
$\ell\in e,\mu,\tau$. Similarly, $Y_5$ has been recast via Eq. \eqref{eq:Y5} with
$\sin\theta = 0.01$.
The corresponding decay width is~\cite{PhysRevD.98.030001},
\begin{align}
\Gamma(h\rightarrow \ell\ell)=\frac{1}{8\pi m_h^2}|Y^h_{\rm eff}|^2\left(m_h^2-4m_\ell^2\right)^{3/2}~,
\end{align} 
where, $Y^h_{\rm eff}= Y^{SM}_{\ell}+{\tilde Y_\ell}^h$. Now, for the same masses and Yukawa couplings as discussed for the flavor violating $Z$ decays (also see Table~\ref{tab:Y}), 
$\Gamma(h\rightarrow ee/\mu\mu)$ has been found to be practically unchanged to
the corresponding SM value. 
\item[4)]
  Contribution to $W^\pm \ell^\mp \nu_\ell$ vertex:
  \begin{figure}[ht]
\begin{center}
\includegraphics[scale=0.5]{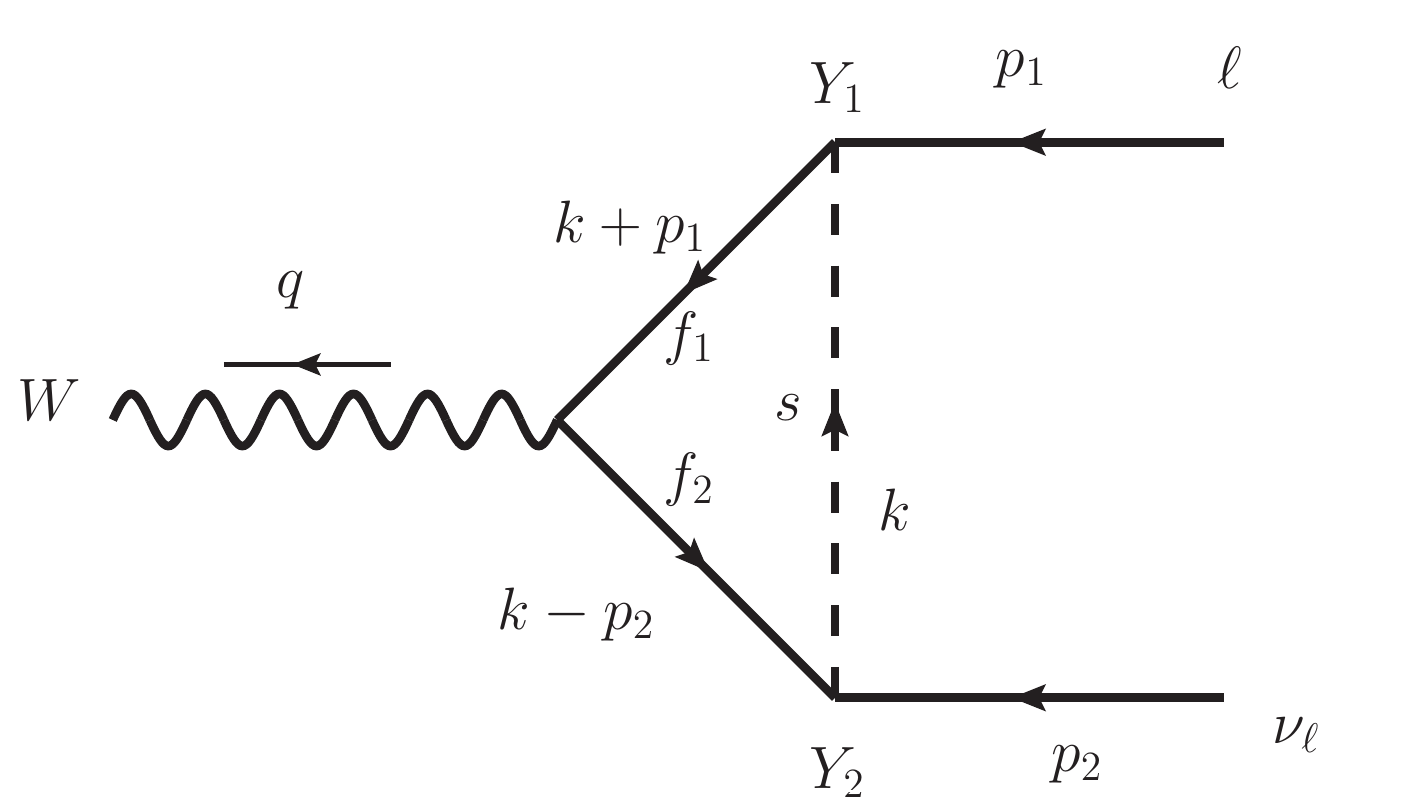}
\end{center}
\caption{Representative diagram for one-loop correction to the $W\rightarrow \ell \nu_\ell$ vertex.}
\label{fig:W}
\end{figure}
The one-loop correction to $W\rightarrow \ell \nu_\ell$ process as shown in Fig.~\ref{fig:W}, results in,
\begin{align}
\label{eq:wlnu}
\tilde  V^{l\nu}=&\frac{Y_1Y_2\,g}{16\sqrt{2}\,\pi^2}\,m_\ell\Big[M_{f_2}
  C_0(0,0,M_W^2,M_s^2,M_{f_1}^2,M_{f_2}^2) + \nonumber\\
  & \frac{(M_{f_2}+M_{f_1})}{M_W^2}
  \Big\{(M_S^2-M^2_{f_2}) C_0(0,0,M_W^2,M_s^2,M_{f_1}^2,M_{f_2}^2)
  +B_0(M_W^2,M_{f_2}^2,M_{f_1}^2) -B_0(0,M_{f_1}^2,M_{S}^2)\Big\}\Big],
\end{align}
where, $C_0,B_0$ are the standard PV integrals. $M_{f_1}$ and $M_{f_2}$ correspond to the masses of VL leptons $f_1$ and $f_2$ respectively, while $m_\ell$ stands for the mass of SM lepton. We are assuming the neutrinos to be massless. \\

Clearly, $\tilde V^{l\nu}$ will include the desired corrections at one loop to
$W^\pm \ell^\mp \nu_\ell$ vertex due to presence
of the BSM states. However, we find the total contribution to be much suppressed.
For having an
estimate about the most significant part in it, we consider $f_1=L_2^\pm$,
$f_2=L_2^0$, $\ell=\mu$ and $s=S$. In this case, the general couplings in
Eq.~\eqref{eq:wlnu} can be read as, $Y_1=Y_2=Y_{3\mu}$. We set the masses and couplings
in accordance with our previous discussion i.e., $M_{L_2}=190$ GeV, $M_S=130$ GeV and $Y_{3\mu}=2.3$. With these choice of parameters, one can directly get, 
$\tilde V^{l\nu} \sim  10^{-6}$, thus smaller than its tree level values.

As evident from the discussion, in our model, gauge boson-leptonic
vertex does not
receive any meaningful
contribution at all. In fact, both $Z\ell^\pm\ell^\mp$ and
$W^\pm \ell^\mp \nu_\ell$ can be considered at their SM values, thus,
processes involving
leptonic or semileptonic decays of mesons, e.g., $K_L\to
\mu\mu$, $K_L\to \pi \nu\nu$, or $B_s \to \mu\mu$,
or precisely measured CKM elements can be completely determined by the SM
physics.
\end{itemize}
 
\section{Dark Matter Phenomenology}\label{sec:DM}
This model may offer a singlet-doublet dark matter; phenomenology of
such scenarios have been studied in detail
\cite{Cohen:2011ec,Cheung:2013dua,Vicente:2014wga,Restrepo:2015ura,
  Calibbi:2015nha,Bhattacharya:2015qpa,Yaguna:2015mva,Arcadi:2018pfo,
  Konar:2020wvl}. Here we would simply check that if all the couplings which
are already constrained by the different precision and collider bounds, can
provide us with an acceptable DM relic density, consistent with SI
DM-nucleon elastic cross section bounds.
After EWSB,
$\chi_0$ --- a dominantly singlet-like state, odd under \sym symmetry
can be considered to be the lightest
particle --- thus a valid DM candidate while the other neutral state
$\chi_1$ carries a strong doublet-like nature for a small mixing angle $\theta$.
In general, the singlet-doublet mixing parameter $\theta$ is completely
controlled by the SI direct detection bounds~(much stronger than the EWPO constraints); usually, only a very tiny $\theta$ is
allowed.
We have fixed all other BSM
particles~($L_1^{\pm}$, $L_2^{\pm,0}$, $\eta$, $S$)
at a heavier mass scale, discussed as in our previous exercises.
Since a small $M_{\chi_0}$ is preferred
from cLFV and $\Delta a_\ell$, we may focus on the parameter space
with a light DM. 

The relic abundance of DM in the universe as obtained from the PLANCK data
is $\Omega_{DM}h^2=0.1198\pm 0.0012$~\cite{Aghanim:2018eyx}. The singlet-like
fermionic DM $\chi_0$, being the lightest odd particle and stable under the
imposed \sym symmetry, was in thermal equilibrium in the early
universe through its interaction with the SM particles. But at a point of
time~(or temperature: $T\leq T_{freeze\,out}$) it gets decoupled from the thermal
bath when the interaction rate fell shorter than the expansion rate of the
universe.  The relic density of the DM can be obtained by solving the
Boltzmann equation, given by, 
\begin{align}
\frac{dn}{dt} + 3 \mathcal H n = - \langle \sigma_{eff}\, v \rangle (n^2 - n_{\rm eq}^2) 
\label{eq:Boltz}
\end{align}
where $\mathcal{H}$ is the Hubble constant, $\langle \sigma_{eff}\, v \rangle$
is the thermal averaged cross section of the DM annihilating to the SM
particles and $n$ signifies the number of interacting particles, with the
subscript `${\rm eq}$' designating its equilibrium value. Though, for doing
the numerical analysis we have used
micrOMEGAs~\cite{Belanger:2006is,Belanger:2008sj}. After implementing the model parameters
in LanHEP~\cite{Semenov_2009}, the output files have been used as the input for
micrOMEGAs, to solve the Boltzmann equation numerically and for
calculating the relic density. Here, the mass parameters have been fixed
at the same values as was done in Sec.~\ref{sec:LFV}, with $M_\eta$ assuming the lower value, i.e. 300 GeV.
For the flavor dependent Yukawa couplings, which are
restricted by the cLFV and $(g-2)_\ell$ bounds, we choose them at the
representative values, shown in Table~\ref{tab:Y}. We also note here
that though the choices for $Y_{4\tau}$ or $Y_{3\tau}$ are somewhat different
than the values in Fig.~\ref{fig:Br_LFV_400} and in Fig.~\ref{fig:Br_LFV_1200},
we have checked that the cLFV constraints are completely unaffected.\\
The other meaningful coupling for DM phenomenology is $Y_{6i}$
(=$Y_{6(1i)}$, as in Eq.~\eqref{eq:massbasis1}) )--- the interaction
between DM, singlet scalar $S$ and the right handed neutrinos
$\nu_{Ri}$. 
The same coupling controls the calculation of neutrino
masses~[see Sec.~\ref{sec:nu_mass}]. Here we set $Y_6$
without affecting the neutrino
masses and mixings, e.g., $Y_{6\tau}=0.13$ is taken.

\begin{figure}[ht]
\begin{center}
\includegraphics[scale=0.5]{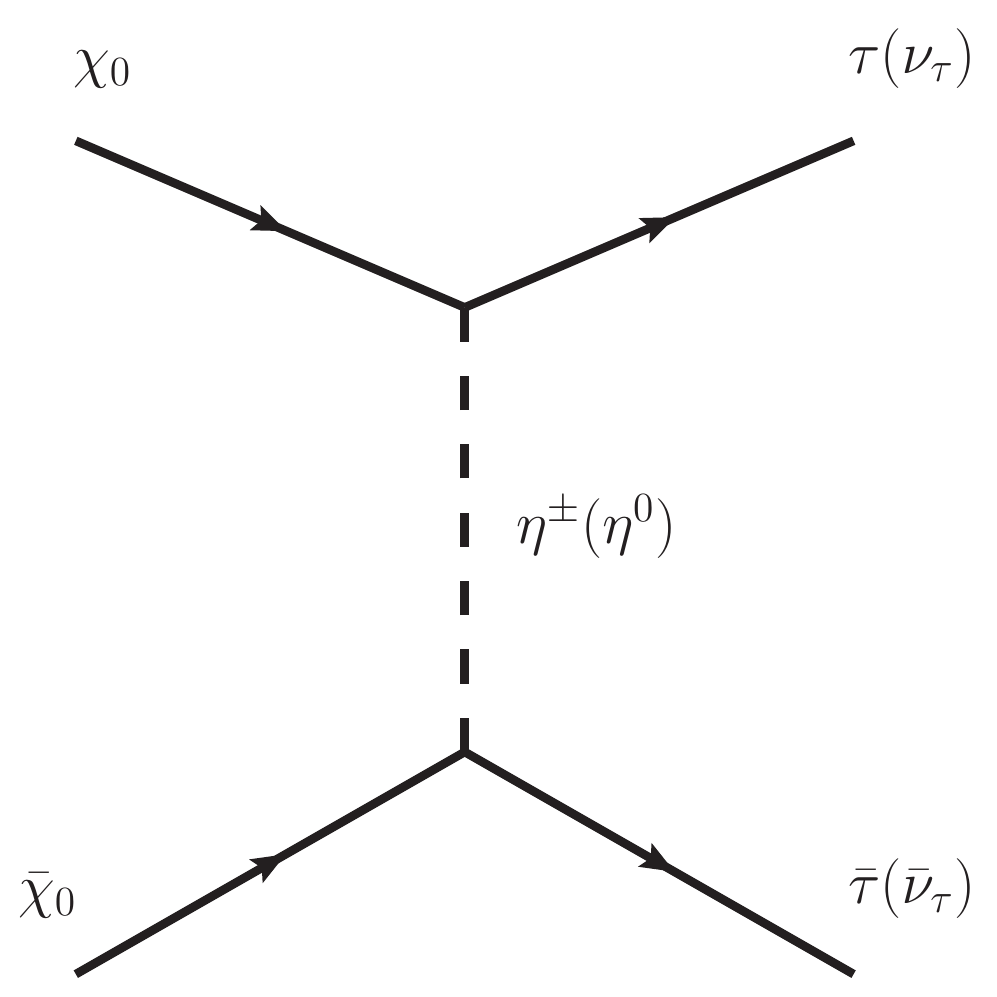}
\end{center}
\caption{The most dominant annihilation channels contributing to the relic density. This is particularly true when the other input parameters are fixed at the values
  shown in Table \ref{tab:Y}.}
\label{fig:relic_feyn}
\end{figure} 
In this model, there may be a number of annihilation channels which can
contribute to the relic density calculation. The order of dominance of
these channels changes with the choice of the other input parameters. Here,
Fig.~\ref{fig:relic_feyn} shows the most dominant annihilation channel
for the chosen parameter space. We have listed the annihilation channels at $M_{\chi_0}=120$ GeV in Table~\ref{tab:anni}.\\
 \begin{table}[H]
  \begin{center}
\begin{tabular}{|c|c|}
\hline 
Parameters & $M_{\chi_0}=120$ GeV~\\ 
\hline 
$M_{\chi_1}=800$ GeV & $\chi_0\bar{\chi}_0\rightarrow \tau\bar{\tau}$~(49\%) \\
$M_\eta=300$ GeV & $\chi_0\bar{\chi}_0\rightarrow \nu_\tau\bar{\nu}_\tau$~(49\%) \\
$M_S=130$ GeV & $\chi_0\bar{\chi}_0\rightarrow \nu_R\nu_R$~(2\%)\\
$M_{L_2}=190$ GeV & \\
$\sin\theta=0.01$ & \\
\hline
$\Omega_{\chi_0}h^2\quad\Rightarrow$  & $0.103$\\
\hline 
\end{tabular} 
  \end{center}
\caption{Dominant~($\geq 1\%$) annihilation channels relevant in
  determining the relic density at $M_{\chi_0}=120$ GeV.}
\label{tab:anni}
\end{table} 

\begin{figure}[ht]
\includegraphics[scale=0.6]{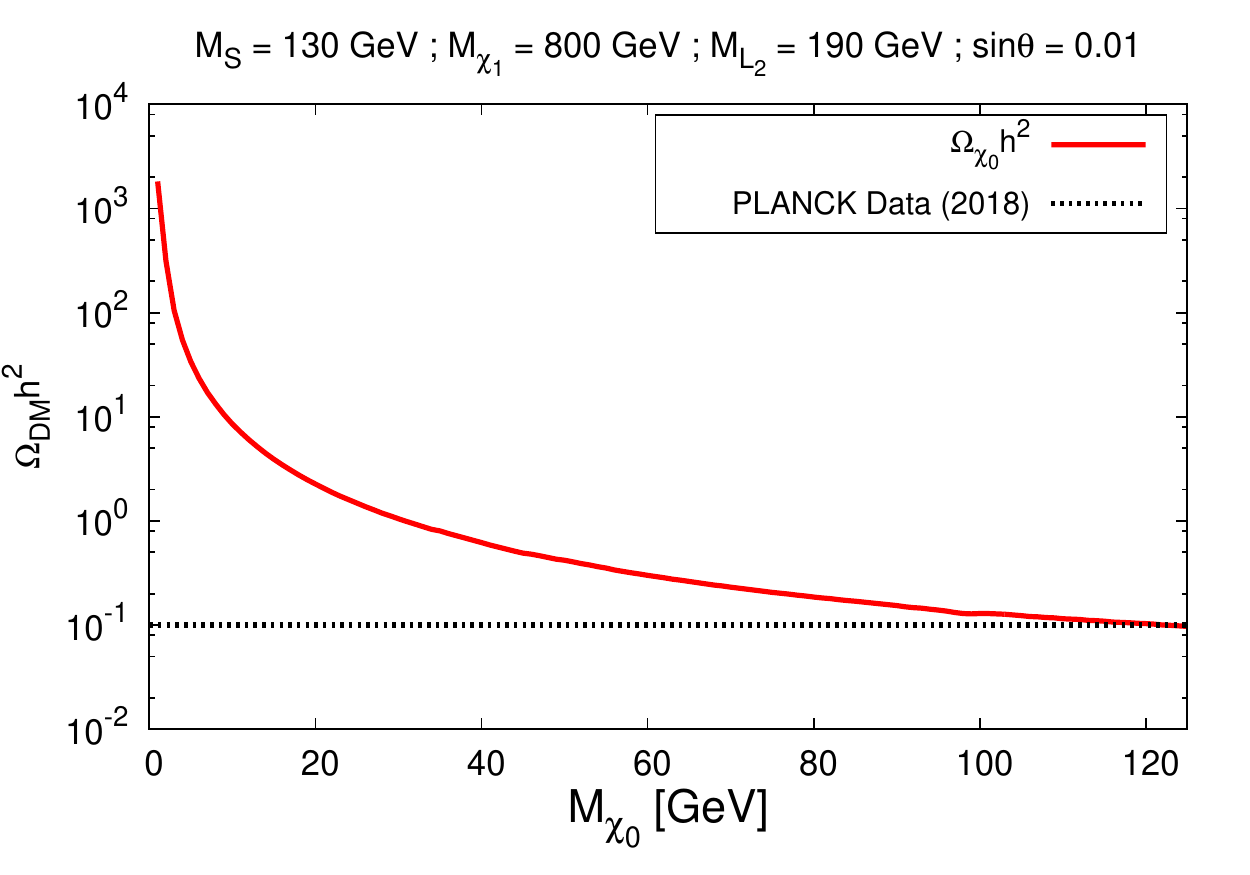}\hspace{1.2 cm}\includegraphics[scale=0.6]{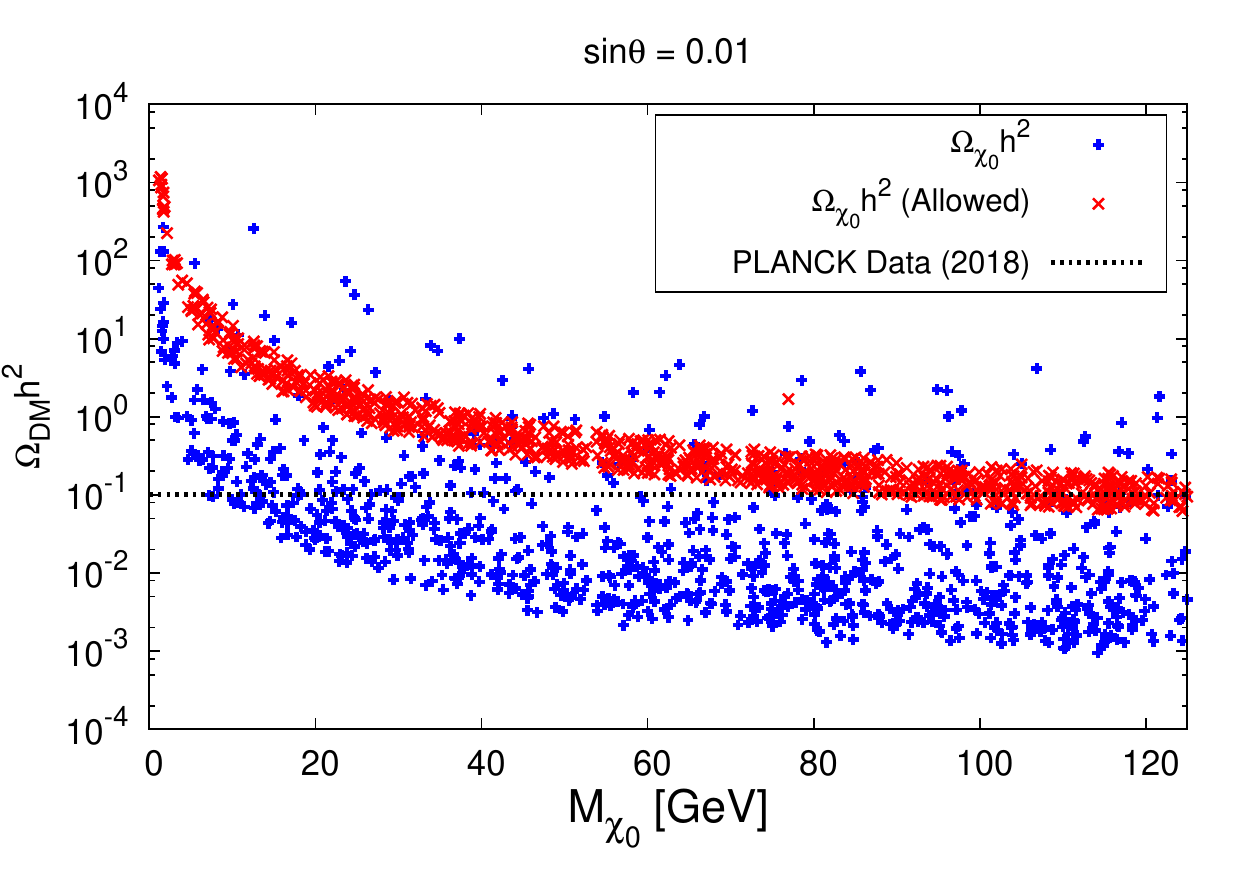}\\
\begin{center}
(a)\hspace{8 cm}(b)
\end{center}
\caption{(a) Variation of Relic density as a function of DM mass $M_{\chi_0}$, when $M_\eta=300$ GeV. (b) Allowed paramter space projected over the relic density plane.}
\label{fig:Relic}
\end{figure}

Fig.~\ref{fig:Relic}~(a) depicts the variation of relic density with respect to
$M_{\chi_0}$ for $\sin\theta=0.01$. The horizontal straight line at
$\Omega h^2\sim 0.12$ is the central value for the acceptable DM relic
abundance, while the red line signifies the calculated relic density in
this model as one varies $M_{\chi_0}$ in the range of $[1-125]$ GeV. Fig.~\ref{fig:Relic}~(b) represents the allowed parameter space projected over the relic density plane. Here, the blue dotted region corresponds to the comeplete parameter space which has been obtained by varying all the parameters randomly, while the red patch stands for the region which is simultaneously allowed from the $(g-2)_\ell$, cLFV, EWPO and neutrino mass constraints. The parameters have been varied within the range of
    $M_{\chi_0} \to [1 : 125],M_{\chi_1} \to [700 : 2000], Y_{1\mu}
    \to [10^{-6} : 10^{-1}], Y_{1\tau}\to[10^{-3} : 2], Y_{4\tau}\to [10^{-3} : 2],
    Y_{6i} \to [0.01 : 1],Y_{1e}\to[0.01 : 5], Y_{4e}\to[0.01 : 5],
    Y_{4\tau}\to[0 : 1]$. 
 
\begin{figure}[ht]
\includegraphics[scale=0.5]{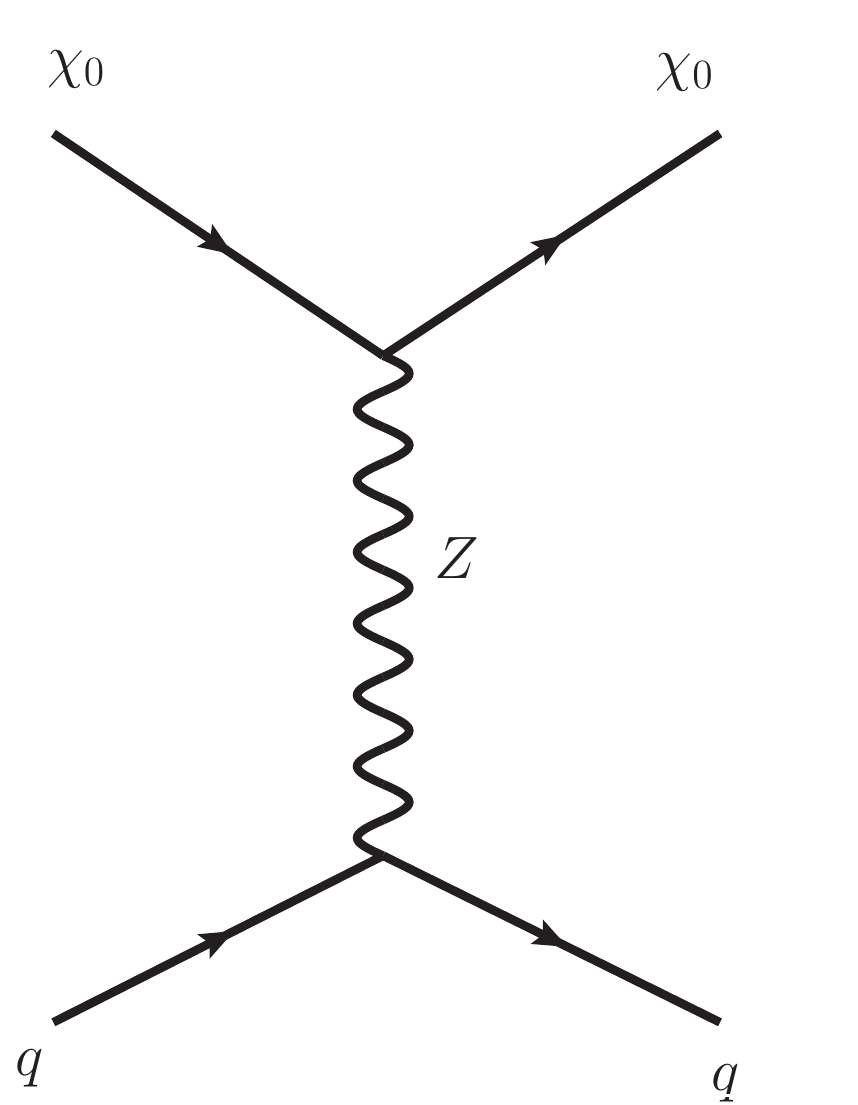}\hspace{3 cm}
\includegraphics[scale=0.5]{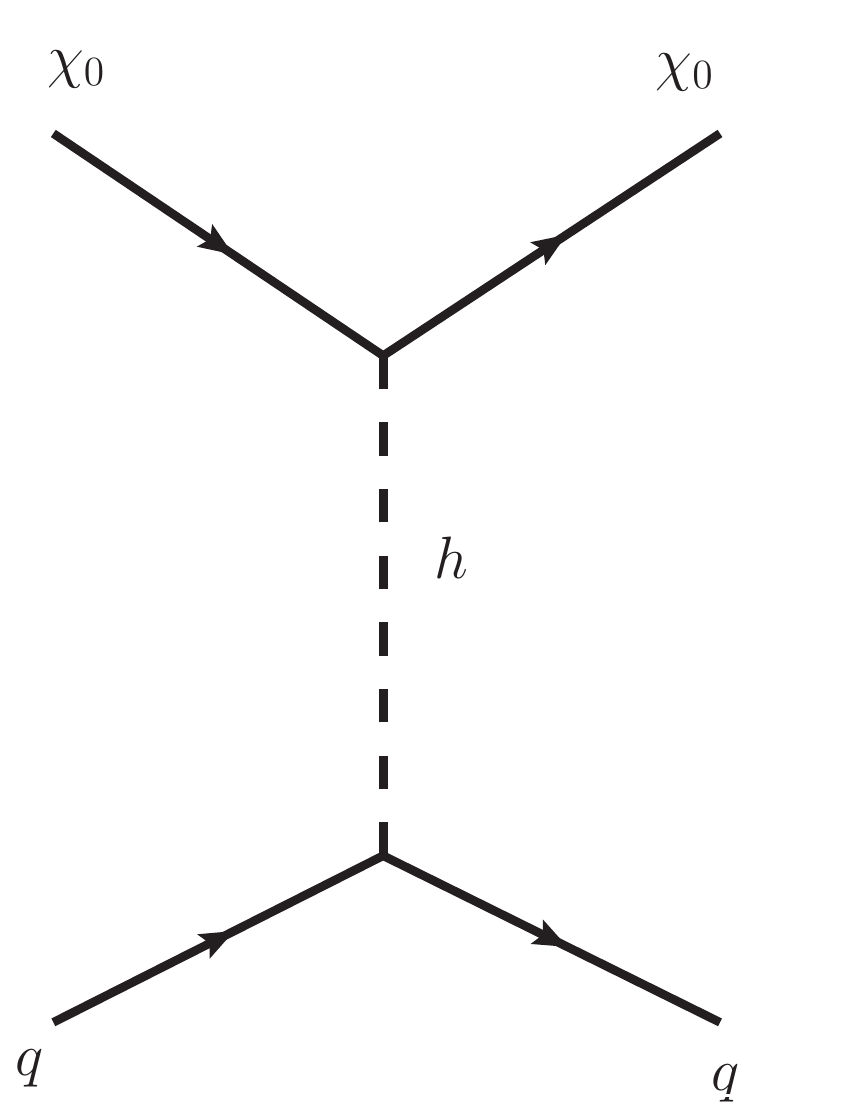}
\caption{Feynman diagrams contributing to the SI direct detection cross section.}
\label{fig:DD}
\end{figure}

 In our model $\chi_0-nucleon$  SI scattering processes,
mediated by the Higgs and $Z$ bosons are shown in Fig.~\ref{fig:DD}. 
The SI scattering cross sections per nucleon corresponding to the
$Z$-mediated diagram of Fig.~\ref{fig:DD} is given by~\cite{Essig_2008,Arcadi:2014lta,Hamaguchi:2015rxa},
\begin{align}
\sigma^{SI}_Z=\frac{G_F^2\mu_r^2}{2\pi A^2}\Big[(1-4\sin^2\theta_W)Z-(A-Z)\Big]^2\sin^4\theta,
\label{eq:sigma_Z}
\end{align}
where, `$A$' and `$Z$' represent the mass number and atomic number of the
target nucleus respectively, $G_F$ is the Fermi's constant,
$\mu_r=\left(\frac{M_{\chi_0}m_N}{M_{\chi_0}+m_N}\right)\approx m_N$ defines the
reduced mass, $m_N$ being the mass of nucleon~(proton or neutron). The second contribution in direct detection comes from the $h$-mediated diagram and the corresponding SI cross section per nucleon is given as,
\begin{align}
\sigma^{SI}_h=&\frac{\mu_r^2}{\pi A^2}\Big[Zf_p+(A-Z)f_n\Big]^2,
\label{eq:sigma_h}
\end{align}  
where the DM-nucleon effective interaction strength can be parameterized as,
\begin{align}
  f_N=\sum_{q=u,d,s}f^{(N)}_{Tq}\alpha_q\frac{m_N}{m_q}+\frac{2}{27}f^{(N)}_{TG}
  \sum_{q=c,t,b}\alpha_q\frac{m_N}{m_q}.
\end{align}
Where $N=n,p$ and
$\alpha_q=\frac{Y_5\sin2\theta}{\sqrt{2}\,m_h^2}\left(\frac{m_q}{v}\right) =
-\frac{(M_{\chi_1}-M_{\chi_0})\sin^2 2\theta}{2v m_h^2}
\left(\frac{m_q}{v}\right)$.
$f^{(N)}_{Tq}$ is the nuclear
matrix element as determined in the chiral perturbation theory from the
pion-nucleon scattering sigma term, and the gluonic part $f^{(N)}_{TG}$ is given by,
\begin{align}
f^{(N)}_{TG}=1-\sum_{q=u,d,s}f^{(N)}_{Tq}\,.
\end{align}
Thus for a fixed $M_{\chi_1}$, the above equation becomes only a function
of $M_{\chi_0}$~(DM mass) and the mixing angle $\theta$. Here we note that,
Higgs contribution to SI scattering can be completely
evaded if one
considers the light-quark Yukawa couplings to assume
non-Standard Model (non-SM)-like values~\cite{Das:2020ozo}\footnote{See Ref.~\cite{Bhaskar:2020kdr} for radiative generation of such non-SM-like Yukawa couplings.}.
For generating the
numerical results we have used the code ``micrOMEGAs",
as was done for studying the relic density, and analysed the variation of SI
scattering cross section as a function of DM mass for $\sin\theta=0.01$.
\begin{figure}[!ht]
\includegraphics[scale=0.6]{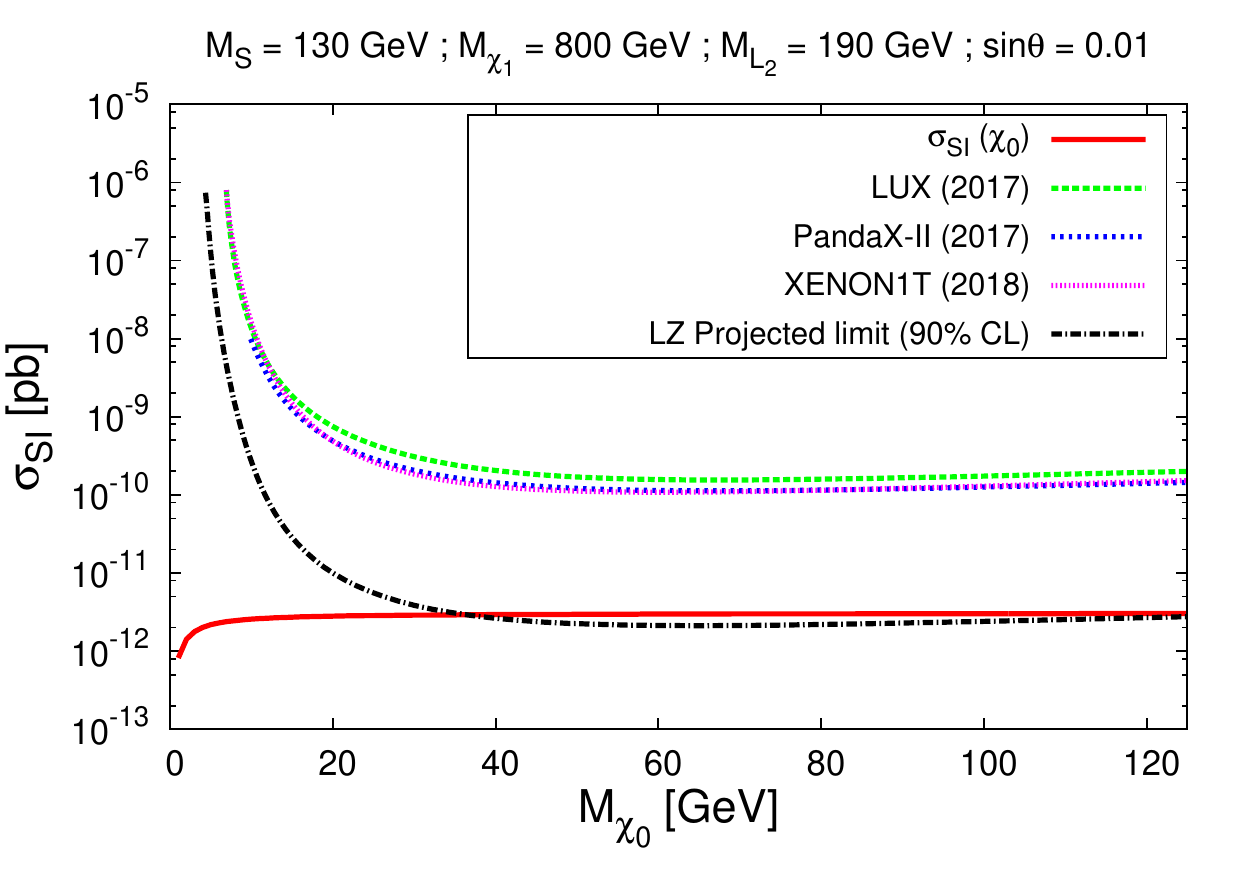}
\caption{Variation of SI scattering cross section as a function of DM mass $M_{\chi_0}$ when $M_\eta=300$ GeV.}
\label{fig:Cross}
\end{figure}
In Fig.~\ref{fig:Cross} the variation of SI cross section with respect to $M_{\chi_0}$ is shown by the red line. All the other mass and coupling parameters are fixed at the same values as was done for the relic density analysis~[see Table~\ref{tab:Y}]. Mostly, for the entire parameter space, the $\sigma_{SI}$ becomes effectively independent of the DM mass, since the $Z$-mediated scattering process~(shown in Fig.~\ref{fig:DD}) appears as the dominant contributor to the total SI cross section over this mass regime. From the observational side we have mainly considered the LUX~\cite{Akerib:2017kat}, PandaX-II~\cite{Cui:2017nnn} and
XENON 1T~\cite{Aprile:2018dbl} limits, which show that the calculated SI cross section, proportional to $\sin^4\theta$, lies much below the present bounds for the entire mass range. However,
the future projected limit coming from LZ collaboration~\cite{Akerib:2018lyp}
may probe only a parts of the parameter space~[see Fig.~\ref{fig:Cross}]. Further, due to $Z$-mediation there is a small amount of SD cross section as well, but it is observed to be far below the existing limits. Moreover, note that, the direct detection cross section has no dependence on the $Y_{1i}$, $Y_{3i}$ and $Y_{4i}$ couplings, which directly govern the $(g-2)_\ell$ and cLFV phenomenology. Therefore, under the variation of different Yukawa couplings~(as was done in Fig.~\ref{fig:Relic}~(b)), the $\sigma_{SI}$ remains mostly unchanged.
\section{Conclusion}\label{sec:conc}
In this paper, we have studied a simple extension where SM is augmented
with a pair of vector like lepton
doublets $L_1$ and $L_2$ , a $SU(2)$ doublet scalar $\eta$ in particular.
Similarly, singlet-like states including a scalar $S$ and a singlet fermion $\psi$
are also considered for specific purposes.
An additional \sym symmetry has been imposed under which all the SM fields
are even while the new fields may be odd under the transformation. Adopting
a bottom-up approach, in
this paper, we systematically scrutinize the parameter space in terms of
the allowed couplings and masses to obtain:
$(i)$ the Dirac masses for the SM neutrinos and mixings through a radiative mechanism,
$(ii)$ electron and muon $(g - 2)$ discrepancy simultaneously while
considering the cLFV and EWPO constraints and finally $(iii)$ a
viable DM candidate, consistent with direct detection observations so far.

We start with our proposed model where the new interactions have been
introduced. 
Subsequently we
discuss about the relevant constraints on the new parameters by reviewing
the different experimental constraints related
to the lepton $(g - 2)$ observations, cLFV bounds, vacuum stability
conditions, electroweak precision constraints and collider observables. 
In our model, $L_1$ and $\psi$ may mix to produce the physical states, and the
lightest state $\chi_0$ can be regarded as the dark matter.
Electroweak precision parameters and, more importantly, the null results from the dark matter
direct detection experiments require
a small mixing between $L_1$ and $\psi$; thus we choose $\sin\theta = 0.01$.

We have shown that in the absence of a tree-level neutrino mass~(being forbidden due to the imposed symmetry), one can generate the correct neutrino mass matrix at one-loop level
if the $\mathcal{Z}^\prime_2$ is allowed to break softly.
The masses and mixings may be controlled by two free parameters $Y_{2(1i)}$ and $Y_{6(1i)}$
which do not have any effect on the charged lepton flavor processes, e.g.,  
$(g - 2)_{\mu/e}$ or different cLFV processes like $\ell_\alpha \rightarrow \ell_\beta \gamma$ and $\ell_\alpha \rightarrow 3\ell_\beta$.
We have performed a comprehensive study to show the interplay between different charged and neutral vector like leptons for satisfying $(g - 2)_e$ and $(g - 2)_\mu$ bounds simultaneously.
A moderately large coupling $Y_{3\mu}$ is required to tune $(g - 2)_\mu$ while $\Delta a_e$ can easily be controlled with other $\mathcal {O}(1)$ couplings.  
Further, the same diagrams are able to generate $\ell_\alpha \rightarrow \ell_\beta \gamma$ processes when $\alpha \neq \beta$. For the 3-body processes like $\ell_\alpha \rightarrow 3\ell_\beta$, we
have considered all the $Z$ and photon penguin diagrams along with the box contributions. Numerically, we have calculated
Br$(\ell_\alpha \rightarrow \ell_\beta \gamma)$ and Br$(\ell_\alpha \rightarrow 3\ell_\beta)$ for different lepton generations and shown their variations as functions of the
relevant couplings for two sets of doublet scalar masses~
($M_\eta \sim$ 300 GeV and 1200 GeV), along with their respective
experimental bounds. These
cLFV constraints, in addition to the lepton $(g - 2)$ results set an important exclusion limit or upper bound for the different Yukawa couplings present in this model. Here, we note that, larger mass value of $\eta$ is not at all disfavoured
in the context of tuning the charged lepton flavor conserving or violating
processes. However, the vector like leptons, especially $L_2$ has to be light ($\le 200$ GeV), otherwise, the relevant coupling $Y_{3\mu}$ may have to be raised to accommodate $(g-2)_\mu$.
Moreover, in the parts of the parameter space, $Z$-dominance over the $\gamma$-penguin in the computation of the 3-body charged lepton processes may be observed. 
Finally, the dark matter phenomenology of the singlet-doublet $\chi_0$ DM has been presented. As shown, a light DM can comply with $(g-2)_\mu$ bound, though in general TeV scale values of DM are allowed in our model. This minimal model can be tested at the LHC. Presently a stringent bound can be realized on the mass of vector like leptons $L_1$ and $L_2$, though the mass difference between the VLs and the inert doublet $\eta$ can be tuned to evade the strong bounds on them. The mass splitting does not have any role on the lepton phenomenology which we have exhaustively studied here. 

\section*{Acknowledgements}\noindent
Our computations were supported in part by SAMKHYA: the High Performance Computing Facility provided by the Institute of Physics (IoP), Bhubaneswar, India. DD likes to thank Subhadeep Mondal for some valuable discussions. NS likes to thank Dr. Prafulla Kumar Panda, Utkal University, for his valuable suggestions. NS acknowledges RUSA 2.0 project,  Ministry of Human Resource Development, India.
\section*{Appendix A}
\label{sec:appA}
In this appendix we list all the Feynman rules required for our calculation. These rules have been expressed in physical
eigen basis for particles: Neutral scalar $s_X=(\eta^0,S)$, charged scalar $\eta^\pm$, neutral VL fermions $\chi_a$~$(a=1,0)$
and charged VL fermions $L_a^-$~$(a=1,2)$.
\subsection*{Scalar interactions}
The Feynman rules for scalar interactions are given by,\\
\includegraphics[scale=0.5]{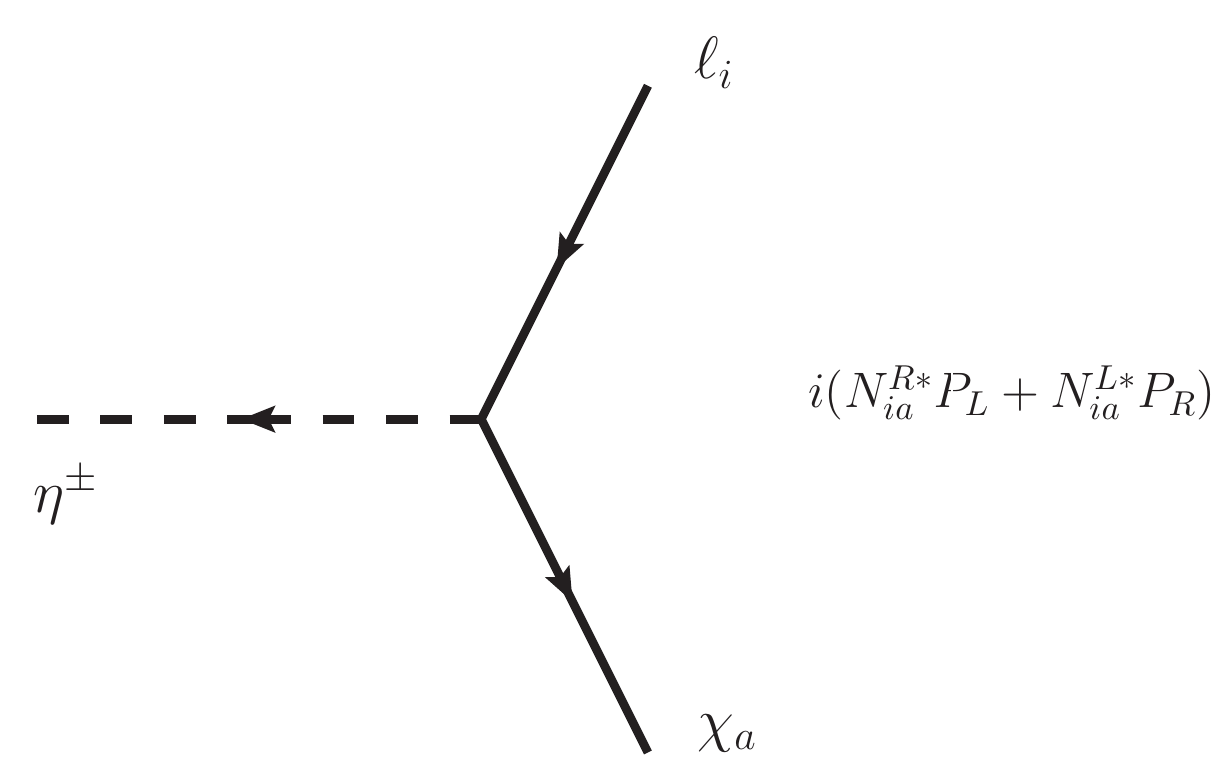}\hspace{3 cm}
\includegraphics[scale=0.5]{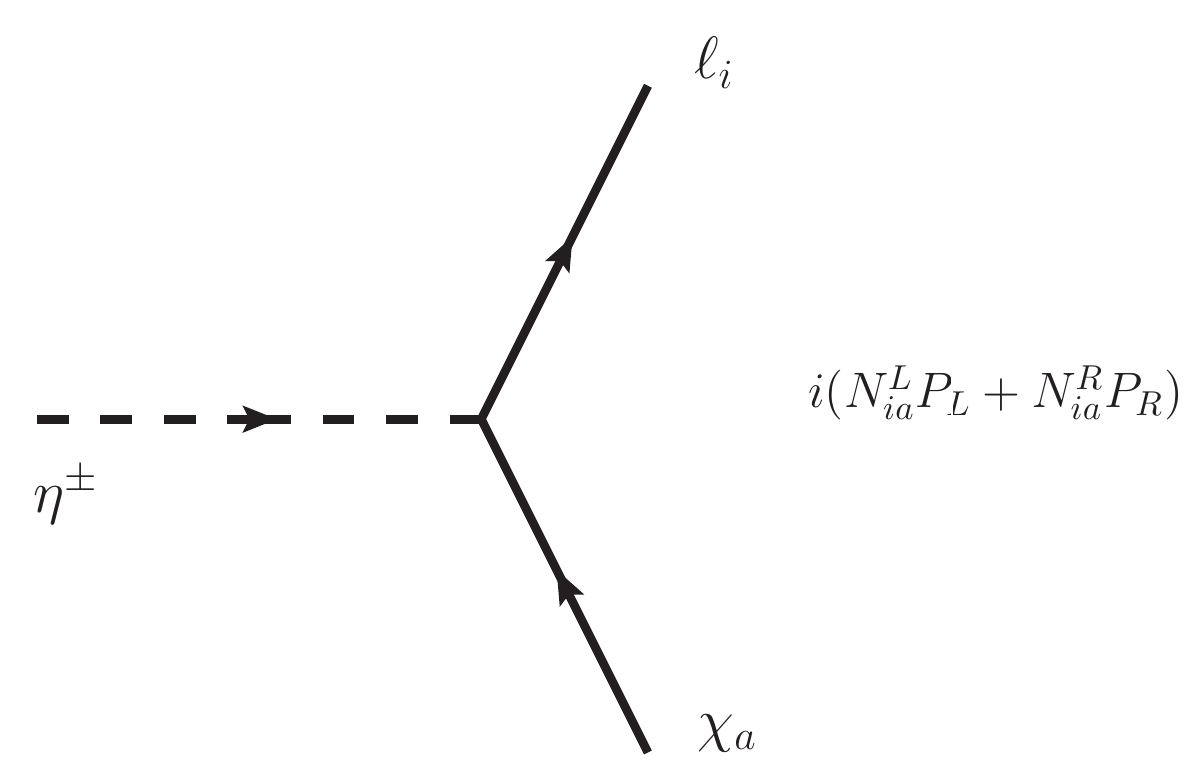}\\
\includegraphics[scale=0.5]{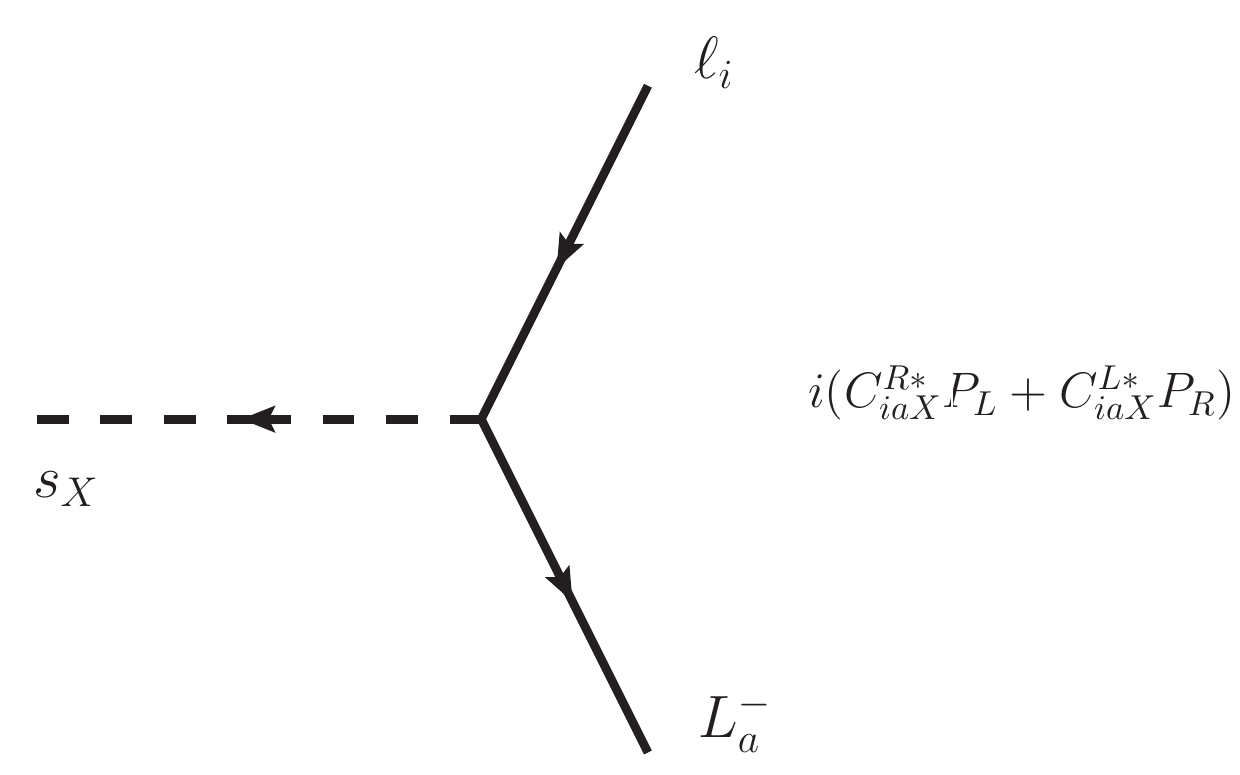}\hspace{3 cm}
\includegraphics[scale=0.5]{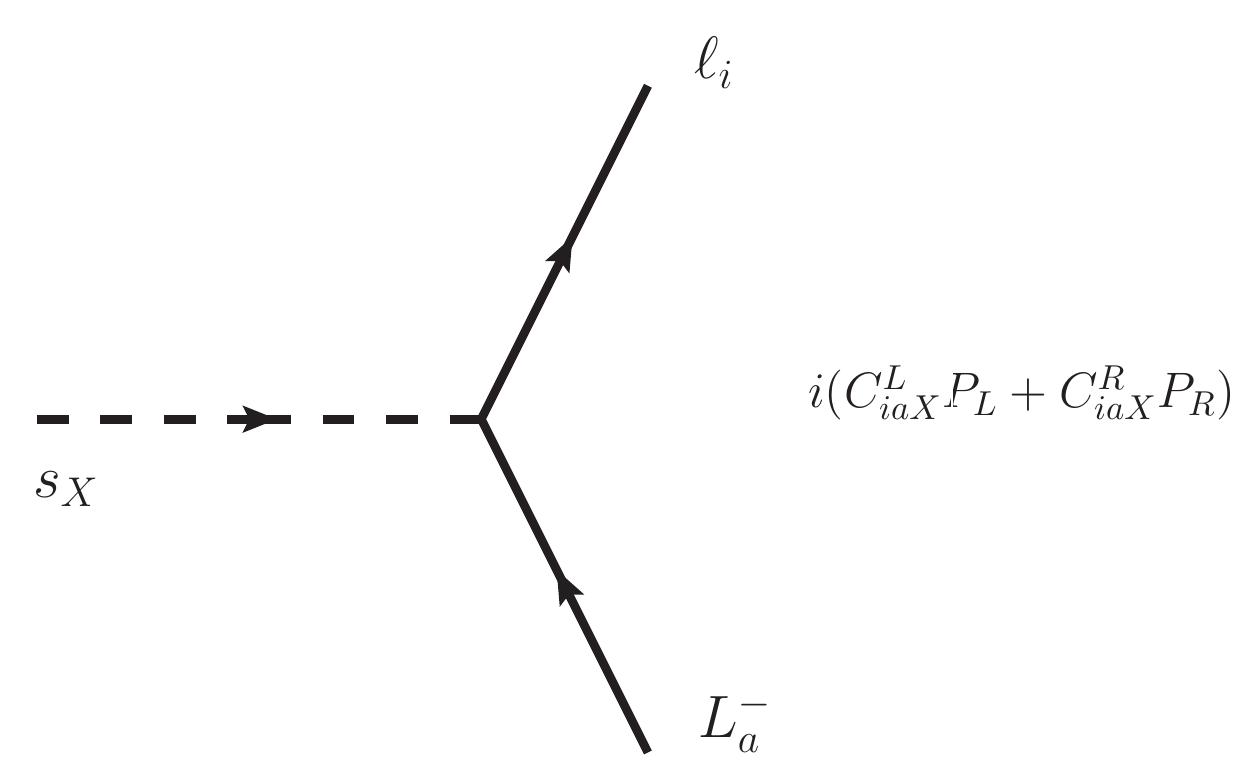}\\

Where, $N^R_{ia}=Y_{4i} U_a$, with $U_1 =\sin\theta~,~U_0 =-\cos\theta$~; $N^L_{ia}=Y^\dagger_{1i} U^\prime_a$,  with
$U^\prime_1 =\cos\theta~,~U^\prime_0 =\sin\theta$ and
$C^L_{i11}=Y^\dagger_{1i}$, $C^R_{i11}=0$, $C^L_{i22}=0$, $C^R_{i22}=Y_{3i}$.

\subsection*{$Z$ boson interactions}
The Feynman rules governing the $Z$ boson interactions are given by,\\
\includegraphics[scale=0.5]{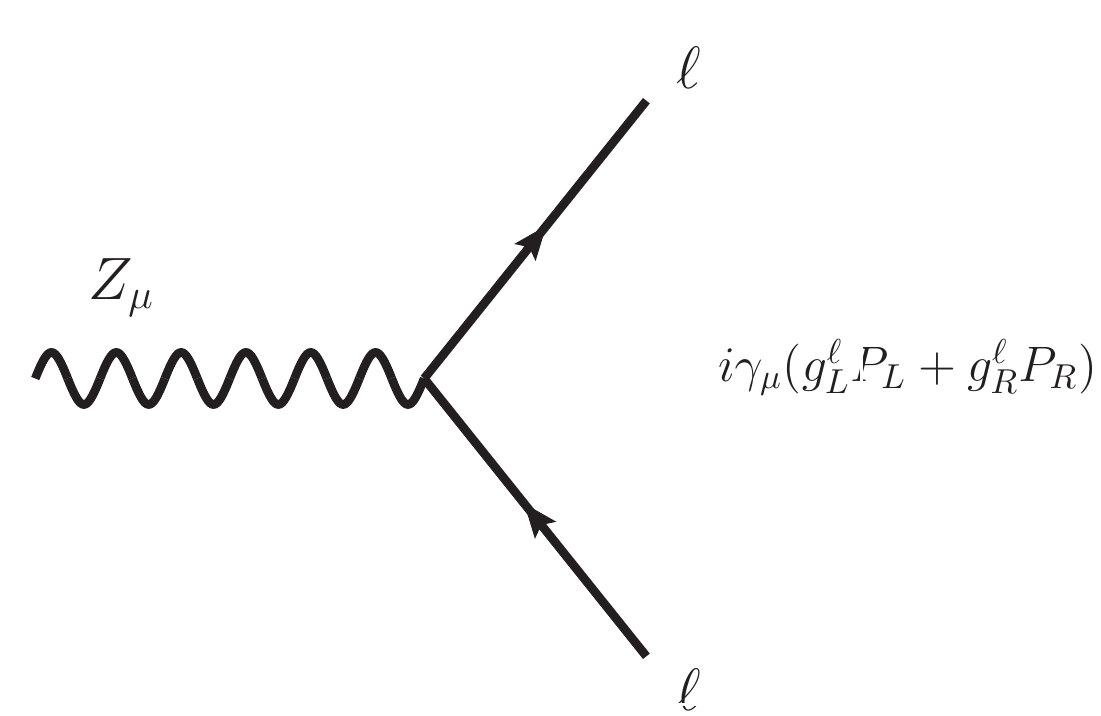}\hspace{3 cm}
\includegraphics[scale=0.5]{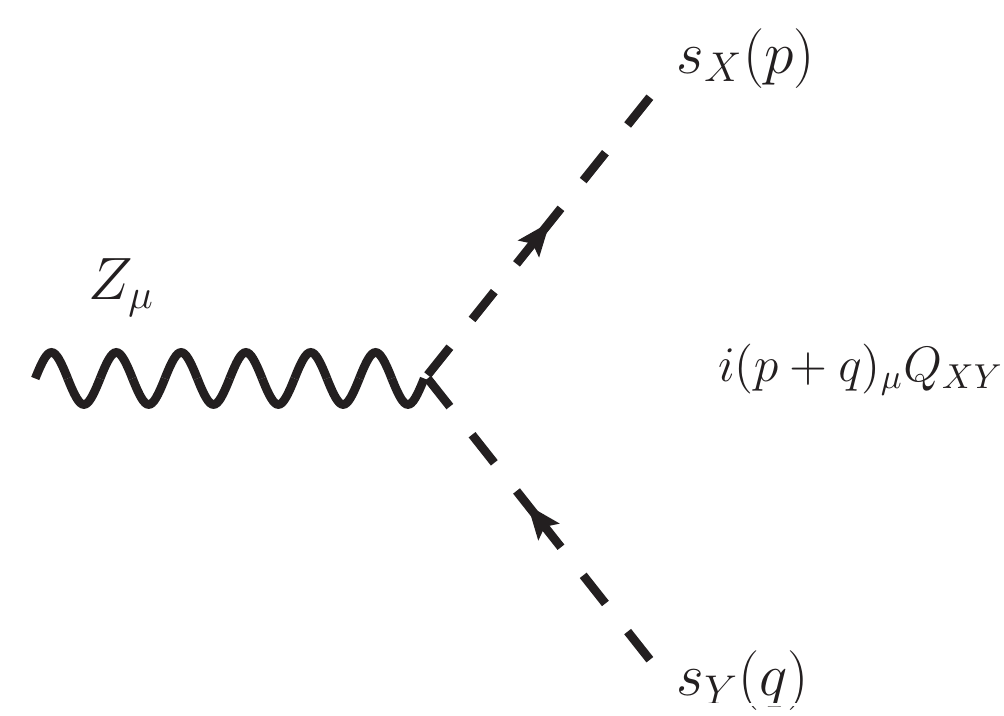}\\
\includegraphics[scale=0.5]{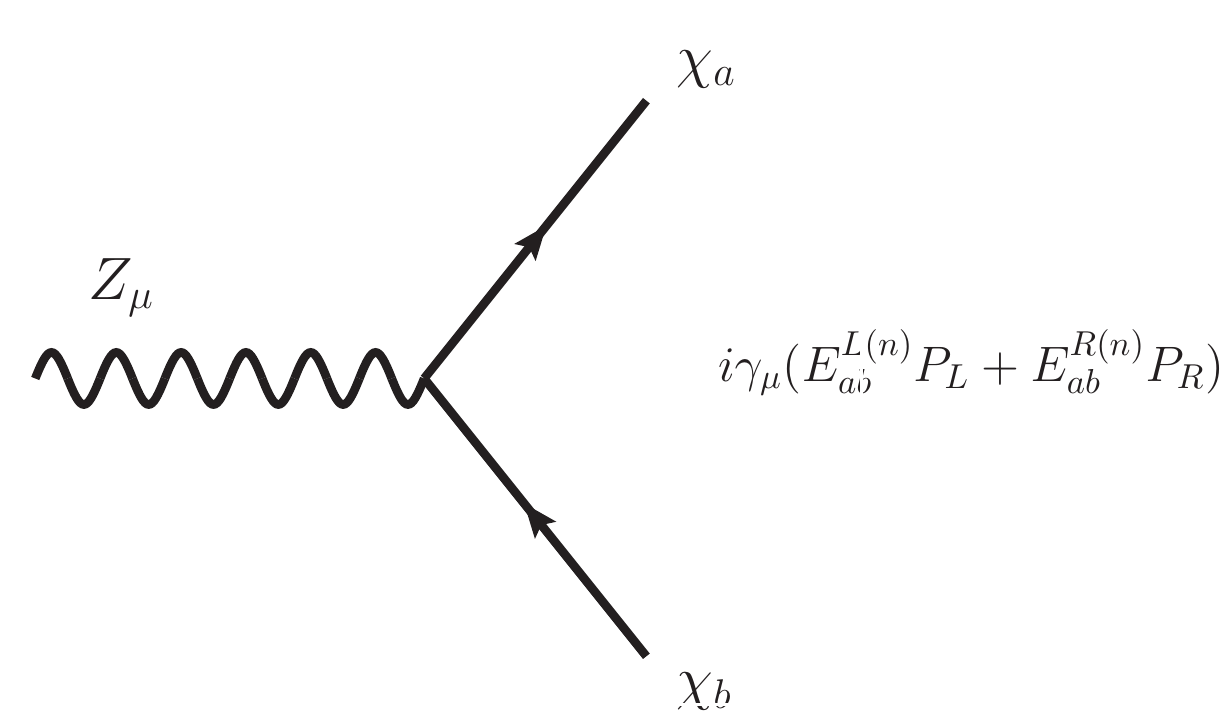}\hspace{3 cm}
\includegraphics[scale=0.5]{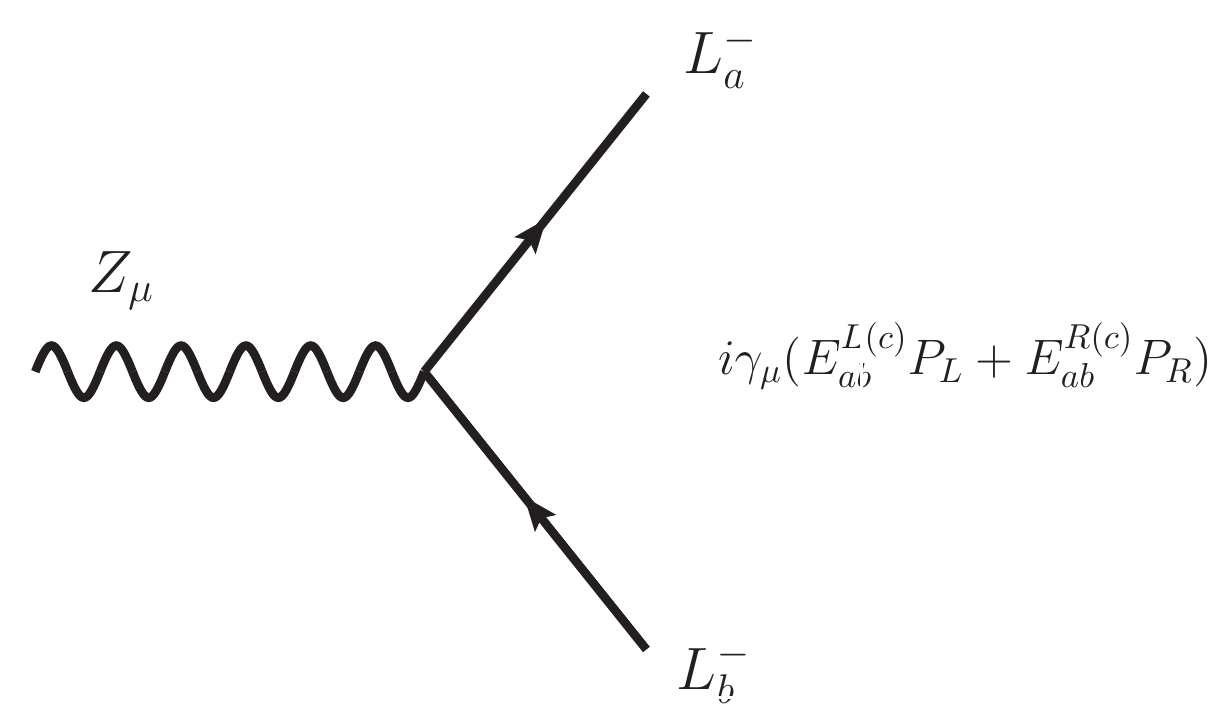}\\
$\bullet$ $g_L^{(\ell)}=-\frac{g}{\cos\theta_W}\left(-\frac{1}{2}+\sin^2\theta_W\right)$, $g_R^{(\ell)}=-\frac{g}{\cos\theta_W}\sin^2\theta_W$ are the left and right chiral couplings among two SM leptons and $Z$ boson respectively, $g$ being the $SU(2)_L$ coupling constant.\\ 
$\bullet$ $Q_{11}=Q_{\eta^0\eta^0}=-\frac{g}{2\cos\theta_W}$ \& $Q_{\eta^+\eta^+}=-\frac{g\,\cos 2\theta_W}{2\cos\theta_W}$, $Q_{22}=Q_{SS}=0$ and $Q_{12}=Q_{21}=0$ are the $Z$-scalar-scalar couplings.\\
$\bullet$ For neutral VL fermions: $E^{L,R(n)}_{11}=-\frac{g}{2\cos\theta_W}\cos^2\theta$, $E^{L,R(n)}_{00}=-\frac{g}{2\cos\theta_W}\sin^2\theta$, $E^{L,R(n)}_{10}=E^{L,R(n)}_{01}=-\frac{g}{2\cos\theta_W}\sin\theta\cos\theta$ and for charged VL fermions: $E^{L,R(c)}_{11}=E^{L,R(c)}_{22}=-\frac{g}{\cos\theta_W}\left(-\frac{1}{2}+\sin^2\theta_W\right)$, $E^{L,R(c)}_{12}=E^{L,R(c)}_{21}=0$ are the $Z$-fermion-fermion couplings. 
\subsection*{Photon interactions}
The Feynman rules for $\gamma$ interactions are given by,\\
\includegraphics[scale=0.5]{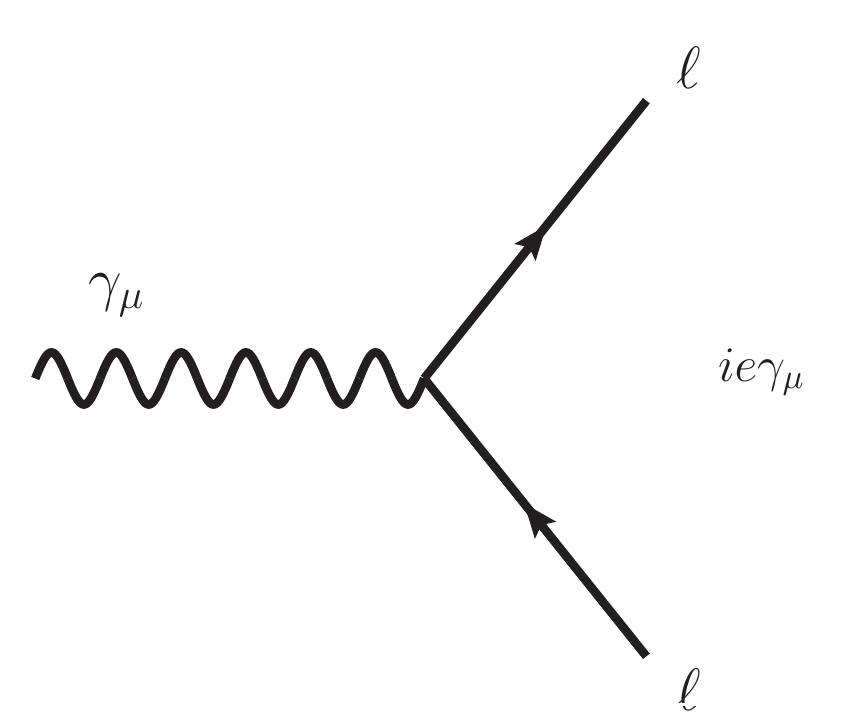}\hspace{2 cm}
\includegraphics[scale=0.5]{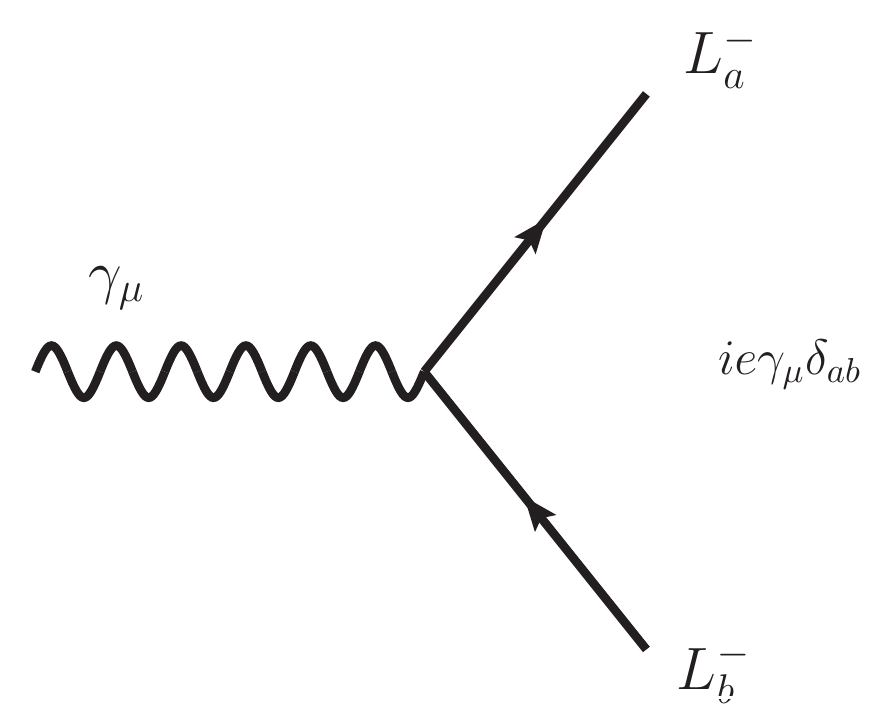}\hspace{2 cm}
\includegraphics[scale=0.5]{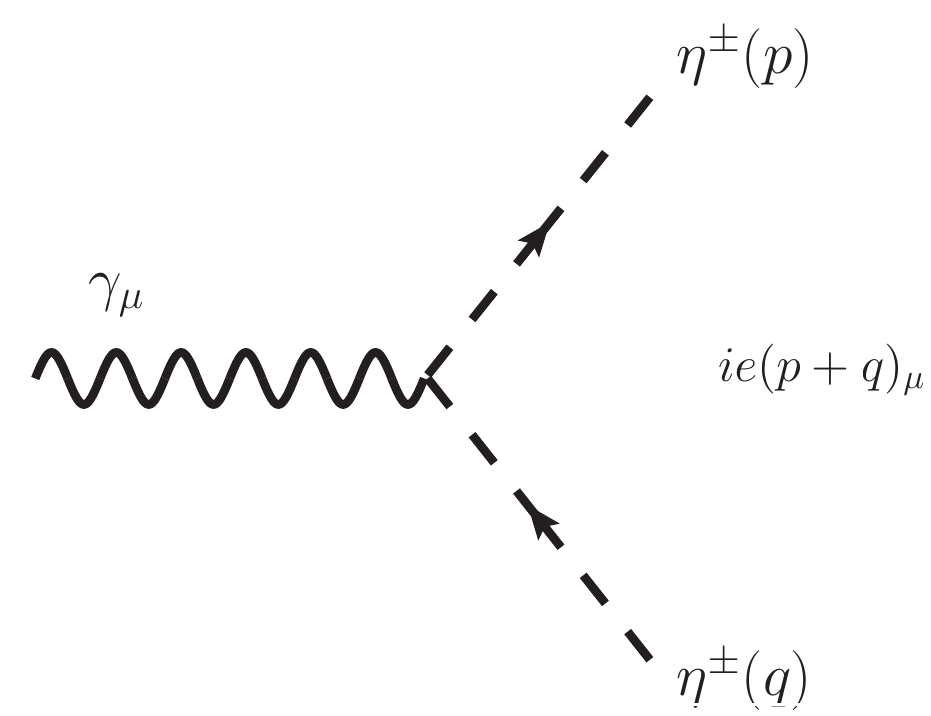}\\

\section*{Appendix B}
\label{sec:appB}
In this appendix, we have listed the explicit forms of all the mass functions. The two point and three functions are defined as,
\begin{align}
B_1(m_1^2,m_2^2)=-\frac{1}{2}+\frac{1}{2}{\rm ln}\,m_2^2-\frac{m_1^2-m_2^2+2m_1^2\,{\rm ln}\left(\frac{m_2^2}{m_1^2}\right)}{4(m_1^2-m_2^2)^2},
\end{align}
\begin{align}
C_0(m_1^2,m_2^2,m_3^2)=-\frac{1}{m_2^2-m_3^2}\Bigg[\frac{m_1^2{\rm ln}\,m_1^2-m_2^2{\rm ln}\,m_2^2}{m_1^2-m_2^2}-\frac{m_1^2{\rm ln}\,m_1^2-m_3^2{\rm ln}\,m_3^2}{m_1^2-m_3^2}\Bigg],
\end{align}
\begin{align}
&4C_{24}(m_1^2,m_2^2,m_3^2)= \tilde{C}_0(m_1^2,m_2^2,m_3^2)+\frac{1}{2}=\frac{3}{2}-\frac{1}{m_2^2-m_3^2}\Bigg[\frac{m_1^4{\rm ln}\,m_1^2-m_2^4{\rm ln}\,m_2^2}{m_1^2-m_2^2}-\frac{m_1^4{\rm ln}\,m_1^2-m_3^4{\rm ln}\,m_3^2}{m_1^2-m_3^2}\Bigg].
\end{align}
The functional forms for the four point functions relevant in case of Box diagrams are given by,
\begin{align}
D_0(m_1^2,m_2^2,m_3^2,m_4^2)=&-\frac{m_1^2\,{\rm ln}\,m_1^2}{(m_1^2-m_2^2)(m_1^2-m_3^2)(m_1^2-m_4^2)} +\frac{m_2^2\,{\rm ln}\,m_2^2}{(m_1^2-m_2^2)(m_2^2-m_3^2)(m_2^2-m_4^2)}\nonumber\\
&-\frac{m_3^2\,{\rm ln}\,m_3^2}{(m_1^2-m_3^2)(m_2^2-m_3^2)(m_3^2-m_4^2)} +\frac{m_4^2\,{\rm ln}\,m_4^2}{(m_1^2-m_4^2)(m_2^2-m_4^2)(m_3^2-m_4^2)},
\end{align}
\begin{align}
\tilde{D}_0(m_1^2,m_2^2,m_3^2,m_4^2)=&-\frac{m_1^4\,{\rm ln}\,m_1^2}{(m_1^2-m_2^2)(m_1^2-m_3^2)(m_1^2-m_4^2)} +\frac{m_2^4\,{\rm ln}\,m_2^2}{(m_1^2-m_2^2)(m_2^2-m_3^2)(m_2^2-m_4^2)}\nonumber\\
&-\frac{m_3^4\,{\rm ln}\,m_3^2}{(m_1^2-m_3^2)(m_2^2-m_3^2)(m_3^2-m_4^2)} +\frac{m_4^4\,{\rm ln}\,m_4^2}{(m_1^2-m_4^2)(m_2^2-m_4^2)(m_3^2-m_4^2)}.
\end{align}
The other functions appearing in the expressions of the dipole and monopole terms of the $\gamma$-penguin are defined as:
\begin{align}
F_1(r)=&\frac{1 - r^2 + 2r{\rm ln}\,r}{2 (1 - r)^3}, \nonumber\\
F_2(r)=&\frac{1-6r+3r^2+2r^3-6r^2{\rm ln}\,r}{6(1-r)^4}, \nonumber\\
F_3(r)=&\frac{2+3r-6r^2+r^3+6r{\rm ln}\,r}{6(1-r)^4}, \nonumber\\
F_4(r)=&\frac{2-9r+18r^2-11r^3+6r^3{\rm ln}\,r}{(1-r)^4}, \nonumber\\
F_5(r)=&\frac{16-45r+36r^2-7r^3+6(2-3r){\rm ln}\,r}{(1-r)^4}.
\end{align}

\section*{Appendix C}
\label{sec:appC}
In this section, we present the general and explicit results for the on-shell and off-shell decays of the charged leptons.

\subsection{$\ell_\alpha\rightarrow \ell_\beta\gamma$}
The on-shell amplitude, mediated by the dipole operators, can be expressed as,
\begin{equation} 
  {\cal L}_{\ell \ell \gamma} \supset e \, \bar u_\beta \left[  i m_{\ell_\alpha}
    \sigma^{\mu \nu} q_\nu \left(A_2^L P_L + A_2^R P_R \right) \right] u_\alpha A_\mu + h.c.
  \label{eq:L_llg}
\end{equation}
Here $e$ is the electric charge, $q$ is the photon momentum,
$P_{L,R}= \frac{1}{2} (1 \mp \gamma_5)$ are the usual chirality projectors and
the lepton spinors are denoted by $u_{\alpha,\beta}$, where $\alpha,\beta$ stand for the flavor indices. The coefficients in Eq.~\eqref{eq:L_llg} can be written as, $A_2^{L,R}=A_2^{(n)L,R}+A_2^{(c)L,R}$, where `$n$' and `$c$' indicate the dipole contributions from neutral and charged fermion loops~[shown in Fig.~\ref{fig:lep_g}] respectively. The general forms for $A_2^{(n)L,R}$ and $A_2^{(c)L,R}$ are given below:
\begin{align}
  A^{(n)L}_{2}=\frac{1}{32\pi^2M^2_\eta}\Bigg[N^L_{\beta a}{N^{R*}_{\alpha a}}
    \left(\frac{2M_{\chi_a}}{m_{\ell_\alpha}}\right)\,F_1\left(\frac{M^2_{\chi_a}}{M^2_\eta}\right)+& 
   N^L_{\beta a}{N^{L*}_{\alpha a}} \,F_2\left(\frac{M^2_{\chi_a}}{M^2_\eta}\right) \nonumber\\
   &+ 
   N^R_{\beta a}{N^{R*}_{\alpha a}} \left(\frac{m_{\ell_\beta}}{m_{\ell_\alpha}}\right)\,
   F_2\left(\frac{M^2_{\chi_a}}{M^2_\eta}\right)\Bigg],
   \label{eq:AnL2}
   \end{align}
   \begin{align}
&A^{(n)R}_{2}=A^{(n)L}_{2}|_{L\leftrightarrow R},
\label{eq:AnR2}
\end{align}
\begin{align}
  &A^{(c)L}_{2}=\frac{1}{32\pi^2M^2_S}C^R_{\beta 22}{C^{R*}_{\alpha 22}}\frac{m_{\ell_\beta}}{m_{\ell_\alpha}}
  F_3\left(\frac{M^2_{L_2}}{M^2_S}\right) 
+\frac{1}{32\pi^2M^2_\eta}C^L_{\beta 11}{C^{L*}_{\alpha 11}} F_3\left(\frac{M^2_{L_1}}{M^2_\eta}\right), \\
&A^{(c)R}_{2}=\frac{1}{32\pi^2M^2_\eta}C^L_{\beta 11}{C^{L*}_{\alpha 11}}\frac{m_{\ell_\beta}}{m_{\ell_\alpha}}
F_3\left(\frac{M^2_{L_1}}{M^2_\eta}\right) 
+\frac{1}{32\pi^2M^2_S}C^R_{\beta 22}{C^{R*}_{\alpha 22}} F_3\left(\frac{M^2_{L_2}}{M^2_S}\right).
\label{eq:AcR2}
\end{align}
\subsection{$\ell_\alpha\rightarrow 3\ell_\beta$}
The amplitude for such a process like $\ell^-_\alpha(p)\rightarrow\ell^-_\beta
(p_1)\ell^-_\beta(p_2)\ell^+_\beta(p_3)$ can be decomposed into three major contributions given by,
\begin{align}
\mathcal{M}(\ell^-_\alpha\rightarrow\ell^-_\beta\ell^-_\beta\ell^+_\beta
) \simeq \mathcal{M}_\gamma+\mathcal{M}_Z+\mathcal{M}_{\rm Box}.
\label{eq:penguin}
\end{align}
In general there should be a contribution from Higgs penguin diagrams~(i.e. $\mathcal{M}_H$) as well, but one can neglect it in most cases, 
in comparison to the other three contributions of Eq.~\eqref{eq:penguin}. Different contributions can be expressed as follows:\\
\begin{itemize}
\item
Photon penguin contribution:  The monopole and dipole contributions can be calculated from,
\begin{align}
\mathcal{M}_\gamma=&\,\bar{u}_{\beta}(p_1)\Bigg[q^2\gamma^\mu(A^L_1P_L+A^R_1P_R)+im_{\ell_\alpha}\sigma^{\mu\nu}q_\nu (A^L_2P_L+A^R_2P_R)\Bigg]u_{\alpha}(p)\nonumber\\
&\times\frac{e^2}{q^2}\bar{u}_{\beta}(p_2)\gamma_\mu v_{\beta}(p_3)-(p_1\leftrightarrow p_2).
\end{align}
The explicit form of the Wilson coefficients $A_2^L$ and $A_2^R$ are already described in Eqs.~\eqref{eq:AnL2}$-$\eqref{eq:AcR2}. 
The coefficients associated with the monopole operator can be calculated as,
\begin{align}
&A_1^{(n)L}=\frac{1}{576\pi^2M_\eta^2}\Bigg[N^R_{\beta a}{N^{R*}_{\alpha a}} \,F_4\left(\frac{M^2_{\chi_a}}{M_\eta^2}\right)\Bigg]~
,~A_1^{(n)R}=A_1^{(n),L}|_{L\leftrightarrow R}\, ,\\
&A_1^{(c)L}=-\frac{1}{576\pi^2M_S^2}\Bigg[C^R_{\beta 22}{C^{R*}_{\alpha 22}} \,F_5\left(\frac{M^2_{L_2}}{M_S^2}\right)\Bigg]~,~
A_1^{(c)R}=-\frac{1}{576\pi^2M_\eta^2}\Bigg[C^L_{\beta 11}{C^{L*}_{\alpha 11}} \,F_5\left(\frac{M^2_{L_1}}{M_\eta^2}\right)\Bigg].
\end{align}
\item
  $Z$ penguin contribution:  Feynman diagrams are shown in Fig.~\ref{fig:Z_peng}. We have calculated the coefficients as follows:
  \begin{align}
    \label{eq:appenZpen}
\mathcal{M}_Z=&\,\frac{1}{M_Z^2}\bar{u}_{\beta}(p_1)\Bigg[\gamma^\mu(F_LP_L+F_RP_R)\Bigg]u_{\alpha}(p)
\bar{u}_{\beta}(p_2)\Bigg[\gamma_\mu(g^{(\ell)}_LP_L+g^{(\ell)}_RP_R)\Bigg] v_{\beta}(p_3)-(p_1\leftrightarrow p_2),
\end{align}
where, as before, $F_{L,R}=F^{(n)}_{L,R}+F^{(c)}_{L,R}$. The expressions for these form factors are given below:
\begin{align}
  F^{(n)}_{L}=&-\frac{1}{16\pi^2}\Bigg[N^R_{\beta b}{N^{R*}_{\alpha a}} \left\lbrace E^{R(n)}_{ba}
    \left( 2C_{24}(M_\eta^2,M_{\chi_a}^2,M_{\chi_b}^2)-\frac{1}{2}\right)-E^{L(n)}_{ba}M_{\chi_a}M_{\chi_b}C_0(M_\eta^2,M_{\chi_a}^2,M_{\chi_b}^2)\right\rbrace \nonumber\\
    &+N^R_{\beta a}{N^{R*}_{\alpha a}} \left\lbrace 2Q_{\eta\eta}C_{24}(M_{\chi_a}^2,M_\eta^2,M_\eta^2)\right\rbrace
    +N^R_{\beta a}{N^{R*}_{\alpha a}} \left\lbrace g_L^{(\ell)}B_1(M_{\chi_a}^2,M_\eta^2)\right\rbrace\Bigg], \\
  F^{(c)}_{L}=&-\frac{1}{16\pi^2}\Bigg[C^R_{\beta aX}{C^{R*}_{\alpha aX}} \left\lbrace E^{R(c)}_{aa}
   \left( 2C_{24}(M_X^2,M_{L_a}^2,M_{L_a}^2)-\frac{1}{2}\right)-E^{L(c)}_{aa}M_{L_a}M_{L_a}C_0(M_X^2,M_{L_a}^2,M_{L_a}^2)\right\rbrace \nonumber\\
    &+C^R_{\beta aX}{C^{R*}_{\alpha aX}} \left\lbrace 2Q_{XX}C_{24}(M_{L_a}^2,M_X^2,M_X^2)\right\rbrace
    +C^R_{\beta aX}{C^{R*}_{\alpha aX}} \left\lbrace g_L^{(\ell)}B_1(M_{L_a}^2,M_X^2)\right\rbrace\Bigg],\\
F^{(n)}_{R}=&F^{(n)}_{L}|_{L\leftrightarrow R}~,~~~~~~~~~~F^{(c)}_{R}=F^{(c)}_{L}|_{L\leftrightarrow R}.
\end{align}
\item
  Box diagram contribution:  Leading contributions are shown in Fig.~\ref{fig:Box}.
\begin{align}
  \mathcal{M}_{\rm Box} \simeq &e^2B^L_1[\bar{u}_\beta(p_1)(\gamma^\mu P_L)u_\alpha(p)][\bar{u}_\beta(p_2)(\gamma_\mu P_L)v_\beta(p_3)]
  \nonumber\\
&+e^2B^R_1[\bar{u}_\beta(p_1)(\gamma^\mu P_R)u_\alpha(p)][\bar{u}_\beta(p_2)(\gamma_\mu P_R)v_\beta(p_3)]\nonumber\\
&+e^2B^L_2\left\lbrace[\bar{u}_\beta(p_1)(\gamma^\mu P_L)u_\alpha(p)][\bar{u}_\beta(p_2)(\gamma_\mu P_R)v_\beta(p_3)]
-(p_1\leftrightarrow p_2)\right\rbrace\nonumber\\
&+e^2B^R_2\left\lbrace[\bar{u}_\beta(p_1)(\gamma^\mu P_R)u_\alpha(p)][\bar{u}_\beta(p_2)(\gamma_\mu P_L)v_\beta(p_3)]-
(p_1\leftrightarrow p_2)\right\rbrace\nonumber \\
&+e^2B^L_3\left\lbrace[\bar{u}_\beta(p_1)( P_L)u_\alpha(p)][\bar{u}_\beta(p_2)( P_L)v_\beta(p_3)]-
(p_1\leftrightarrow p_2)\right\rbrace\nonumber\\
&+e^2B^R_3\left\lbrace[\bar{u}_\beta(p_1)(P_R)u_\alpha(p)][\bar{u}_\beta(p_2)( P_R)v_\beta(p_3)]-
(p_1\leftrightarrow p_2)\right\rbrace\nonumber\\
&+e^2B^L_4\left\lbrace[\bar{u}_\beta(p_1)(\sigma^{\mu\nu}P_L)u_\alpha(p)][\bar{u}_\beta(p_2)( \sigma_{\mu\nu}P_L)v_\beta(p_3)]-
(p_1\leftrightarrow p_2)\right\rbrace\nonumber\\
&+e^2B^R_4\left\lbrace[\bar{u}_\beta(p_1)(\sigma^{\mu\nu}P_R)u_\alpha(p)][\bar{u}_\beta(p_2)( \sigma_{\mu\nu}P_R)v_\beta(p_3)]-
(p_1\leftrightarrow p_2)\right\rbrace,
\end{align}
where, $B_i^{L,R}=B_i^{(n)L,R}+B_i^{(c)L,R}\quad[i=1,2,3,4]$. The $B$-factors for the neutral fermions~($\chi_1$, $\chi_0$)
can be calculated as,
\begin{align}
  &e^2B_1^{(n)L}=\frac{1}{16\pi^2}\Bigg[\frac{\tilde{D}_0}{2}N^R_{\beta a}{N^{R*}_{\alpha a}} N^R_{\beta b}{N^{R*}_{\beta b}} +
    D_0\,M_{\chi_a}M_{\chi_b}N^R_{\beta b}N^R_{\beta b}{N^{R*}_{\alpha a}} {N^{R*}_{\beta a}}\Bigg],
\end{align}
\begin{align}
  e^2B_2^{(n)L}=\frac{1}{16\pi^2}\Bigg[\frac{\tilde{D}_0}{4}N^R_{\beta a}{N^{R*}_{\alpha a}} N^L_{\beta b}{N^{L*}_{\beta b}} -
    \frac{D_0}{2}\,M_{\chi_a}M_{\chi_b}N^L_{\beta a}{N^{R*}_{\alpha a}} N^R_{\beta b} {N^{L*}_{\beta b}} \nonumber\\
    -\frac{\tilde{D}_0}{4}N^L_{\beta b}N^R_{\beta b} {N^{R*}_{\alpha a}} {N^{L*}_{\beta a}} +\frac{\tilde{D}_0}{4}N^R_{\beta b}N^L_{\beta b}
    {N^{R*}_{\alpha a}} {N^{L*}_{\beta a}} \Bigg],
\end{align}
\begin{align}
  &e^2B_3^{(n)L}=\frac{1}{16\pi^2}\Bigg[D_0\,M_{\chi_a}M_{\chi_b}N^L_{\beta a}{N^{R*}_{\alpha a}} N^L_{\beta b}{N^{R*}_{\beta b}}
    + \frac{D_0}{2}\,M_{\chi_a}M_{\chi_b}N^L_{\beta b}N^L_{\beta b}{N^{R*}_{\alpha a}} {N^{R*}_{\beta a}}\Bigg],\\
  &e^2B_4^{(n)L}=\frac{1}{16\pi^2}\Bigg[\frac{D_0}{8}\,M_{\chi_a}M_{\chi_b}{N^{R*}_{\alpha a}} {N^{R*}_{\beta a}} N^L_{\beta b}
    N^L_{\beta b}\Bigg],\\
&B_i^{(n)R}=B_i^{(n)L}|_{L\leftrightarrow R},
\end{align}
where,
\begin{align}
D_0=D_0(M^2_{\chi_a},M^2_{\chi_b},M_\eta^2,M_\eta^2),\qquad\qquad\tilde{D}_0=\tilde{D}_0(M^2_{\chi_a},M^2_{\chi_b},M_\eta^2,M_\eta^2).
\end{align}
And for the charged fermions~($L_1^\pm$, $L_2^\pm$),
\begin{align}
  &e^2B_1^{(c)L}=\frac{1}{16\pi^2}\Bigg[\frac{\tilde{D}_0}{2}\,C^R_{\beta aX} {C^{R*}_{\alpha aX}}
    C^R_{\beta aX} {C^{R*}_{\beta aX}} \Bigg],\\
  &e^2B_2^{(c)L}=\frac{1}{16\pi^2}\Bigg[\frac{\tilde{D}_0}{4}\,C^R_{\beta aX} {C^{R*}_{\alpha aX}}
    C^L_{\beta aX} {C^{L*}_{\beta aX}} -\frac{D_0}{2}\,M_{L_a}M_{L_a}C^L_{\beta aX} {C^{R*}_{\alpha aX}} C^R_{\beta aX} {C^{L*}_{\beta aX}}\Bigg],\\
&e^2B_3^{(c)L}=\frac{1}{16\pi^2}\Bigg[D_0\,M_{L_a}M_{L_a}C^L_{\beta aX} {C^{R*}_{\alpha aX}} C^L_{\beta aX} {C^{R*}_{\beta aX}}\Bigg],\\
&e^2B_4^{(c)L}=0,\\
&B_i^{(c)R}=B_i^{(c)L}|_{L\leftrightarrow R},
\end{align}
where,
\begin{align}
D_0=D_0(M^2_{L_a},M^2_{L_a},M_X^2,M_X^2),\qquad\qquad\tilde{D}_0=\tilde{D}_0(M^2_{L_a},M^2_{L_a},M_X^2,M_X^2),
\end{align}
with $M_1=M_\eta$ and $M_2=M_S$. The generic functional forms for these $D_0$ and $\tilde{D}_0$ are
again available in Appendix B.
  
The decay width for $\ell_\alpha^-\rightarrow \ell_\beta^-\ell_\beta^-\ell_\beta^+$ can be obtained by considering all the possible contributions coming from photon and $Z$ penguins in addition to the box diagrams and can be expressed as~\cite{Hisano_1996,Arganda:2005ji},
\begin{align}
\Gamma(\ell_\alpha^-\rightarrow \ell_\beta^-\ell_\beta^-\ell_\beta^+)=&\frac{e^4\,m^5_{\ell_\alpha}}{512\pi^3}\Bigg[|A^L_1|^2+|A^R_1|^2-2\left(A^L_1{A^R_2}^* +A^L_2{A^R_1}^* +h.c.\right)+\left(|A^L_2|^2+|A^R_2|^2\right)\times\nonumber\\
&\left\lbrace\frac{16}{3}{\rm ln}\left(\frac{m_{\ell_\alpha}}{m_{\ell_\beta}}\right)-\frac{22}{3}\right\rbrace+\frac{1}{6}\left(|B^L_1|^2+|B^R_1|^2\right)+\frac{1}{3}\left(|B^L_2|^2+|B^R_2|^2\right)\nonumber\\
&+\frac{1}{24}\left(|B^L_3|^2+|B^R_3|^2\right)+6\left(|B^L_4|^2+|B^R_4|^2\right)-\frac{1}{2}\left(B^L_3{B^L_4}^* +B^R_3{B^R_4}^* +h.c.\right)\nonumber\\
&+\frac{1}{3}\left(A^L_1{B^L_1}^* +A^R_1{B^R_1}^* +A^L_1{B^L_2}^* +A^R_1{B^R_2}^* + h.c.\right)\nonumber\\
&-\frac{2}{3}\left(A^R_2{B^L_1}^* +A^L_2{B^R_1}^* +A^L_2{B^R_2}^* +A^R_2{B^L_2}^* + h.c.\right)\nonumber\\
&+\frac{1}{3}\Bigg\{2\left(|F_{LL}|^2+|F_{RR}|^2\right)+|F_{LR}|^2+|F_{RL}|^2\nonumber\\
&+\left(B^L_1F_{LL}^* +B^R_1F_{RR}^* +B^L_2F_{LR}^* +B^R_2F_{RL}^* +h.c.\right)\nonumber\\
&+2\left(A^L_1F_{LL}^* +A^R_1F_{RR}^* +h.c.\right)+\left(A^L_1F_{LR}^* +A^R_1F_{RL}^* +h.c.\right)\nonumber\\
&-4\left(A^R_2F_{LL}^* +A^L_2F_{RR}^* +h.c.\right)-2\left(A^L_2F_{RL}^* +A^R_2F_{LR}^* +h.c.\right)\Bigg\}\Bigg],
\end{align}
where, 
\begin{align}
F_{LL}=&\frac{F_Lg^{(\ell)}_L}{g^2\sin^2\theta_W M^2_Z},\qquad\qquad F_{RR}=F_{LL}|_{L\leftrightarrow R},\nonumber\\
F_{LR}=&\frac{F_Lg^{(\ell)}_R}{g^2\sin^2\theta_W M^2_Z},\qquad\qquad F_{RL}=F_{LR}|_{L\leftrightarrow R}.
\end{align}
The corresponding branching ratio can be directly calculated as ${\rm Br}(\ell_\alpha^-\rightarrow \ell_\beta^-\ell_\beta^-\ell_\beta^+)=\tau_\alpha\Gamma(\ell_\alpha^-\rightarrow \ell_\beta^-\ell_\beta^-\ell_\beta^+)$, $\tau_\alpha$ being the lifetime of $\ell_\alpha$.
\end{itemize}
\bigskip
\bibliographystyle{JHEPCust.bst}
\bibliography{VLDM}

\end{document}